\documentclass[a4paper,11pt]{article}
\usepackage{graphicx}
\usepackage{float}
\usepackage[T1]{fontenc}
\usepackage{epsfig}
\usepackage{color}
\usepackage{amsmath,jheppub,mathtools}
\usepackage{subfloat}
\usepackage{amsfonts}
\usepackage{braket}
\usepackage{cleveref}
\usepackage{epstopdf}
\usepackage{caption}
\usepackage{subcaption}
\usepackage{enumitem}
\usepackage[numbers]{natbib}
\usepackage[titletoc,toc,title]{appendix}
\usepackage{hyperref}
\hypersetup{
	colorlinks=true,
	linkcolor=blue,
	filecolor=red,      
	urlcolor=blue,
	citecolor=blue
} 
\usepackage{comment}

\title{\boldmath Holographic Reflected Entropy and Islands in Interface CFTs}

\author[a,a']{ Jaydeep Kumar Basak}
\author[b]{ Debarshi Basu}
\author[c]{ Vinay Malvimat}
\author[d,d']{ Himanshu Parihar}
\author[b]{ Gautam Sengupta}

\affiliation[a]{Department of Physics,
	National Sun Yat-Sen University,
	Kaohsiung 80424, Taiwan}
\affiliation[a']{Center for Theoretical and Computational Physics,
	Kaohsiung 80424, Taiwan}
\affiliation[b]{Department of Physics,
	Indian Institute of Technology Kanpur,
	208016, India}
\affiliation[c]{Asia Pacific Center for Theoretical Physics, 77 Cheongam-ro, Nam-gu, Pohang-si, Gyeongsangbuk-do, 37673, Korea.}
\affiliation[d]{Center of Theory and Computation, National Tsing-Hua University, Hsinchu 30013, Taiwan}
\affiliation[d']{Physics Division, National Center for Theoretical Sciences, Taipei 10617, Taiwan}
\emailAdd{jkb.hep@gmail.com, debarshi@iitk.ac.in, vinay.malvimat@apctp.org, himansp@phys.ncts.ntu.edu.tw, sengupta@iitk.ac.in}

\date{}

\abstract{We investigate the reflected entropy for various mixed state configurations in the two dimensional holographic conformal field theories sharing a common interface (ICFTs).   In the AdS$_3$/ICFT$_2$ framework, we compute the  holographic reflected entropy for the required configurations in the vacuum state of the ICFT$_{\text{2}}$ which is given by twice the  entanglement wedge cross section (EWCS) in a spacetime involving two AdS$_3$ geometries glued along a thin interface brane. Subsequently, we evaluate the EWCS in the bulk geometry involving  eternal BTZ black strings with an AdS$_2$ interface brane, which is dual to an ICFT$_2$ in the thermofield double (TFD) state.  We explore the system from a doubly holographic perspective and determine the island contributions to the reflected entropy in the two dimensional semi-classical description involving two CFT$_{\text{2}}$s coupled to an AdS$_2$ brane. We demonstrate that the results from the island formula match precisely with the bulk AdS$_3$ results in the large tension limit of the interface brane.		We illustrate that the phase structure of the reflected entropy is quite rich involving many novel induced island phases and demonstrate that it obeys the expected Page curve for the reflected entropy  in a radiation bath coupled to the AdS$_2$ black hole.  }

\begin{document} 
	\maketitle

	\flushbottom
	\section{Introduction}

	In recent years,  the measure of entanglement entropy has been central to a novel resolution of the black hole information loss puzzle. This new resolution involves the appearance of certain regions  termed  "islands" in the late time entanglement wedge of a bath collecting the Hawking radiation. This in turn results in  a particular formula to obtain the fine grained entropy of the bath by including the contributions from the island regions  \cite{Penington:2019npb,Almheiri:2019psf,Almheiri:2019hni,Almheiri:2019yqk,Almheiri:2020cfm}. Furthermore, it has been demonstrated that the island formula leads to the expected Page curve for the entanglement entropy of the bath/radiation subsystem and hence indicates towards the unitarity of black hole evaporation process. The  island formula has been demonstrated to naturally arise in the context of doubly holographic models and the holographic duals of  conformal field theories with boundaries (AdS/BCFT scenarios) in certain limits. This AdS/BCFT construction involves a $d$-dimensional strongly coupled conformal field theory with a boundary (BCFT$_d$) which is dual to a bulk AdS$_{d+1}$ spacetime truncated by an end of the world (EOW) brane \cite{Takayanagi:2011zk,Fujita:2011fp}. The holographic entanglement entropy in the AdS/BCFT scenario was demonstrated to naturally contain the island contributions whenever the RT surfaces end on the EOW brane \cite{Chen:2020uac,Chen:2020hmv,Deng:2020ent,Grimaldi:2022suv,Chu:2021gdb,Suzuki:2022xwv}.  
	
	Another interesting related system consists of two conformal field theories that share a common boundary. If the boundary is also conformally invariant then such a quantum system is termed as Interface Conformal Field Theory (ICFT). Furthermore, as described in \cite{Anous:2022wqh} the holographic dual of such a two dimensional ICFT is described by two AdS$_3$ geometries sharing a common AdS$_2$ brane at which the Israel junction conditions are satisfied. 
	Considering the bulk to be semi-classical, it is possible to describe the above model in a two dimensional effective field theory picture.
	
	 Furthermore, the ratio of central charges of the two CFTs plays a crucial role, and in the limit in which the ratio vanishes, the ICFT reduces to a  BCFT \footnote{Note that if the ICFT itself is holographic both the central charges are large c$_{\text{I}}$,c$_{\text{II}}$. In this scenario, the BCFT limit is defined by considering c$_{\text{I}}$$<<$c$_{\text{II}}$ or vice versa such that the ratio goes to zero.}. In \cite{Anous:2022wqh}, the authors determined the entanglement entropy of a subsystem described by two semi-infinite intervals (one in each CFT) of such a holographic ICFT$_2$. They obtained the entanglement entropy by computing the length of the appropriate geodesics in the bulk AdS$_3$ geometry and demonstrated that there are various novel phases, such as the one in which the geodesic double crosses the bulk interface and is partially in both the AdS$_3$ geometries. Following this, the authors also obtained the entanglement entropy in the semi-classical picture in two dimensions using the island formalism. In the large tension limit, the results from the 3d bulk computation and the 2d computation from the island formula match precisely.
	 
	  Furthermore, the entanglement entropy of the semi-infinite intervals in a holographic ICFT in the thermo-field double state obeys the expected Page curve in the context of a 2d black hole on the AdS$_2$ brane induced by an eternal black hole in AdS$_3$. An intriguing feature of this construction involves RT surfaces which cross the interface  AdS$_2$ brane and return to the original AdS$_3$ geometry. These RT surfaces unique to the AdS/ICFT correspondence were demonstrated to be derived from novel replica wormhole saddles for the entanglement entropy which results in what are known as {\textit induced islands} in one of the CFT$_2$  \cite{Afrasiar:2023nir}.

	In the context of the above mentioned AdS/ICFT correspondence, it would be quite interesting to probe further aspects of entanglement and correlations in ICFTs through various other measures described in quantum information theory. An interesting measure in this regard is the reflected entropy which characterizes the correlations between subsystems in holographic quantum theories\footnote{ Recently in \cite{Hayden:2023yij,Basak:2023uix} it was shown that the reflected entropy for certain states does not obey a desired property for any correlation measure which is the monotonicity under partial trace. However, for holographic states, it has been demonstrated to obey the above mentioned property through the entanglement wedge nesting of the dual EWCS\cite{Wall:2012uf,Takayanagi:2017knl,Dutta:2019gen}. So although it might not serve as a correlation measure for generic quantum systems, it is still useful to characterize correlations between subsystems in the context of holography. }. This quantity, introduced in \cite{Dutta:2019gen}, is holographically dual to the cross-section of the entanglement wedge (EWCS) in the dual bulk AdS geometry. Furthermore, the difference between the reflected entropy and the mutual information known as the Markov gap is expected to contain information about tripartite entanglement in the system \cite{Hayden:2021gno,Zou:2020bly,BasakKumar:2022stg}. Hence, this measure is crucial to understanding the deeper entanglement structure of holographic quantum systems especially in the context of black hole information loss paradox. The island contributions to the reflected entropy and the Markov gap have been studied in various interesting scenarios \cite{Chandrasekaran:2020qtn,Li:2020ceg,Li:2021dmf,Lu:2022cgq}. In the present article, we compute the holographic reflected entropy for various mixed state configurations involving adjacent and disjoint intervals in the vacuum and the TFD state of an ICFT$_2$ \footnote{Note that recently reflected entropy of various configurations has been investigated for interface CFTs in a slightly different context in \cite{Kusuki:2022bic,Tang:2023chv}.}. Furthermore, we will also compute the island contributions to the reflected entropies of the above mentioned configurations in the two dimensional semi-classical effective field theory picture and demonstrate that the results obtained match exactly with the corresponding bulk computation in the large tension limit of the interface brane. We will demonstrate that the phase structure of reflected entropy is much richer than that of the entanglement entropy of the corresponding subsystems. Quite interestingly, we will see that in the 3d bulk geometry whenever the RT surfaces cross the interface brane to the second side and then return back to the original AdS$_3$ geometry it leads to induced reflected entropy islands for one of the CFT, similar to the induced entanglement entropy islands resulting from the novel saddles mentioned earlier.
	In the two dimensional effective theory, we will show that these induced reflected entropy islands always correspond to certain asymetric factorizations of the twist correlation functions. Finally, for the configurations involving the TFD state we determine the analogues of the Page curves for the reflected entropy of the bath coupled to the AdS$_2$ black hole induced by the eternal black hole in the three dimensional bulk.

	The paper is organized as follows: In \cref{sec:ICFT} we present a short review of the holographic ICFT$_2$ model considered in this article.  
	Following this, in \cref{sec:EWCS0} we compute the holographic reflected entropy for the adjacent and disjoint intervals in the vacuum state of an  ICFT$_2$ by determining the entanglement wedge cross section (EWCS) in the dual bulk pure AdS$_3$ geometries glued at the interface. 
	Subsequently, in \cref{sec:SRIs0} we obtain the reflected entropy of various configurations explained above in effective two dimensional island perspective and demonstrate that the results from the bulk and the island formulation match precisely in the large brane tension limit. 
	In \cref{sec:EWCST} we compute the holographic reflected entropy of adjacent intervals by analyzing the EWCS in the geometry involving two eternal black hole geometries sharing an AdS$_2$ brane dual to the TFD state of an ICFT. Subsequently we obtain the analogues of the Page curves of the reflected entropy for mixed states in the bath collecting the Hawking radiation from the AdS$_2$ black hole. 
	Furthermore, in \cref{sec:SRIsT} we determine the island contributions for the reflected entropy of the above mentioned subsystems which match exactly with the results from the bulk computations in the large brane tension limit. 
	Finally in \cref{sec:summary} we summarize and present our conclusions.

	\section{Review: Holographic ICFT$_2$}\label{sec:ICFT} 
	From AdS/CFT dictionary it is well known that the vacuum state of a  CFT$_2$  is dual to a pure AdS$_3$   spacetime. Similarly as described in \cite{Anous:2022wqh}, the vacuum state of an   ICFT$_2$  is dual to two pure AdS$_3$   geometries ( with different AdS length scales L$_{\text{I}}$ and L$_{\text{II}}$ ) that are smoothly glued along a thin interface brane with appropriate  Israel–Lanczos junction conditions imposed. Here we briefly review the details of the bulk AdS$_3$   geometry. The bulk action in this scenario is given by
	\begin{align}
		I & =\frac{1}{16 \pi G_N}\left[\int_{\mathcal{B}_{\mathrm{I}}} \mathrm{d}^3 x \sqrt{-g_{\mathrm{I}}}\left(R_{\mathrm{I}}+\frac{2}{L_{\mathrm{I}}^2}\right)+\int_{\mathcal{B}_{\mathrm{II}}} \mathrm{d}^3 x \sqrt{-g_{\mathrm{II}}}\left(R_{\mathrm{II}}+\frac{2}{L_{\Pi}^2}\right)\right] \\
		& +\frac{1}{8 \pi G_N}\left[\int_{\Sigma} \mathrm{d}^2 y \sqrt{-h}\left(K_{\mathrm{I}}-K_{\mathrm{II}}\right)-T \int_{\Sigma} \mathrm{d}^2 y \sqrt{-h}\right]
	\end{align}
	where $G_N$ corresponds to the AdS$_3$   Newton's constant, ${\cal B}_{\text{I},\text{II}}$ denote the bulk AdS$_3$ geometries and $\Sigma$ denotes the EOW brane with tension T. Note that in the above equation g$_{\text{I}}$ and g$_{\text{II}}$ are the metric determinants of the two AdS$_{\text{3}}^{\text{I},\text{II}}$ geometries with AdS length scales given by L$_{\text{I}}$ and L$_{\text{II}}$ respectively and  R$_{\text{I}}$, R$_{\text{II}}$ are the corresponding Ricci scalars. The determinant of the induced metric on the interface brane  is denoted by $h$ and K$_{\text{I}}$, K$_{\text{II}}$ correspond to the extrinsic curvature on either side of the  brane. The two AdS$_3$   geometries have to be joined smoothly at the interface. This is imposed through the standard  Israel junction conditions given by
	\begin{align}\label{juncon}
		\left(K_{\mathrm{I}, a b}-K_{\mathrm{II}, a b}\right)-h_{a b}\left(K_{\mathrm{I}}-K_{\mathrm{II}}\right)=-T h_{a b}.
	\end{align}
	where T is the tension on the interface brane. The second Israel junction condition ensures that  the induced metric $h_{\text{ab}}$ on the interface brane derived from the two AdS$_3$  geometries to be the same. The AdS$_3$  geometries on either side may be expressed as foliation of  AdS$_2$ metrics as described below
	\begin{align}
		\mathrm{d} s_{\mathcal{B}_{i}}^{2}=\mathrm{d} \rho_{i}^{2}+L_{i}^{2} \cosh ^{2}\left(\frac{\rho_{i}}{L_{i}}\right)\left(\frac{\mathrm{d} y_{i}^{2}+\mathrm{d} \tau_{i}^{2}}{y_{i}^{2}}\right),
	\end{align}
	where $i=\text{I},\text{II}$. It may then be shown that the locations of the brane denoted by $\rho_{\mathrm{I},II}^*$ in the two geometries, are related to the tension by the junction condition given in \cref{juncon}. The authors in \cite{Anous:2022wqh} showed that the junction conditions lead to the following relations
	\begin{align}
		\tanh \left(\frac{\rho_{\mathrm{I}}^*}{L_{\mathrm{I}}}\right)=\frac{L_{\mathrm{I}}}{2 T}\left(T^2+\frac{1}{L_{\mathrm{I}}^2}-\frac{1}{L_{\mathrm{II}}^2}\right), \quad \tanh \left(\frac{\rho_{\mathrm{II}}^*}{L_{\mathrm{II}}}\right)=\frac{L_{\mathrm{II}}}{2 T}\left(T^2+\frac{1}{L_{\mathrm{II}}^2}-\frac{1}{L_{\mathrm{I}}^2}\right).
	\end{align}
	As the range of the \textit{tanh} function lies between -1 and 1, the tension is bounded from above and below as follows
	\begin{align}
		T_{\min }=\left|\frac{1}{L_{\mathrm{I}}}-\frac{1}{L_{\mathrm{II}}}\right|<T<\frac{1}{L_{\mathrm{I}}}+\frac{1}{L_{\mathrm{II}}}=T_{\max }
	\end{align}
	Furthermore, the angles made by the brane with the verticals perpendicular to the two boundaries are related to the brane's location $\rho_*$ as follows
	\begin{align}
		\sin \left(\psi_{\mathrm{I}, \mathrm{II}}\right)=\tanh \left(\frac{\rho_{\mathrm{I}, \mathrm{II}}^*}{L_{\mathrm{I}, \mathrm{II}}}\right).
	\end{align}
	In order to take the large tension limit of the brane appropriately, T is parametrized by $\delta$ as described below
	\begin{align}\label{T-max}
		T^2=\frac{1}{L_{\mathrm{I}}^2}+\frac{1}{L_{\mathrm{II}}^2}+\frac{2-\delta^2}{L_{\mathrm{I}} L_{\mathrm{II}}}
	\end{align}
	It is clear from the above expression that the limit $\delta\to 0$ the tension is maximum $T\to T_{max}$. In this limit, the angles $\psi_{\text{I,II}}$ may be expanded as  \cite{Anous:2022wqh}
	\begin{align}\label{angle-small}
		\psi_\text{I}=\frac{\pi}{2}-\frac{L_\text{I}}{L_\text{I}+L_{\text{II}}}\delta+\mathcal{O}\left(\delta^2\right)~~,~~\psi_{\text{II}}=\frac{\pi}{2}-\frac{L_{\text{II}}}{L_\text{I}+L_\text{II}}\delta+\mathcal{O}\left(\delta^2\right)
	\end{align}
	Furthermore, as described in \cite{Anous:2022wqh}, the holographic dual of the bulk geometry described above is given by two 2D conformal field theories with large central charges ($c_{\text{I},\text{II}}>>1$) interacting via a quantum dot ( holographically dual to the theory on brane ). The central charges in such an interface CFT (ICFT$_2$) are related to the bulk Newton's constant via the Brown Hennaux formula \cite{Brown:1986nw}
	\begin{align}
		c_{i}=\frac{3L_{i}}{2 G_{N}},
	\end{align}
	where $i=\text{I},\text{II}$.
	In the intermediate picture the two non-gravitating CFTs are coupled to a gravitating theory on the brane and the entanglement entropy in such an effective picture is described by the island formalism. The large tension limit discussed above is significant because it is in this limit the holographic entanglement entropy of any subsystem computed from the bulk geometry precisely matches with the corresponding result obtained from the island formula in the two dimensional effective field  theory involving two  CFTs coupled to the brane.

	\section{EWCS in Poincar\'e AdS$_3$ dual to vacuum state of ICFT$_2$}\label{sec:EWCS0}
	In this section, we determine the minimal (extremal) entanglement wedge cross section (EWCS) in the above described bulk AdS$_3$ geometry dual to half of the reflected entropy  of mixed state configurations involving adjacent and disjoint subsystems in an ICFT$_{\text{2}}$\footnote{Note that for a different set of mixed state configurations, quite recently EWCS dual to the vacuum state of an ICFT$_{\text{2}}$ on a circle has been determined  in \cite{Tang:2023chv}.}. Furthermore, we will determine the corresponding expressions for EWCS in the large tension limit of the interface brane. In the subsequent sections we will demonstrate that the results derived from the island formula for the reflected entropy match precisely with the corresponding expressions for twice the area of EWCS obtained in the large tension limit.  
	
	\subsection{Adjacent Subsystems }
	In this section, we compute the minimal (extremal) entanglement wedge cross section corresponding to two adjacent subsystems in a holographic ICFT$_2$. Consider the bipartite mixed state configuration of two adjacent subsystems $A$ and $B$ at a constant time slice $\tau=\tau_0$, described by
	\begin{align}
		A=\big[\tilde{b}_1,\tilde{b}_2\big]_{\text{I}}\cup \big[b_1,b _2\big]_{\text{II}} ~~\text{and} ~~B=\big[\tilde{b}_2,\tilde{b}_3\big]_{\text{I}}\cup \big[b_2,b_3\big]_{\text{II}}\notag,
	\end{align}
	where the subscripts $\text{I,II}$ denote whether the subsystem resides in the CFT$_\text{I}$ or CFT$_\text{II}$. The schematics of this configuration is depicted in \cref{Adjacent-bulk-I}. 
	%
	The computation of the minimal EWCS for $A\cup B$ consists of two parts. As there are many choices for the RT surface corresponding to a subsystems $A\cup B$, the first step involves determining all such RT saddles and their  corresponding entanglement wedges. Now, depending on the size of the subsystems and their distances from the interface, EWCS can have many different phases within each RT saddle or the entanglement wedge of $A\cup B$. In the following, we will divide the possible configurations of RT saddles into two sub-classes, namely those corresponding to the RT surfaces crossing the EOW brane once and those where multiple crossovers are possible \cite{Anous:2022wqh}. For each phase of the RT saddle we will construct the bulk entanglement wedge and subsequently compute the corresponding minimal (extremal) cross-section dual to the reflected entropy $S_R(A:B)$.

	\subsubsection{Configurations involving single crossover of RT surfaces}\label{311}
	In this subsection, we consider the cases where the RT surface crosses the EOW brane once and subsequently compute the minimal EWCS for various phases in that RT saddle. In particular, the RT surfaces connecting the ends points of the subsystems on both CFTs consists of circular segments\footnote{Such RT saddles have already been considered in \cite{Anous:2022wqh}, where the authors utilized techniques from hyperbolic geometry to obtain the lengths of these surfaces. In the following, however, we will use an alternative method more suited to our purpose and find agreement with earlier results.} which cross the EOW brane at the points $y_1$ and $y_3$. The entanglement wedge  and the RT surfaces are depicted by the shaded region and the green curves respectively in \cref{Adjacent-bulk-I}. Note that, according to the Israel junction conditions \cite{Anous:2022wqh}, the distances $y_{1,3}$ along the EOW brane are identical as seen from either AdS$_3$ geometry.

	To find to the length of the RT surface, we utilize the fact that the length of a geodesic connecting two bulk points $(\tau_1,x_1,z_1)$ and $(\tau_2,x_2,z_2)$ in the Poincar\'e AdS$_3$ geometry is given by 
	\begin{align}\label{Poincare-geod}
		d=L\cosh^{-1}\left[\frac{(x_1-x_2)^2+(\tau_1-\tau_2)^2+z_1^2+z_2^2}{2z_1z_2}\right],
	\end{align}
	where $L$ is the AdS$_3$ length scale. The Poincar\'e coordinates of the points $y_i$ on the EOW brane as seen from the AdS$_3^{\text{II}}$ and AdS$_3^{\text{I}}$ geometry respectively, are given by
	\begin{align}
		\left(\tau_0,y_i\sin\psi_{\text{II}},y_i\cos\psi_{\text{II}}\right)&~~ \text{from AdS$_3^\text{II}$}\,,\notag\\
		\left(\tau_0,y_i\sin\psi_{\text{I}},y_i\cos\psi_{\text{I}}\right) &~~\text{from AdS$_3^\text{I}$}\,,\label{coord-1}
	\end{align}
	where $\psi_\text{I,II}$ are the angles made by the EOW brane with the holographic directions $z_i$ in each AdS$_3$ geometry. Therefore, the total length of the geodesic segments connecting the points $\tilde{b}_{1,3}$ and $b_{1,3}$ may be obtained using \cref{Poincare-geod} as
	
	\begin{equation}\label{Adj-bulk-EE-single}
		\begin{aligned}
			d=&L_{\text{I}}\log\left[\frac{\big(\tilde{b }_1+y_1\sin \psi_\text{I}\big)^2+\left(y_1 \cos \psi_\text{I}\right)^2}{y_1 \cos \psi_\text{I}}\right]+L_{\text{II}} \log \left[\frac{\left(b _1+y_1 \sin \psi_{\text{II}}\right)^2+\left(y_1 \cos \psi_{\text{II}}\right)^2}{y_1 \cos \psi_{\text{II}}}\right]\\
			&+L_{\text{I}}\log\left[\frac{\big(\tilde{b }_2+y_2\sin \psi_\text{I}\big)^2+\left(y_2 \cos \psi_\text{I}\right)^2}{y_2 \cos \psi_\text{I}}\right]+L_{\text{II}} \log \left[\frac{\left(b _2+y_2 \sin \psi_{\text{II}}\right)^2+\left(y_2 \cos \psi_{\text{II}}\right)^2}{y_2 \cos \psi_{\text{II}}}\right].
		\end{aligned}
	\end{equation}
	Extremizing with respect to $y_{1,3}$, the locations of the crossing points may be expressed as\footnote{To extremize the above expression, we are required to impose the Israel-Lanczos junction condition \cite{Anous:2022wqh} ${L_\text{I}\,\sec\psi_\text{I}}={L_{\text{II}}\,\sec\psi_\text{II}}$.}
	
	\begin{equation}\label{y*-entropy}
		\begin{aligned}
			y_{i}^*= \frac{\big(b _{i}-\tilde{b}_{i}\big) \sin \left(\frac{\psi_\text{I}-\psi_{\text{II}}}{2} \right)+\sqrt{(b _{i}-\tilde{b }_{i})^2 \sin ^2\left(\frac{\psi_\text{I}-\psi_{\text{II}}}{2} \right)+4 b _{i} \tilde{b }_{i}\cos^2\left(\frac{\psi_\text{I}+\psi_{\text{II}}}{2}\right)}}{2\cos\left(\frac{\psi _\text{I}+\psi _{\text{II}}}{2} \right)}\,,
		\end{aligned}
	\end{equation}
	where $i=1,3$. Substituting the above extremal values in \cref{Adj-bulk-EE-single} and subsequently utilizing the Ryu-Takayanagi prescription \cite{Ryu:2006bv}, we may obtain the entanglement entropy for $A\cup B$ when the single crossing RT saddles dominate.
	\subsubsection*{Phase-I}\label{sec:adj-I}
	\begin{figure}[ht]
		\centering
		\includegraphics[scale=0.55]{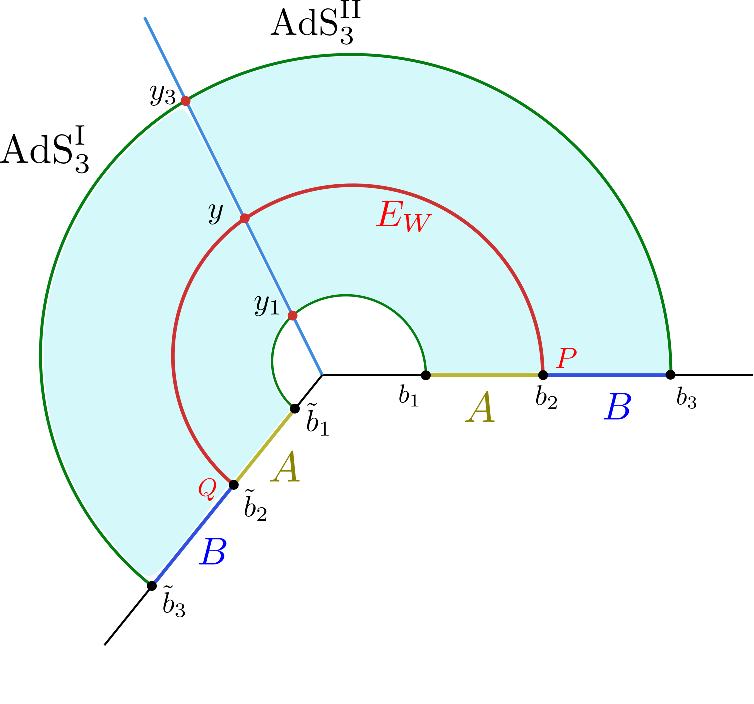}
		\caption{Configuration of single crossing RT saddles for adjacent subsystems in the boundary  CFT$_2$ s. }
		\label{Adjacent-bulk-I}
	\end{figure} 
	In phase-I the subsystems $A$ and $B$ are comparable and close to the interface. The candidate for the minimal EWCS depicted by the red curve in \cref{Adjacent-bulk-I}, is given by two circular geodesic segments connecting the points $P=(\tau_0,b_2)$ and $Q=(\tau_0,\tilde{b}_2)$ on both sides of the interface which meet smoothly\footnote{The smoothness of the geodesics segments across the EOW brane is a consequence of the Israel-Lanczos gluing conditions \cite{Anous:2022wqh}.} at the EOW brane at the point which is at a distance $y$ from the interface. The total length of the geodesic segments connecting the points $P$ and $Q$ may now be obtained using \cref{Poincare-geod} and \cref{coord-1} as
	\begin{equation}\label{geod-1}
		d_{PQ}=L_{\text{I}}\log\left[\frac{\big(\tilde{b }_2+y\sin \psi_\text{I}\big)^2+\left(y \cos \psi_\text{I}\right)^2}{\epsilon \, y \cos \psi_\text{I}}\right]+L_{\text{II}} \log \left[\frac{\left(b _2+y \sin \psi_{\text{II}}\right)^2+\left(y \cos \psi_{\text{II}}\right)^2}{\epsilon \, y \cos \psi_{\text{II}}}\right]\,.
	\end{equation}
	Extremizing the total length with respect to $y$, the location of the crossing point is given by \cref{y*-entropy} with $i=2$. We now consider the large tension limit $T\to T_{\text{max}}$ described in \cref{T-max} where the EOW brane is pushed towards the asymptotic boundary.  We may now utilize \cref{angle-small} to obtain the minimal EWCS in the large tension limit $\delta\to 0$, for this phase as follows
	\begin{equation}\label{Adj-I-EW}
		E_W(A:B)=\frac{L_{ I}}{4G_N} \log \left[ \frac{(y^* + \tilde{b}_2)^2}{2y^* \epsilon} \right]+ \frac{L_{ \text{II}}}{4G_N} \log \left[ \frac{(y^* + b_2)^2}{2y^* \epsilon}  \right]+S_{\text{int}}^{(\delta)}\,,
	\end{equation}
	where the location of the intersection point $y^*$ is now given by
	\begin{equation}\label{Adj-y*-I-EW}
		y^*=\frac{\left(L_{\text{II}}-L_\text{I}\right) \big(b _2-\tilde{b }_2\big)+\sqrt{\left(L_{\text{II}}-L_\text{I}\right)^2 \big(b _2-\tilde{b }_2\big)^2+4b_2\tilde{b}_2\left(L_{\text{I}}+L_\text{II}\right)^2}}{2\left(L_{\text{I}}+L_\text{II}\right)}\,,
	\end{equation}
	and $S^{(\delta)}_\text{int}$ is the large tension limit of the interface entropy $S_\text{int}=\frac{\rho_{\text{I}}^*+\rho_{\text{II}}^*}{4 G_N}$, defined as 
	\begin{align}
		S^{(\delta)}_\text{int}&=\frac{c_\text{I}}{6}\log\left[\frac{2(L_\text{I}+L_\text{II})}{L_\text{I}\,\delta}\right]+\frac{c_\text{II}}{6}\log\left[\frac{2(L_\text{I}+L_\text{II})}{L_\text{II}\,\delta}\right]+\mathcal{O}\left(\delta\right)\,.\label{S-int}
	\end{align}
	
	\subsubsection*{Phase-II}
\begin{figure}
		\centering
		\begin{subfigure}[b]{0.45\textwidth}
			\centering
			\includegraphics[width=\textwidth]{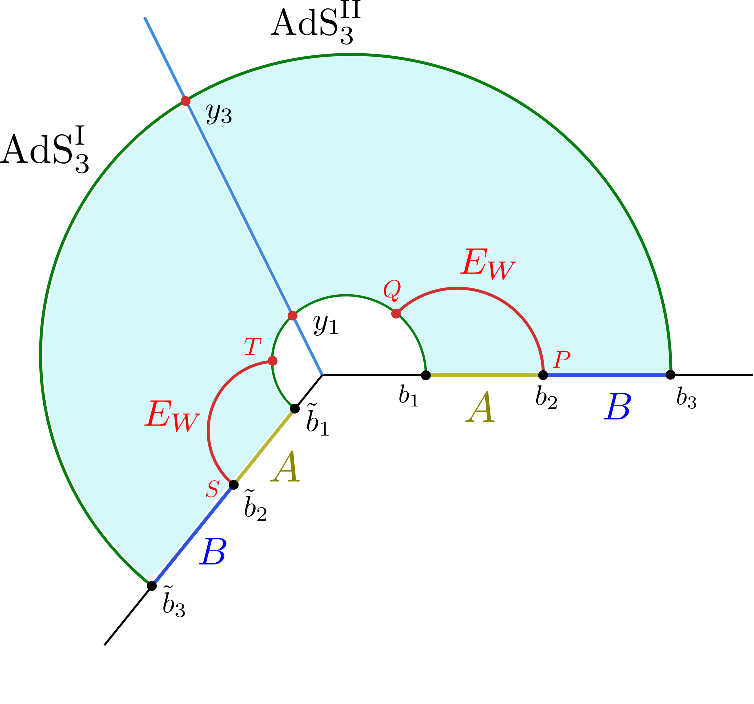}
			\caption{}
			\label{Adjacent-bulk-II}
		\end{subfigure}
		\hfill
		\begin{subfigure}[b]{0.45\textwidth}
			\centering
			\includegraphics[width=\textwidth]{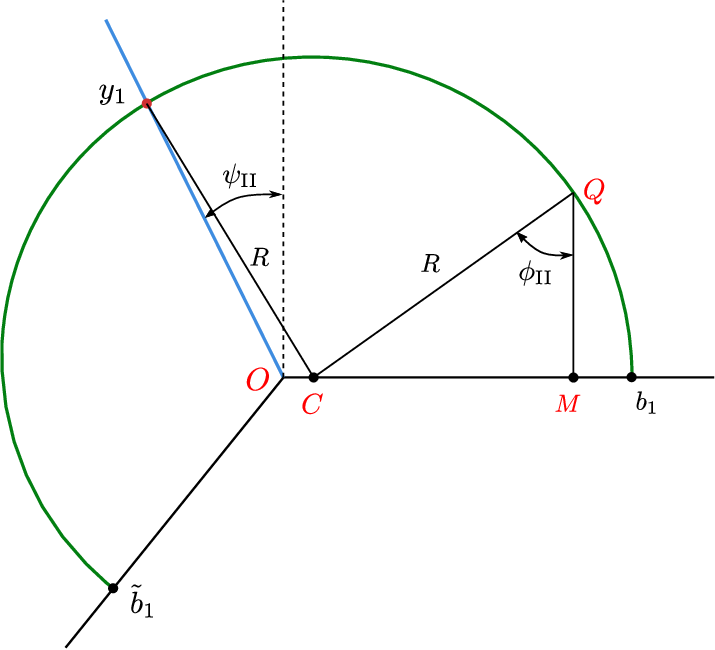}
			\caption{}
			\label{Adjacent-bulk-II-a}
		\end{subfigure}
		\caption{Adjacent subsystems: phase-II.}
		
	\end{figure}

	In phase-II, we consider the subsystem $A$ to be smaller compared to $B$. The minimal EWCS in this phase depicted by red curve in \cref{Adjacent-bulk-II},  consists of two circular arcs each connecting the common boundary of $A$ and $B$ and the smaller RT surface joining $b_1$ and $\tilde{b}_1$ on either side of the EOW brane. 
	%
	%
	The coordinates of the end point $Q$ of the geodesic segment $PQ$ in the AdS$_3^{\text{II}}$ may be parametrized by an angle $\phi_\text{II}$ as follows
	\begin{equation}\label{cord-b2-y}
		Q:(\tau_\text{II},x_\text{II},z_\text{II})=(\tau_0,\overline{OM},\overline{MQ})=\left(\tau_0,b_1-R+R\sin \phi_\text{II},R\cos \phi_\text{II}\right),
	\end{equation}
	where the overline in $OM$ and $MQ$ simply denote that they are Euclidean distances (here we have followed the notation in \cite{Anous:2022wqh}), $R$ is the radius of the circular arc joining the points $b_1$ and $y_1^*$ on the AdS$_3^{\text{II}}$ side as shown in \cref{Adjacent-bulk-II-a}.  Note that, as described earlier, the location of the point $y_1^*$ is given in \cref{y*-entropy}.
	
	%
	%
	Similarly for the AdS$_3^{\text{I}}$ region, the coordinates of the point $T$ may be parametrized by an arbitrary angle $\phi_\text{I}$ as
	\begin{equation}
		T:(\tau_\text{I},x_\text{I},z_\text{I})=\big(\tau_0,\tilde{b}_1-r+r\sin \phi_\text{I},r\cos \phi_\text{I}\big)\,,
	\end{equation}
	where, $r$ is the radius of the circular arc joining the points $\tilde{b}_1$ and $y_1^*$ on the AdS$_3^{\text{II}}$ side. To compute the radii $R$ and $r$, we utilize the equations of the circular arc as follows
	\begin{equation}
		\begin{aligned}
			R^2=(b_1-R)^2-2 y_1^* (b_1-R) \cos \left(\psi_{\text{II}}+\frac{\pi }{2}\right)+y_1^{*^2}\,,\\
			r^2=\big(r-\tilde{b}_1\big)^2-2 y_1^* \big(r-\tilde{b}_1\big)\cos \left(\frac{\pi }{2}-\psi_\text{I}\right)+y_1^{*^2}\,,
		\end{aligned}
	\end{equation}
	which may be solved to obtain
	\begin{equation}\label{R-r}
		\begin{aligned}
			R= b_1+\frac{y_1^{*^2}-b_1^2}{2 (b_1+y_1 \sin \psi_{\text{II}})}~~,~~
			r= \tilde{b}_1+\frac{y_1^{*^2}-\tilde{b}_1^2}{2 (\tilde{b}_1+y_1 \sin \psi_{\text{I}})}.
		\end{aligned}
	\end{equation}

	Now using \cref{Poincare-geod}, the total length of the circular arcs is obtained to be
	\begin{equation}
		\begin{aligned}
			d&=d_{PQ}+d_{ST}\\
			&=L_\text{I} \log \left[\frac{\big(\tilde{b}_2-\big(\tilde{b}_1+r \sin \phi_\text{I}-r\big)\big)^2+(r \cos \phi_\text{I})^2}{r \epsilon \cos \phi_\text{I}}\right]\\
			&\hspace{3cm}+L_{\text{II}} \log \left[\frac{\left(b_2-\left(b_1+R \sin \phi_\text{II}-R\right)\right)^2+(R \cos \phi_\text{II})^2}{R \epsilon \cos \phi_\text{II}}\right]\,.
		\end{aligned}
	\end{equation}
	Extremizing the above length with respect to the arbitrary angles $\phi_\text{I}$ and $\phi_\text{II}$, we have
	\begin{align}
		\phi_\text{I}=\tan ^{-1}\left[\frac{2 r \big(r+\tilde{b}_2-\tilde{b}_1\big)}{\big(\tilde{b}_2-\tilde{b}_1\big) \big(2 r+\tilde{b}_2-\tilde{b}_1\big)}\right]~~,~~\phi_\text{II}=\tan ^{-1}\left[\frac{2 R \left(R-b_1+b_2\right)}{\left(b_2-b_1\right) \left(2 R+b_2-b_1\right)}\right]\,.
	\end{align}
	Substituting the above extremal values, the total minimal EWCS may be obtained as
	\begin{align}\label{geod-length-phase-II}
		E_W(A:B)&=\frac{L_\text{I}}{4G_N}\log \left[\frac{\big(\tilde{b}_2-\tilde{b}_1\big) \big(2 r+\tilde{b}_2-\tilde{b}_1\big)}{\epsilon\,r}\right]+\frac{L_{\text{II}}}{4G_N}\log \left[\frac{\left(b_2-b_1\right) \left(2 R+b_2-b_1\right)}{\epsilon\,R}\right]\notag\\
		&=\frac{L_\text{I}}{4G_N}\log \left[\frac{2 \big(\tilde{b}_2-\tilde{b}_1\big) \big(y_1^{*^2}+\tilde{b}_1 \tilde{b}_2+y_1^* \big(\tilde{b}_1+\tilde{b}_2\big) \sin \psi _\text{I}\big)}{\epsilon \left(y_1^{*^2}+\tilde{b}_1^2+2 y_1^* \tilde{b}_1 \sin \psi _\text{I}\right)}\right]
		\notag\\
		&\hspace{3cm}+\frac{L_{\text{II}}}{4G_N}\log \left[\frac{2 \left(b_2-b_1\right) \left(y_1^{*^2}+b_1 b_2+ y_1^*\left(b_1+b_2\right) \sin\psi _\text{II}\right)}{\epsilon \left(y_1^{*^2}+b_1^2+2 b_1 y_1^* \sin \psi _\text{II}\right)}\right]\,.
	\end{align}
	
	We now utilize \cref{angle-small} to obtain the minimal EWCS in the large tension limit $\delta\to 0$, as follows
	\begin{equation}\label{Adj-sing-II-EW}
		E_W(A:B)=\frac{L_\text{I}}{4G_N} \log \left[\frac{2 \big(\tilde{b}_2-\tilde{b}_1\big) \big(\tilde{b}_2+y_1^*\big)}{\epsilon\big(\tilde{b}_1+y_1^*\big)}\right]+\frac{L_{\text{II}}}{4G_N}  \log \left[\frac{2 (b_2-b_1) (b_2+y_1^*)}{\epsilon(b_1+y_1^*)}\right].
	\end{equation}

	\subsubsection*{Phase-III}
\begin{figure}[ht]
		\centering
		\includegraphics[scale=0.52]{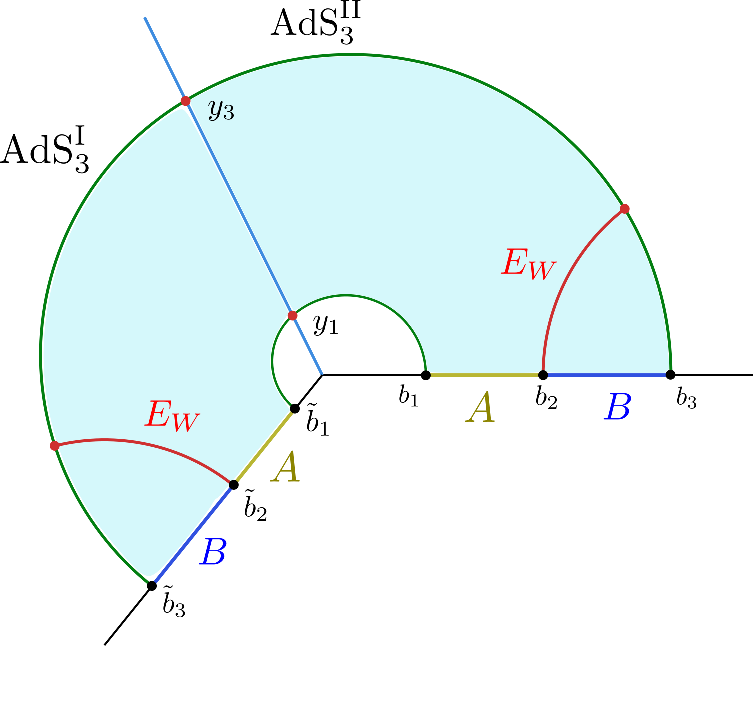}
		\caption{Adjacent subsystems: phase-III.}
		\label{Adjacent-bulk-III}
	\end{figure}

	In phase III, the subsystem $B$ is small compared to $A$ and the minimal EWCS lands on the outer RT surface on both AdS$_3$ regions as depicted in \cref{Adjacent-bulk-III}. 
	The computation of the minimal EWCS for this phase follows a procedure similar to the previous subsection. The total length of the candidate EWCS in this case is given by
	\begin{equation}\label{length-3}
		\begin{aligned}
			d&=L_\text{I} \log \left[\frac{\big(\tilde{b}_2-\big(\tilde{b}_3+r \sin \phi_\text{I}-r\big)\big)^2+(r \cos \phi_\text{I})^2}{\epsilon\,r \cos \phi_\text{I}}\right]\\
			&\hspace{3cm}+L_{\text{II}} \log \left[\frac{\left(b_2-\left(b_3+R \sin \phi_\text{II}-R\right)\right)^2+(R \cos \phi_\text{II})^2}{\epsilon\,R \cos \phi_\text{II}}\right]\,.
		\end{aligned}
	\end{equation}
	In the above expression, we have parametrized two arbitrary points on the RT surface connecting $\tilde{b}_3$ and $b_3$ on the asymptotic boundary. As earlier, $r$ and $R$ denote the radii of the circular geodesic segments joining the set of points $(\tilde{b}_3,y_3^*)$ and $({b}_3,y_3^*)$ respectively. These radii may be obtained by utilizing the equations of the respective circular segments, similar to the previous subsection (cf. \cref{R-r}).
	
	Now extremizing \cref{length-3} with respect to the arbitrary angles $\phi_\text{I}$ and $\phi_\text{II}$, the minimal EWCS in phase III may be obtained by the following replacements: $y_1^*$ with $y_3^*$, and $b_1\,, \tilde{b}_1$ with $b_3\,, \tilde{b}_3$ in \cref{geod-length-phase-II}. In the large tension limit, the EWCS reduces to
	\begin{equation}\label{Adj-sing-III-EW}
		E_W(A:B)=\frac{L_\text{I}}{4G_N}\log \left[\frac{ 2(\tilde{b}_3-\tilde{b}_2) (\tilde{b}_2+y_3^*)}{\epsilon (\tilde{b}_3+y_3^*)}\right]+\frac{L_{\text{II}}}{4G_N}\log \left[\frac{ 2(b_3-b_2) (b_2+y_3^*)}{\epsilon(b_3+y_3^*)}\right].
	\end{equation}
	\subsubsection{Configurations involving double crossing of RT surfaces}
	\begin{figure}[ht]
		\centering
		\includegraphics[scale=0.55]{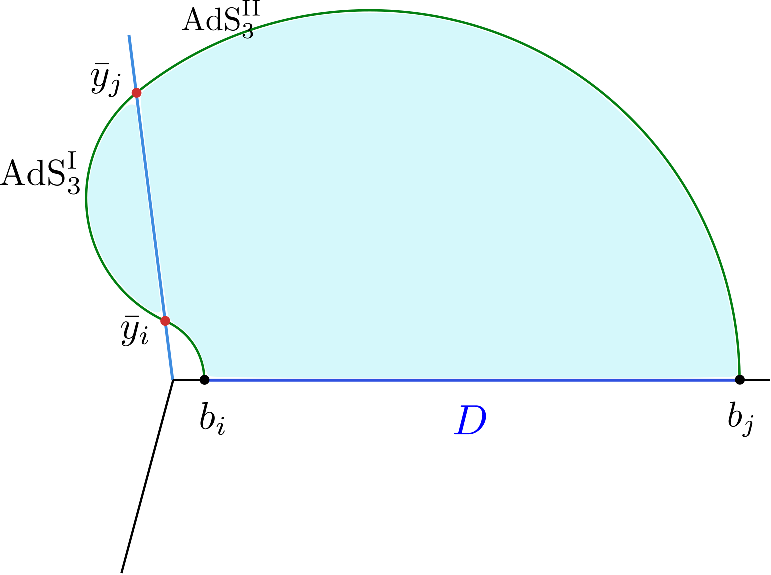}
		\caption{Schematics of the (double-crossing) bulk geodesic homologous to the subsystem $D$ described by a finite subsystem $[b_i,b_j]$ in dual CFT$_\text{II}$. }
		\label{Adjacent-bulk-crossing-RT}
	\end{figure} 
	We now consider the RT saddles homologous to $A\cup B$ which cross the EOW brane multiple times before ending on either of the asymptotic boundaries. Recall that, following the convention in \cite{Anous:2022wqh}, we have set $c_\text{I}<c_\text{II}$. With this convention, it was demonstrated in \cite{Anous:2022wqh} that for a sufficiently large subsystem in the CFT$^\text{II}_2$, there exists at least one such geodesic homologous to the subsystem which finds it more efficient to cross the EOW brane, traverse a finite distance in the AdS$^\text{I}_3$ geometry and then returns to the AdS$^\text{II}_3$ geometry\footnote{Note that, it was further argued in \cite{Anous:2022wqh} that the RT saddles crossing the brane more than twice always have greater length and hence do not contribute to the correlation functions or the entanglement entropy at the leading order.}. The computation of the length of such ``double-crossing'' geodesics was outlined in the appendix of \cite{Anous:2022wqh} utilizing purely geometrical methods. In the following, however, we pursue a different route more suited to our purpose and find agreement with their result. 
	
	Consider a subsystem $D=[b_i,b_j]_\text{II}$ entirely in the CFT$^\text{II}_2$. The double crossing RT saddle homologous to $D$ consists of three semi-circular geodesic arcs as sketched in \cref{Adjacent-bulk-crossing-RT}; two of them connect $b_{i,j}$ with arbitrary bulk points $\bar{y}^{}_{i,j}$ on the brane\footnote{We have denoted the locations of the points where the geodesics cross the brane by $\bar{y}_k$ to emphasize that these points are, in principle, different than those corresponding to a pair of single crossing geodesics emanating from $b_k$.} and the third arc connecting $\bar{y}^{}_i$ with $\bar{y}^{}_j$ residing entirely in the AdS$^\text{I}_3$ geometry. The Poincar\'e coordinates of the bulk points are same as given in \cref{coord-1}. Using the geodesic length formula in \cref{Poincare-geod}, we may obtain the total length of these three circular arcs as follows
	\begin{align}
		d_{PQ}=& L_{\text{II}} \log \left[\frac{\left(b_j+\bar{y}_j \sin\psi_{\text{II}}\right)^2+\left(\bar{y}_j \cos\psi_{\text{II}} \right)^2}{\epsilon_2 \, \bar{y}_j \cos\psi_{\text{II}}}\right]+L_{\text{II}} \log \left[\frac{\left(b_i+\bar{y}_i \sin\psi_{\text{II}}\right)^2+\left(\bar{y}_i \cos\psi_{\text{II}}\right)^2}{\epsilon_2 \, \bar{y}_i \cos\psi_{\text{II}}}\right]\notag\\
		&+L_{\text{I}} \cosh^{-1}\left[\frac{\left(\bar{y}_j-\bar{y}_i\right)^2 \sin ^2\psi_{\text{I}}+\left(\bar{y}_i^2+\bar{y}_j^2\right) \cos^2\psi_{\text{I}}}{2 \bar{y}_i \bar{y}_j \cos ^2\psi_{\text{I}}}\right]\label{d_doublecrossing}
	\end{align}
	We are required to extremize the above length over the arbitrary locations $\bar{y}^{}_{i,j}$. To this end, we make the following change of variables $b_j,\bar{y}_j$ to $\Theta_D, k_D$  :
	\begin{align}
		{b_j}=\Theta_D{b_i}~~,~~\bar{y}_j=k_D^2\Theta_D \bar{y}_i\label{new-variables}
	\end{align}
	Extremization of the length with respect to $\bar{y}_i$ leads to
	\begin{align}
		\bigg(k_D^2 \bar{y}_i^2-b_i^2\bigg) \bigg(\left(k_D^2+1\right) \bar{y}_i b_i \sin\psi_{\text{II}}+k_D^2 \bar{y}_i^2+b_i^2\bigg)=0\,,\label{doublecrossing-ext-y1}
	\end{align}
	and the only real non-negative solution is given by $\bar{y}_i=\frac{b_i}{k_D}$. Substituting this in \cref{d_doublecrossing} and furthermore extremizing over the remaining variable $k_D$, we obtain the following algebraic equation
	\begin{align}
		\cos\psi_{\text{I}} \sec \psi_{\text{II}}\left(1+k_D^2\Theta_D\right)\frac{  \left(1+k_D^2+2 k_D \sin\psi_{\text{II}}\right)}{\sqrt{1+ k_D^4\Theta_D^2+2   k_D^2\Theta_D \cos 2\psi_{\text{I}}}}=1-k_D^2\label{doublecrossing-ext-k}
	\end{align}
	The above eighth order polynomial equation may be readily solved for $k_D$. However, the solutions are not very illuminating and we will omit the details here. Substituting the extremal value $k_D=k_D^*$ (corresponding to the extremal locations $\bar{y}_{i,j}^*$ on the brane) in \cref{d_doublecrossing} we may now obtain the length of the RT saddle connecting $b_i$ and $b_j$ on the right boundary as sketched in \cref{Adjacent-bulk-crossing-RT}. Utilizing the RT prescription \cite{Ryu:2006bv}, the holographic entanglement entropy of subsystem $D$ for the double crossing configuration  is given by
	\begin{align}
		S(\rho_{D})\equiv S_\text{double}\left([b_i,b_j]\right)=&\frac{L_\text{I}}{4G_N}\cosh ^{-1}\left[\frac{ 1+k_D^*{}^2\Theta_D  \cos \left(2 \psi_\text{I}\right)+  k_D^*{}^2\Theta_D \left( k_D^*{}^2\Theta_D-1\right)}{2 k_D^*{}^2\Theta_D \cos ^2\psi_\text{I}}\right]\notag\\
		&\hspace{1cm}+\frac{L_\text{II}}{2G_N}\log\left[\frac{\sqrt{b_i b_j} \left(1+k_D^*{}^2 +2k_D^* \sin \psi _{\text{II}}\right)}{k_D^* \cos \psi _{\text{II}}}\right]\label{EE-double-cross}
	\end{align}
	
	\paragraph{Large Tension Limit:} In the large tension limit, the extremal value of $\bar{y}_1$ remains the same while the extremization conditions in \cref{doublecrossing-ext-k} reduce to the cubic equation
	\begin{align}
		L_\text{II}\left(k_D-1\right) \left(k_D^2\Theta_D-1\right)+L_\text{I}(k_D+1) \left(k_D^2\Theta_D+1\right)=0\,.\label{Double-crossing-EE-ext}
	\end{align}
	
	In the following, we will consider various phases for  the entanglement entropy of $A\cup B$ such that the corresponding RT surface homologous to $[b_1,b_3]$ in AdS$_{\text{II}}$ geometry crosses the EOW brane twice at the points $\bar{y}_1^*$ and $\bar{y}_3^*$. It consists of three semi-circular arcs, one of which resides solely in the AdS$_{\text{I}}$ geometry. Note that, the extremal values of the locations $\bar{y}_{1,2}$ may be obtained from \cref{doublecrossing-ext-k} (or, from \cref{Double-crossing-EE-ext} in the large tension limit.). On the other hand, the RT surface on the AdS$_{\text{I}}$ side is a semi-circle depicted in \cref{Adjacent-bulk-V}.  by the green curve which connects $\tilde{b}_1$ and $\tilde{b}_3$, that does not cross the brane. The bulk entanglement wedge is now the region bounded by these geodesics and the corresponding subsystems as depicted by the shaded regions in \cref{Adjacent-bulk-V}. Furthermore, we will systematically investigate the phase transitions of the minimal EWCS for different subsystem sizes and geometry.
	\subsubsection*{Phase-I}
\begin{figure}
		\centering
		\begin{subfigure}[h]{0.45\textwidth}
			\centering
			\includegraphics[width=\textwidth]{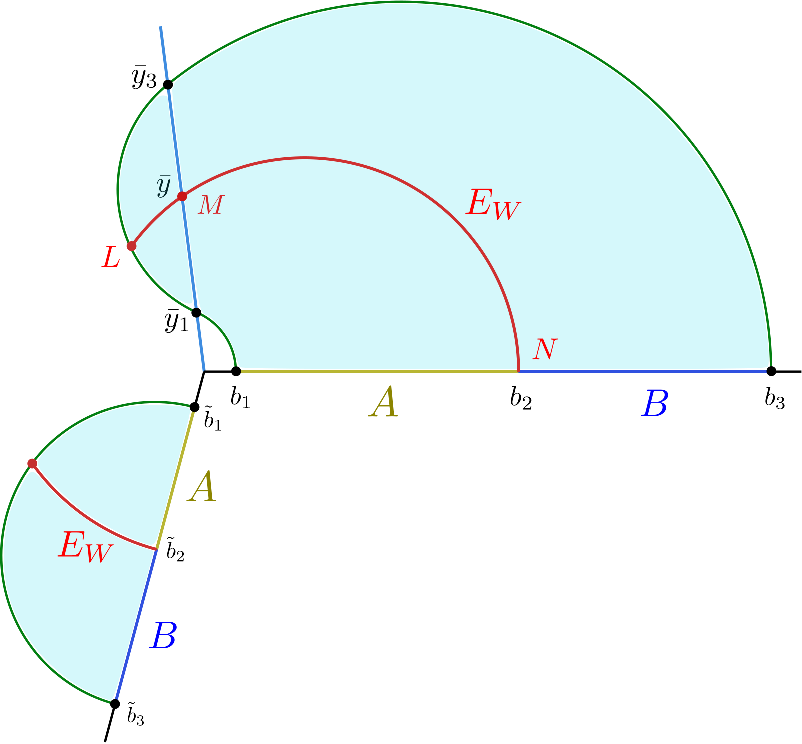}
			\caption{}
			\label{Adjacent-bulk-V}
		\end{subfigure}
		\hfill
		\begin{subfigure}[h]{0.35\textwidth}
			\centering
			\includegraphics[width=\textwidth]{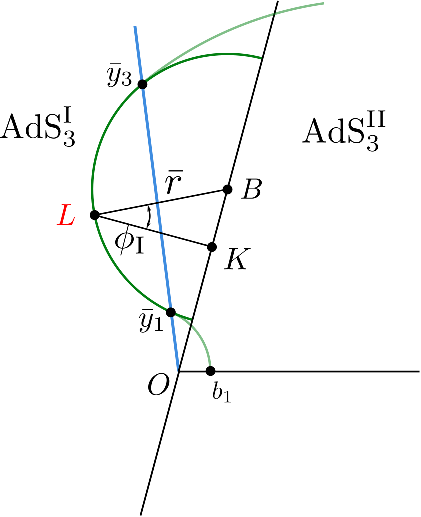}
			\caption{}
			\label{Adjacent-bulk-V-a}
		\end{subfigure}
		\caption{(a) Schematics of the EWCS in phase-I when the RT surface crosses the brane twice. (b) Schematic of circular arc joining $y_1$ and $y_3$ in the AdS$_\text{I}$ geometry.}
		\label{fig:Adj-phase-V}
	\end{figure}

	We begin with the phase where the subsystems in CFT$_2^\text{II}$ are comparable in size and the minimal EWCS consists of two extremal curves as shown by the  red curves in \cref{Adjacent-bulk-V}.
	%
	The minimal EWCS $E_W^I$ residing entirely in the AdS$_\text{I}$ geometry may be computed using standard AdS$_3$/CFT$_2$ techniques \cite{Takayanagi:2017knl,Nguyen:2017yqw} as
	\begin{align}
		E_W^\text{I}(A:B)=\frac{L_\text{I}}{4G_N}\log\left[\frac{2(\tilde{b}_2-\tilde{b}_1)(\tilde{b}_3-\tilde{b}_2)}{\epsilon_1(\tilde{b}_3-\tilde{b}_1)}\right]\,.\label{EW-adj-usual}
	\end{align}
	The minimal EWCS $E_W^{\text{II}}$ in the AdS$_\text{I}$ geometry consists of two circular geodesic segments $NM$ and $ML$ as shown by the red curve  in \cref{Adjacent-bulk-V}. The segment $NM$ starts from the point $b_2$ and ends at the point M on the EOW brane which is at a distance $\bar{y}$ from the interface O. The other circular arc $ML$ connects the point $\bar{y}$ on the EOW brane and ends on the geodesic segment connecting $\bar{y}_1^*$ and $\bar{y}_3^*$ in the AdS$_\text{I}$ region. Hence, the total length of these curves is given by $d_{NL}=d_{NM}+d_{ML}$. The Poincar\'e coordinates of the point $L$ are similar to that given in \cref{cord-b2-y}.
	%
	%
	From \cref{Adjacent-bulk-V-a}, the coordinates of  $L$ can be parametrized as
	\begin{align}
		L:(x_\text{I},z_\text{I})=\left(\overline{OK},\overline{LK}\right)=\left(\overline{OB}-\overline{BK},\overline{LK}\right)=\left(x_0-\bar{r}\sin\phi_{\text{I}},\bar{r}\cos\phi_{\text{I}}\right)
	\end{align}
	where overline on $OK,LK$ once again denote that they are Euclidean distances, $\overline{BL}=\bar{r}$ corresponds to the radius of the circular arc connecting $\bar{y}_1^*$ and $\bar{y}_3^*$ and $K$ is a point where the perpendicular dropped from $L$ intersects $\overline{OB}$. $x_0$ is the center coordinate of the circular arc, and the arbitrary angle $\phi_\text{I}$ parametrizes the position of $L$ on this circular arc. The center and the radius of the circular arc are given by
	\begin{align}
		\bar{r}= \frac{\sqrt{\bar{y}_1^{*2}+\bar{y}_3^{*2}+2\bar{y}_1^*\bar{y}_3^* \cos\left(2 \psi _\text{I}\right)}}{2\sin\psi_\text{I}}~~ ,~~x_{0}= \frac{\bar{y}_1^*+\bar{y}_3^*}{2 \sin \psi_\text{I}}.\label{r-x0}
	\end{align}
	We may now obtain the total length of the two circular geodesic arcs using \cref{Poincare-geod} as 
	\begin{equation}
		\begin{aligned}
			d_{NL}=L_\text{I}\cosh^{-1}\left[\frac{(x_0-\bar{r}\sin\phi_\text{I}+\bar{y}\sin\psi_{\text{I}})^2+(\bar{y}\cos\psi_{\text{I}})^2+(\bar{r}\cos\phi_{\text{I}})^2}{2\bar{y}\cos\psi_{\text{I}}\,\bar{r}\cos\phi_{\text{I}}}\right]\\
			+L_{\text{II}} \log \left[\frac{(b_2+\bar{y} \sin \psi_{\text{II}})^2+(\bar{y} \cos \psi_{\text{II}})^2}{\bar{y} \epsilon_2 \cos \psi_{\text{II}}}\right]\label{d-adj-doubleI}
		\end{aligned}
	\end{equation}
	On extremizing with respect to $\phi_\text{I}$, we obtain 
	\begin{align}
		&\phi_{\text{I}}=\sin ^{-1}\left[\frac{2 \bar{r} (x_0+\bar{y}\sin\psi_\text{I})}{\bar{r}^2+x_0^2+2 x_0 \bar{y} \sin\psi_\text{I}+\bar{y}^2}\right]\,.
	\end{align}
	Now, substituting the value of $\phi_\text{I}$ in \cref{d-adj-doubleI} followed by extremizing over $\bar{y}$, we obtain a polynomial equation in $\bar{y}$ whose physical solution leads to the minimal EWCS 
	\begin{align}
		E_W^\text{II}(A:B)=&L_\text{I} \cosh ^{-1}\left[\frac{\sqrt{\left(\bar{r}^2+x_0^2+2 x_0 \bar{y}^* \sin \psi_\text{I}+\bar{y}^*{}^2\right)^2-4 \bar{r}^2 (x_0+\bar{y}^* \sin \psi_\text{I})^2}}{2 \bar{r} \bar{y}^* \cos \psi_\text{I}}\right]\notag\\
		&\hspace{4.95cm}+L_\text{II} \log \left[\frac{b_2^2+\bar{y}^{*2}+2 b_2 \bar{y}^* \sin \psi_\text{II}}{\epsilon\, \bar{y}^* \cos \psi_\text{II} }\right]\,.
	\end{align}
	
	In the large tension limit $\delta\to 0$, using \cref{r-x0}, the result simplifies to
	\begin{equation}
		E_W^{\text{II}}(A:B)=\frac{L_\text{I}}{4G_N} \log \left[\frac{(\bar{y}^*-\bar{y}_1^*) (\bar{y}^*-\bar{y}_3^*)}{\bar{y}^* (\bar{y}_1^*-\bar{y}_3^*)}\right]+\frac{L_{\text{II}}}{4G_N}  \log \left[\frac{(b_2+\bar{y}^*)^2}{ 2\bar{y}^* \,\epsilon}\right]+S^{(\delta)}_\text{int}\,,\label{EW-double1}
	\end{equation}
	where $\bar{y}^*$ corresponds to the solution of the  extremization condition in the $\delta\to 0$ limit which is given by
	\begin{align}
		L_\text{I} \left(\frac{\bar{y}^*_1}{\bar{y}-\bar{y}^*_1}+\frac{\bar{y}}{\bar{y}-\bar{y}^*_2}\right)+L_\text{II}\, \frac{\bar{y}-b_2}{\bar{y}+b_2}=0\,.\label{Adj-double-I-delta-extr}
	\end{align}
	Note that, in \cref{EW-double1}, $S^{(\delta)}_\text{int}$ denotes the large tension limit of the interface entropy given in \cref{S-int}.

	\subsubsection*{Phase-II}
\begin{figure}
		\centering
		\begin{subfigure}[h]{0.4\textwidth}
			\centering
			\includegraphics[width=\textwidth]{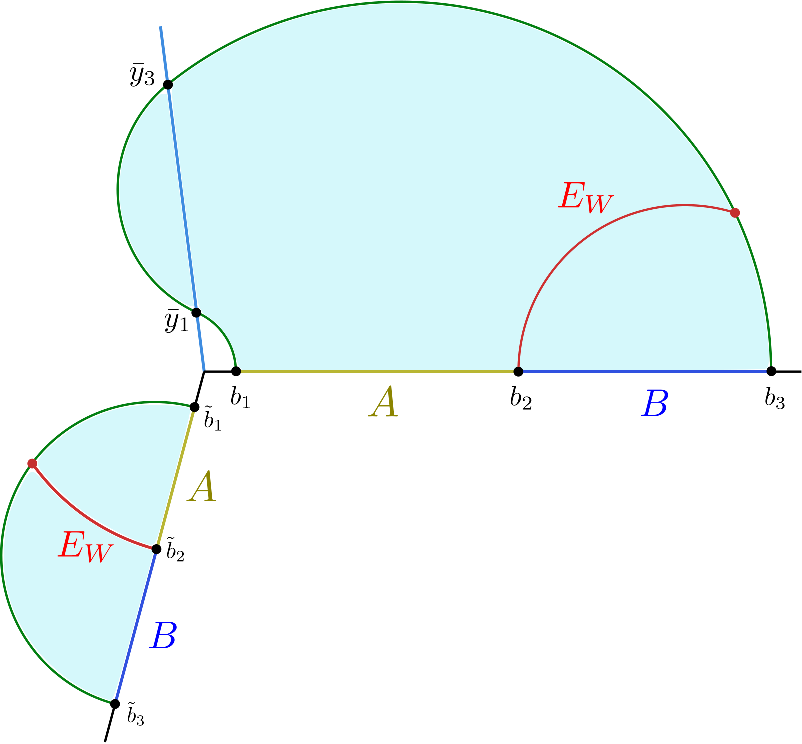}
			\caption{}
			\label{Adjacent-bulk-VI}
		\end{subfigure}
		\hfill
		\begin{subfigure}[h]{0.4\textwidth}
			\centering
			\includegraphics[width=\textwidth]{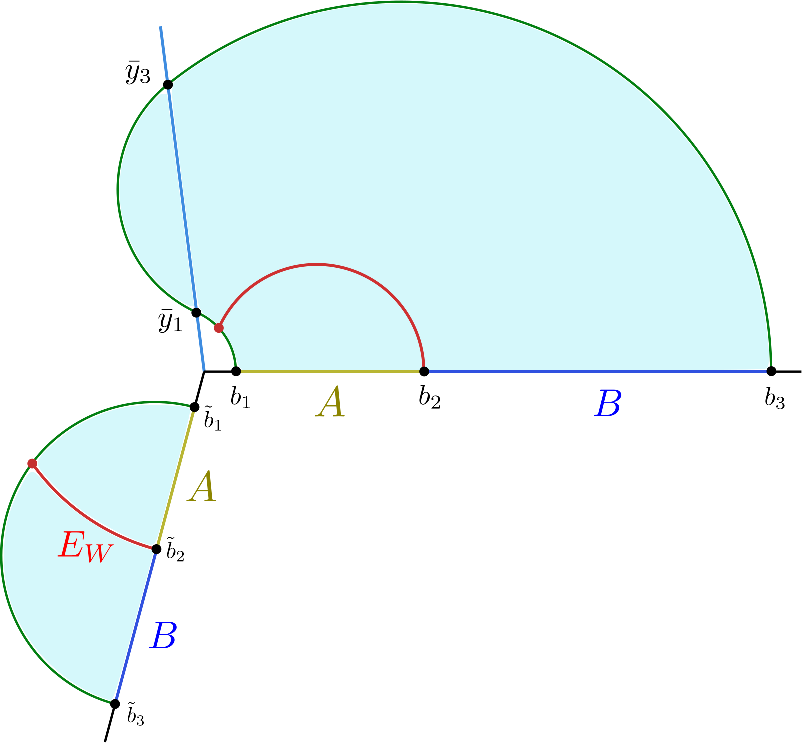}
			\caption{}
			\label{Adjacent-bulk-VII}
		\end{subfigure}
		\caption{Adjacent subsystems: double-crossing phase-II and III.}
		\label{fig:Adj-phaseVI-VII}
	\end{figure}

	In phase-II, the subsystem $B$ is smaller compared to $A$ in CFT$_2^\text{II}$ and the minimal EWCS consists of two circular geodesic segments, one of which is similar to the previous subsection. The other geodesic starts from $b_2$ and ends on the outer RT surface in AdS$_{\text{II}}$ geometry. Both the segments are depicted by the red curves  in \cref{Adjacent-bulk-VI}. 
	The portion of the minimal EWCS in AdS$_3^\text{I}$ geometry is again given by \cref{EW-adj-usual}. The endpoint of the other portion in AdS$_3^\text{II}$ on the outer RT surface may be parametrized by an arbitrary angle $\phi_\text{II}$, similar to \cref{cord-b2-y}. Using \cref{Poincare-geod} the length of this geodesic segment may be expressed as 
	\begin{equation}
		\begin{aligned}
			d_{\text{II}}=L_{\text{II}} \log \left[\frac{\left(b_2-\left(b_3+\bar{R} \sin \phi_\text{II}-\bar{R}\right)\right)^2+(\bar{R} \cos \phi_\text{II})^2}{\bar{R} \epsilon_2 \cos \phi_\text{II}}\right]
		\end{aligned}
	\end{equation}
	where $\bar{R}$ is the radius of the outer RT surface,
	\begin{align}
		\bar{R}= b_3+\frac{\bar{y}_3^{*^2}-b_3^2}{2 (b_3+\bar{y}_3^* \sin \psi_{\text{II}})}\,.
	\end{align}
	
	After extremizing over $\phi_2$, we may readily obtain the minimal EWCS for this phase, with $\bar{y}_3^*$ obtained from \cref{doublecrossing-ext-k}. In the limit of large brane tension $\delta\to 0$, the minimal EWCS reduces to
	\begin{equation}\label{adj-EW-doubleII}
		E_W(A:B)=\frac{L_\text{I}}{4G_N}\log\left[\frac{2(\tilde{b}_2-\tilde{b}_1)(\tilde{b}_3-\tilde{b}_2)}{\epsilon_1(\tilde{b}_3-\tilde{b}_1)}\right]+\frac{L_{\text{II}}}{4 G_N}\log \left[\frac{2(b_3-b_2)(b_2+\bar{y}_3^*)}{\epsilon_2 (b_3+\bar{y}_3^*)}\right].
	\end{equation} 
	with $\bar{y}_3^*$ now given by the physical solution to \cref{Double-crossing-EE-ext}.

	\subsubsection*{Phase-III}
	For phase-III, we consider the subsystem $A$ in CFT$_2^\text{II}$ to be smaller than $B$ such the  minimal EWCS lands on the smaller RT surface crossing the brane at $\bar{y}_1$ as depicted in \cref{Adjacent-bulk-VII}. 
	
	The computation for the EWCS in this phase is similar to the previous subsection and in the large tension limit, it reduces to the following expression
	\begin{equation}
		E_W(A:B)=\frac{L_\text{I}}{4G_N}\log\left[\frac{2(\tilde{b}_2-\tilde{b}_1)(\tilde{b}_3-\tilde{b}_2)}{\epsilon_1(\tilde{b}_3-\tilde{b}_1)}\right]+\frac{L_{\text{II}}}{4G_N}\log \left[\frac{2(b_2-b_1)(b_2+\bar{y}^*_1)}{\epsilon_2 (b_1+\bar{y}^*_1)}\right],\label{EW-adj-double-III}
	\end{equation}
	where $\bar{y}^*_1=\frac{b_1}{k_D^*}$, $k_D^*$ being the solution of the extremization equation in \cref{Double-crossing-EE-ext}.
	\subsubsection{RT saddles with no brane crossing}
\begin{figure}[ht]
		\centering
		\includegraphics[scale=0.4]{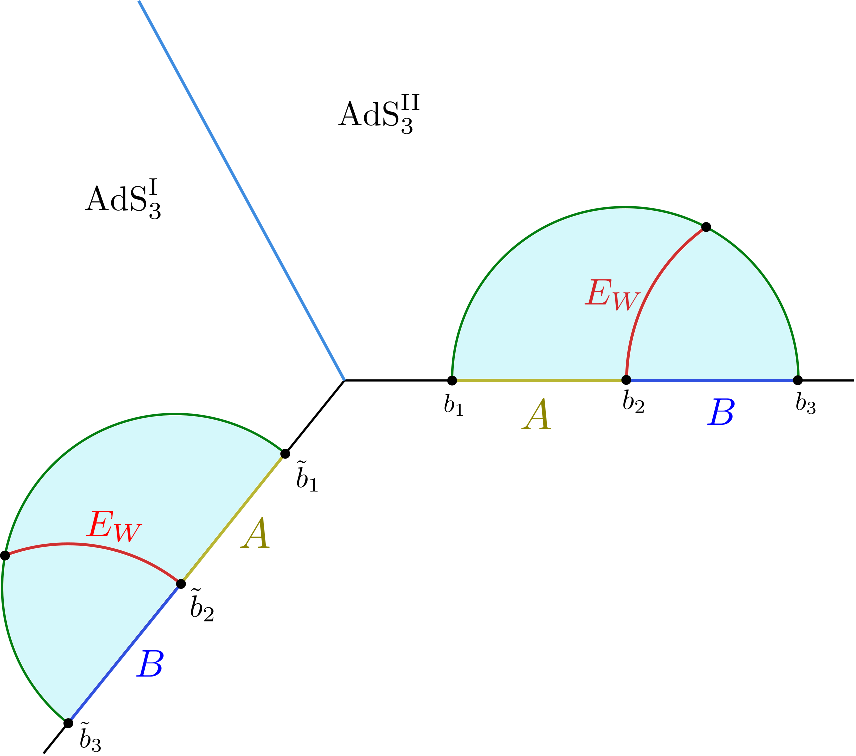}
		\caption{Adjacent subsystems: no crossing.}
		\label{Adjacent-bulk-IV}
	\end{figure}

	When the total system $A\cup B$ is small compared to their distance from the interface, the corresponding RT surface becomes disconnected as shown in \cref{Adjacent-bulk-IV}. 
	The minimal EWCS consist of two circular arcs which correspond to the EWCS of two adjacent subsystems in AdS$_3^{\text{I}}$ and AdS$_3^{\text{II}}$ regions respectively. So, the minimal EWCS for this phase may be expressed as \cite{Takayanagi:2017knl,Nguyen:2017yqw}
	\begin{equation}
		E_W(A:B)=\frac{L_\text{I}}{4G_N}\log\left[\frac{2\big(\tilde{b}_2-\tilde{b}_1\big)\big(\tilde{b}_3-\tilde{b}_2\big)}{\epsilon\big(\tilde{b}_3-\tilde{b}_1\big)}\right]
		+\frac{L_{\text{II}}}{4G_N}\log\left[\frac{2(b_2-b_1)(b_3-b_2)}{\epsilon(b_3-b_1)}\right].\label{EW-adj-no-crossing}
	\end{equation}

	\subsection{Disjoint Subsystems }
	In this section, we consider the bipartite mixed state configuration $\rho_{AB}$ described by unions of two disjoint intervals $\left[{b }_1,{b }_2\right]$ and $\left[b _3,b _4\right]$ on the two  CFT$_2$ s at a constant time slice $\tau=\tau_0$ on either side of the interface. In particular, we take the bipartition of $\rho_{AB}$ as follows:
	\begin{align}
		A=\left[{b }_1,{b }_2\right]_{\text{I}}\cup \left[b _1,b _2\right]_{\text{II}} ~~\text{and} ~~B=\left[{b }_3,{b }_4\right]_{\text{I}}\cup \left[b _3,b _4\right]_{\text{II}}\notag.
	\end{align}
	The schematics of this configuration is sketched in \cref{fig:SingleEEdisj}. The computation of the holographic entanglement entropy for $\rho_{AB}$ consists of different saddles of the bulk Ryu-Takayanagi (RT) surfaces homologous to $A\cup B$ which we investigate systematically in the following. For each phase of the RT saddle we will construct the bulk entanglement wedge and subsequently compute the corresponding minimal (extremal) cross-section which is dual to the reflected entropy $S_\text{R}(A:B)$. Note that, within each phase of the entanglement entropy for $A\cup B$, the minimal cross section dividing the entanglement wedge of $A\cup B$ experiences phase transitions depending upon the subsystem sizes as well as their distances from the interface. Besides we will disregard all possible RT saddles for which the bulk entanglement wedge is disconnected and consequently the EWCS is vanishing. In the following, we will further divide the possible configurations of RT saddles into two sub-classes, namely those corresponding to the RT surfaces crossing the EOW brane once and those where multiple crossovers are possible for a single RT surface as described in \cite{Anous:2022wqh}.
	
	\subsubsection{Configurations involving single crossover of RT surfaces}
\begin{figure}[ht]
		\centering
		\includegraphics[scale=0.45]{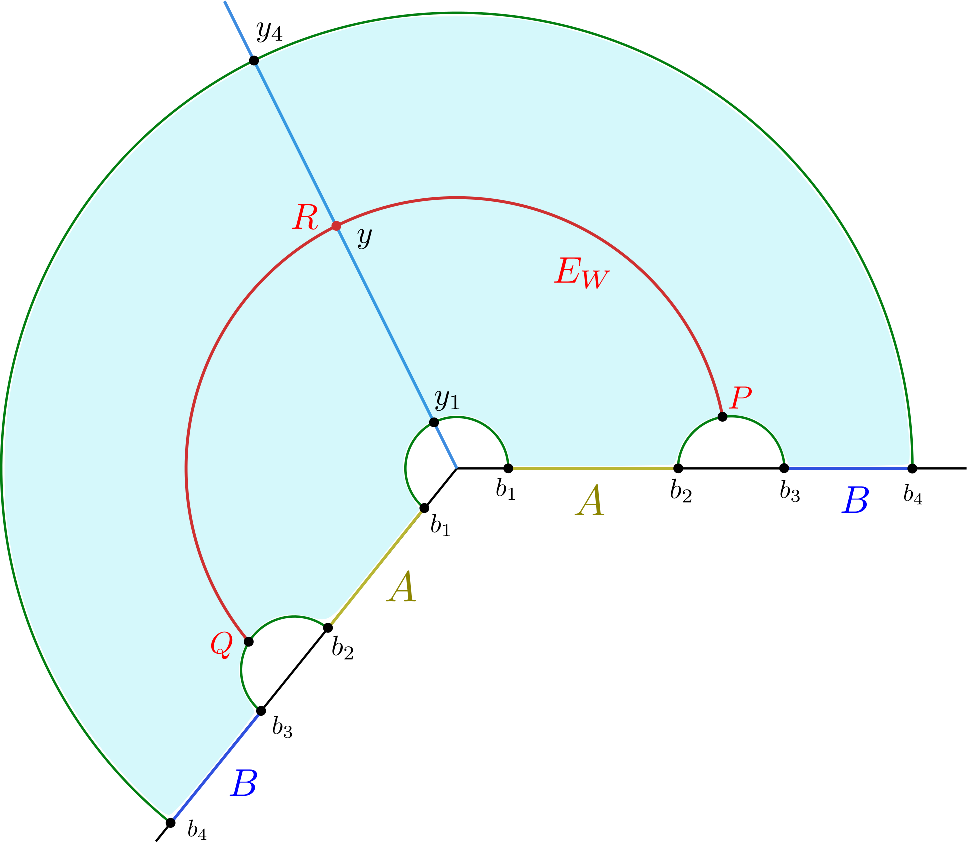}
		\caption{Configurations of single crossing RT saddles for two disjoint subsystems in the boundary CFT$_2$s.}
		\label{fig:SingleEEdisj}
	\end{figure}

	We begin by considering the configurations of bulk extremal surfaces $\Gamma_{AB}$ homologous to $A\cup B$ which cross the EOW brane at the points $y^{}_1$ and $y^{}_4$ along with two usual dome shaped geodesics connecting $b_2$ and $b_3$ in each of the AdS$_3$ geometries as sketched in \cref{fig:SingleEEdisj}. The Poincar\'e coordinates of the points on the brane are given as
	\begin{align}
		&\left(\tau_0,y^{}_k\sin\psi_{\text{I}},y^{}_k\cos\psi_{\text{I}}\right) ~~,~~\text{from AdS}_3^\text{I}\notag\\
		&\left(\tau_0,y^{}_k\sin\psi_{\text{II}},y^{}_k\cos\psi_{\text{II}}\right) ~~,~~\text{from AdS}_3^\text{II}\label{y-parametrization}
	\end{align}
	with $k=\text{I},\text{II}$. In the above parametrization, we have used the fact that the Israel junction conditions enforce
	the distances $y_k$ along the EOW brane to be identical as seen from either side of the geometry. Note that the locations $y_k$ of the bulk points along the EOW brane are chosen arbitrarily.
	The total length of the geodesics homologous to $A\cup B$ may now be computed\footnote{Here we are using the same technique described in \cref{311}.} using \cref{Poincare-geod} as follows
	\begin{align}
		d=&L_\text{I}\log \left[\frac{\left(b_1+y_1 \sin\psi_{\text{I}}\right)^2+\left(y_1 \cos\psi_{\text{I}}\right)^2}{\epsilon\, y_1  \cos \psi_{\text{I}}}\right]+L_\text{II}\log \left[\frac{\left(b_1+y_1 \sin\psi_{\text{II}}\right)^2+\left(y_1 \cos\psi_{\text{II}}\right)^2}{\epsilon\, y_1  \cos \psi_{\text{II}}}\right]\notag\\
		&+L_\text{I}\log \left[\frac{\left(b_4+y_4 \sin\psi_{\text{I}}\right)^2+\left(y_4 \cos\psi_{\text{I}}\right)^2}{\epsilon\, y_4  \cos \psi_{\text{I}}}\right]+L_\text{II}\log \left[\frac{\left(b_4+y_4 \sin\psi_{\text{II}}\right)^2+\left(y_4 \cos\psi_{\text{II}}\right)^2}{\epsilon\, y_4  \cos \psi_{\text{II}}}\right]\notag\\
		&+2\left(L_\text{I}+L_\text{II}\right) \log \left(\frac{b_3-b_2}{\epsilon }\right)\,.\label{EEdisj-1}
	\end{align}
 Extremizing \cref{EEdisj-1} over the bulk points $y_1$ and $y_4$, we obtain the extremal values to be
	\begin{align}
		y_1^*=b_1~~,~~	y_4^*=b_4\,.\label{ext-single-crossing}
	\end{align}
	Substituting these in the expression for the geodesic length and subsequently using the RT formula, the entanglement entropy for the mixed state $\rho_{AB}$ reads
	\begin{align}
		S\left(A\cup B\right)=\frac{L_\text{I}+L_\text{II}}{4G_N}\left[\log\left(\frac{2b_1}{\epsilon}\right)+\log\left(\frac{2b_4}{\epsilon}\right)+2\log \left(\frac{b_3-b_2}{\epsilon }\right)\right] +\frac{\rho_{\text{I}}^*+\rho_{\text{II}}^*}{2 G_N}\,,\label{EE_disj}
	\end{align}
	where we have utilized the following relation between the angles $\psi_k$ and the brane location $\rho_k^*$ in the Poincar\'e slicing coordinates\cite{Anous:2022wqh},
	\begin{align}
		\sec\psi_k=\cosh\left(\frac{\rho_k^*}{L_k}\right)~~,~~(k=\text{I,II})\label{Brane-location-angle}\,.
	\end{align}
	Once we have obtained the RT saddles corresponding to the configuration of disjoint intervals, the entanglement wedge dual to the reduced density matrix can be constructed as the codimension one bulk region bounded by the RT surfaces and the subsystems on the boundary. This is shown by the shaded region in \cref{fig:SingleEEdisj}. As we shall see below, there are three possible phases of the minimal EWCS for this configuration of the RT saddles corresponding to $\rho_{AB}$. In the following, we will investigate the phase transition of the minimal EWCS for different sizes of $A$ and $B$.
	
	\subsubsection*{Phase-I}
	The first phase of EWCS corresponds to sufficiently small separations between rather large subsystems $A$ and $B$ in both CFT$_2$s (recall that, we have considered configurations of $A$ and $B$ to be symmetric with respect to the interface). In this case, as depicted in \cref{fig:SingleEEdisj}, the minimal EWCS connects the dome-shaped geodesics joining $b_2$ and $b_3$ on both sides of the interface by crossing the EOW brane once at the point $R$.
	\begin{figure}[ht]
		\centering
		\includegraphics[scale=0.45]{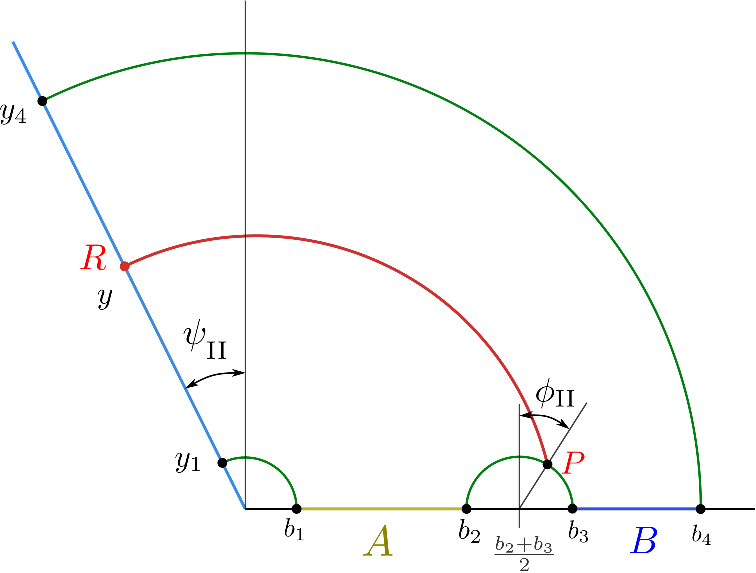}
		\caption{disjoint intervals : phase-I }
		\label{fig:disj1calc}
	\end{figure}
	The candidate EWCS depicted by the red curve in \cref{fig:SingleEEdisj}, consists of two circular geodesic arcs emanating respectively from the points $P$ and $Q$ on the geodesic connecting $b_2$ and $b_3$ on the CFT$_2^\text{II/I}$ and landing on the EOW brane at the common\footnote{Note that the two geodesic segments on either sides of the brane must join smoothly at the location of the brane as discussed in \cite{Anous:2022wqh}. We observed that the smoothness of the geodesic crossing the EOW brane is achieved naturally through the extremization of the total geodesic length with respect to the point $R$.} location denoted by $R$. Therefore, the total length of the red curve is given by the sum of geodesic lengths as $d_{PQ}=d_{PR}+d_{QR}$. Note that the points $P$ and $Q$ are only constrained to be on the geodesics connecting $b_2$ and $b_3$ and hence possess a degree of arbitrariness. We set the location of $P$ by introducing the (arbitrary) angle $\phi_{\text{II}}$ as sketched in \cref{fig:disj1calc} and a similar parametrization of the point $Q$ on the other geodesic is dependent on an angle $\phi_{\text{I}}$. Therefore, in the Poincar\'e AdS$_3^\text{II}$ geometry on the right side of the brane, the coordinate of $P$ are obtained as
	\begin{align}
		P:(\tau_\text{II},x_\text{II},z_\text{II})=\left(\tau_0,\frac{b_3+b_2}{2}+\frac{b_3-b_2}{2}\sin\phi_{\text{II}},\frac{b_3-b_2}{2}\cos\phi_{\text{II}}\right)\,.\label{ccordinate-P}
	\end{align}
	The coordinates of $Q$ in AdS$_3^\text{I}$ may be found similarly.
	Furthermore, the Poincar\'e coordinates of the point $R$ on the brane may be written as
	\begin{align}
		R: \left(\tau_0,y\sin\psi_{\text{I}},y\cos\psi_{\text{I}}\right) ~~,&~~\text{from AdS}_3^\text{I}\notag\\
		\left(\tau_0,y\sin\psi_{\text{II}},y\cos\psi_{\text{II}}\right) ~~,&~~\text{from AdS}_3^\text{II}\label{coordinate-R}
	\end{align}
	Therefore, utilizing \cref{Poincare-geod},we obtain the length of the candidate EWCS as follows
	\begin{align}
		d_{PQ}&=L_{\text{I}}\cosh^{-1}\left[\frac{\left(\frac{b_3+b_2}{2}+\frac{b_3-b_2}{2}\sin\phi_{\text{I}}+y\sin\psi_{\text{I}}\right)^2+\left(\frac{b_3-b_2}{2}\cos\phi_{\text{I}}\right)^2+(y \cos \psi_{\text{I}})^2}{2 \left(\frac{b_3-b_2}{2}\cos\phi_{\text{I}}\right) y\cos\psi_{\text{I}}}\right]\notag\\
		&+L_{\text{II}}\cosh^{-1}\left[\frac{\left(\frac{b_3+b_2}{2}+\frac{b_3-b_2}{2}\sin\phi_{\text{II}}+y\sin\psi_{\text{II}}\right)^2+\left(\frac{b_3-b_2}{2}\cos\phi_{\text{II}}\right)^2+(y \cos \psi_{\text{II}})^2}{2 \left(\frac{b_3-b_2}{2}\cos\phi_{\text{II}}\right) y\cos\psi_{\text{II}}}\right].
	\end{align}
	Extremizing the above expression over the arbitrary angles $\phi_{\text{I}}$ and $\phi_{\text{II}}$, we obtain
	\begin{align}
		&\phi_{\text{I}}=\sin^{-1}\left[\frac{(b_2-b_3) (b_2+b_3+2y\sin\psi_{\text{I}})}{b_2^2+b_3^2+2 y^2+2 y (b_2+b_3) \sin\psi_{\text{I}}}\right]\,,\notag\\
		&\phi_{\text{II}}=\sin^{-1}\left[\frac{(b_2-b_3) (b_2+b_3+2y\sin\psi_{\text{II}})}{b_2^2+b_3^2+2 y^2+2 y (b_2+b_3) \sin\psi_{\text{II}}}\right]\,.
	\end{align}
	Substituting these values back and subsequently extremizing over the location $y$ along the EOW brane, we obtain
	\begin{align}
		\partial_y\,d_{PQ}=0\implies y=\sqrt{b_2 b_3}\,.\label{y-ext}
	\end{align}
	Finally, the minimal EWCS is obtained using \cref{y-ext} as follows
	\begin{align}
		E_W(A:B)&=\frac{L_{\text{I}}}{4 G_N} \cosh ^{-1}\left[\frac{ \left(b_2+b_3+2\sqrt{b_2b_3}\sin\psi_{\text{I}}\right)}{(b_3-b_2)\cos\psi_{\text{I}}}\right]\nonumber\\&\quad +\frac{L_{\text{II}}}{4 G_N} \cosh ^{-1}\left[\frac{ \left(b_2+b_3+2\sqrt{b_2b_3}\sin\psi_{\text{II}}\right)}{(b_3-b_2)\cos\psi_{\text{II}}}\right]
	\end{align}
	Using standard trigonometric identities, the above result may be re-expressed as
	\begin{align}
		E_W(A:B)=&\frac{L_{\text{I}}+L_{\text{II}}}{4 G_N}\log\left(\frac{b_2+b_3+2\sqrt{b_2b_3}}{b_3-b_2}\right)+\frac{L_{\text{I}}}{4 G_N}\cosh^{-1}\left(\frac{1}{\cos\psi_{\text{I}}}\right)\nonumber\\&+\frac{L_{\text{II}}}{4 G_N}\cosh^{-1}\left(\frac{1}{\cos\psi_{\text{II}}}\right)
	\end{align}
	Utilizing the relation between the angles $\psi_k$ and the location of the brane $\rho_k^*$ given in \cref{Brane-location-angle}, the above minimal EWCS may be written in the following instructive form
	\begin{align}
		E_W(A:B)&=\frac{L_{\text{I}}+L_{\text{II}}}{4 G_N}\log\left(\frac{b_2+b_3+2\sqrt{b_2b_3}}{b_3-b_2}\right)+\frac{\rho_{\text{I}}^*+\rho_{\text{II}}^*}{4 G_N}\notag\\
		&=\frac{c_{\text{I}}+c_{\text{II}}}{6}\log\left(\frac{\sqrt{b_3}+\sqrt{b_2}}{\sqrt{b_3}-\sqrt{b_2}}\right)+S_{\text{int}}\,,\label{EW-phase1}
	\end{align}
	where $S_{\text{int}}=\frac{\rho_{\text{I}}^*+\rho_{\text{II}}^*}{4 G_N}$ is termed the interface entropy \cite{Anous:2022wqh}.
	
	We now take the large tension limit, $T\to T_{\text{max}}$,
	and expand around $\delta\to 0$ as described in \cref{angle-small}. Consequently the leading order term in the interface entropy $S_\text{int}$ is given by \cref{S-int}.

	\subsubsection*{Phase-II}
\begin{figure}[ht]
		\centering
		\includegraphics[scale=0.45]{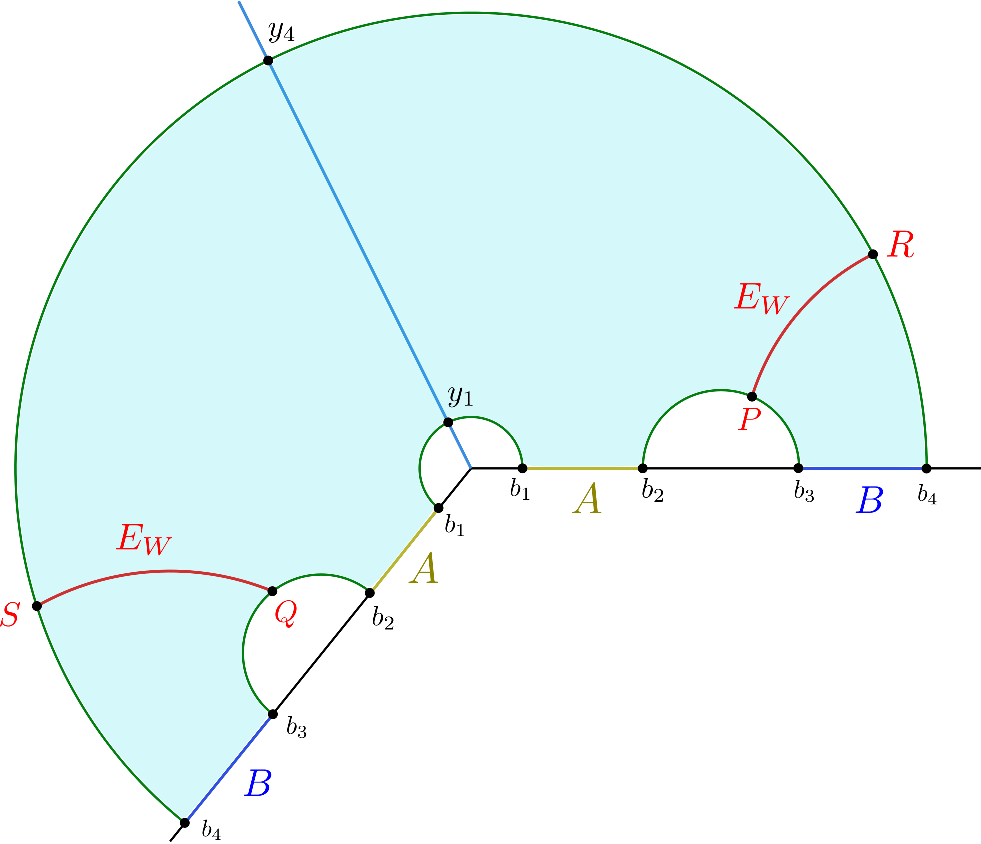}
		\caption{disjoint intervals : phase-II }
		\label{fig:disj2}
	\end{figure}

	Next we consider the situation when the subsystem $B$ is small compared to both $A$ and the separation between $A$ and $B$. In this case, the minimal EWCS comprises of two separate semi-circular arcs emanating from the geodesics connecting $b_2$ and $b_3$ which lands on the RT surface connecting $b_4$ on either side. The schematics of the configuration is sketched in \cref{fig:disj2}. 
	
\begin{figure}[ht]
		\centering
		\includegraphics[scale=0.7]{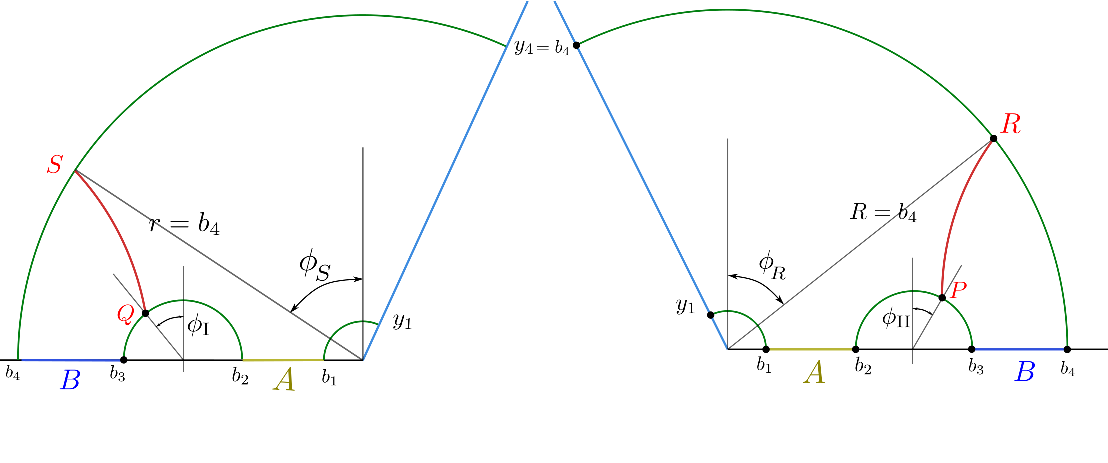}
		\caption{Calculation of the minimal EWCS in phase-II }
		\label{fig:disj2calc}
	\end{figure}

	As seen from \cref{fig:disj2calc}, the Poincar\'e coordinates of the points $P$ and $Q$ are given in \cref{ccordinate-P} while those for the
	point $R$ in the AdS$_3^\text{II}$ geometry are parametrized by the angle $\phi_{R}$ as follows
	\begin{align}
		R:(\tau_\text{II},x_\text{II},z_\text{II})=\left(\tau_0,b_4\sin\phi_R,b_4\cos\phi_R\right)\,,\label{parametrization-R}
	\end{align}
	where we have utilized the fact that the radius of the circular geodesic connecting $b_4$ from either side is $R=b_4$, as seen from \cref{ext-single-crossing}. Similarly, the Poincar\'e coordinates of the point $S$ on the AdS$_3^\text{I}$ geometry is given in terms of the angle $\phi_{S}$ as 
	\begin{align}
		S:(\tau_\text{I},x_\text{I},z_\text{I})=\left(\tau_0,b_4\sin\phi_S,b_4\cos\phi_S\right)\,.\label{parametrization-S}
	\end{align}
	
	Now utilizing \cref{Poincare-geod}, the total length of the geodesics may be obtained as follows
	\begin{align}
		d_{\text{II}}&=d_{PR}+d_{QS}\notag\\
		=&L_{\text{I}} \cosh^{-1}\left[\frac{\left(b_4\sin\phi_S-\left(\frac{b_2+b_3}{2}+\frac{b_3-b_2}{2}\sin\phi_{\text{I}}\right)\right)^2+\left(\frac{b_3-b_2}{2}\cos\phi_{\text{I}}\right)^2+\left(b _4 \cos\phi_S\right)^2}{2\left(\frac{b_3-b_2}{2}\cos \phi_{\text{I}}\right) b_4\cos\phi_S}\right]\notag\\
		&+L_{\text{II}}\cosh^{-1}\left[\frac{\left(b_4\sin\phi_R-\left(\frac{b_2+b_3}{2}+\frac{b_3-b_2}{2}\sin\phi_{\text{II}}\right)\right)^2+\left(\frac{b_3-b_2}{2}\cos\phi_{\text{II}}\right)^2+\left(b _4 \cos\phi_R\right)^2}{2\left(\frac{b _3-b _2}{2}\cos \phi_{\text{II}}\right) b_4\cos\phi_R}\right]\label{d-disj2}\nonumber\\
	\end{align}
	Extremizing over the arbitrary angles $\phi_{\text{I}}\,,\,\phi_{\text{II}}\,,\,\phi_R\,,\,\phi_S$ we obtain
	\begin{align}
		\phi_\text{I}=\phi_\text{II}=\sin^{-1}\left[\frac{b_2^2-b_3^2}{b_2^2+b_3^2-2 b_4^2}\right]~~,~~\phi_R=\phi_S=\sin^{-1}\left[\frac{\left(b _2+b _3\right) b _4}{b _4^2+b _2 b _3}\right]
	\end{align}
	Substituting the above extremal values in \cref{d-disj2}, we obtain the minimal EWCS to be
	\begin{align}
		E_W(A:B)&=\frac{L_\text{I}+L_\text{II}}{4G_N}\cosh^{-1}\left[\frac{b_4^2-b_2b_3}{b_4(b_3-b_2)}\right]\notag\\
		&=\frac{L_\text{I}+L_\text{II}}{4G_N}\log\left[\frac{b_4^2-b_2b_3+\sqrt{\displaystyle(b_4^2-b_2^2)(b_4^2-b_3^2)}}{b_4(b_3-b_2)}\right]\label{EW-disj-II}
	\end{align}
Note that the above expression does not contain any contribution from the brane as the EWCS land on the RT surface ($R$ and $S$ in \cref{fig:disj2}) connecting the $b_4$ points. As a result we observe the absence of any $\delta$ dependent term in the expression of minimal EWCS.
	
	\subsubsection*{Phase-III}
\begin{figure}[h!]
		\centering
		\includegraphics[scale=0.45]{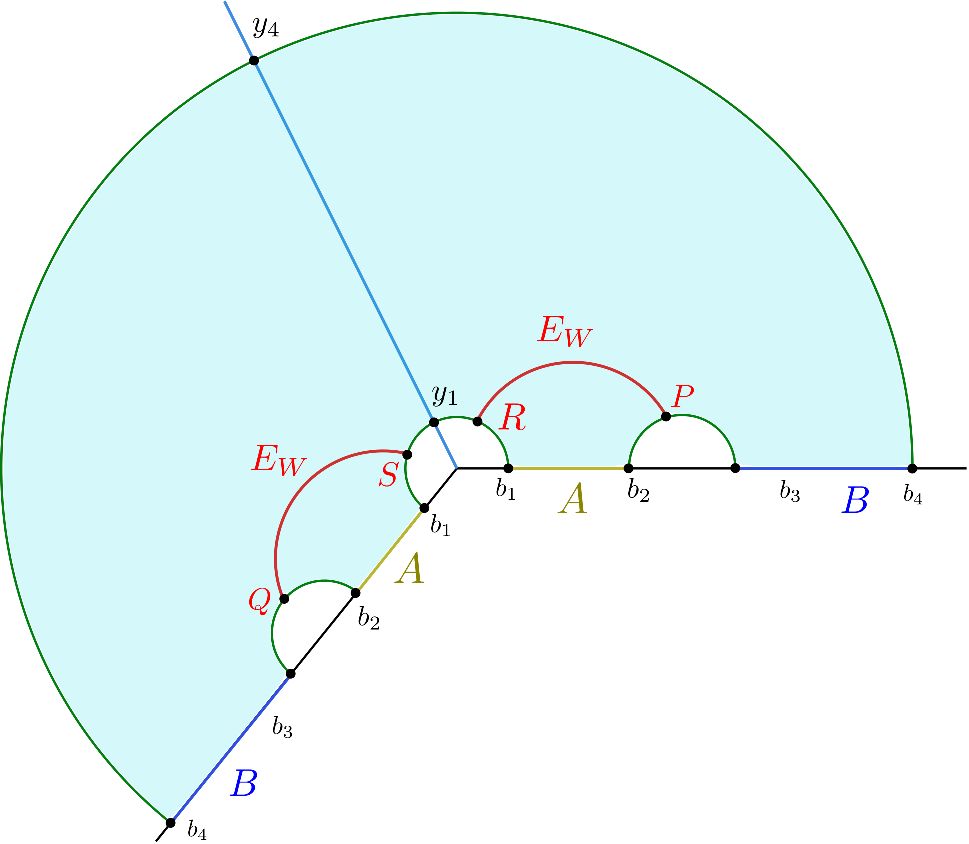}
		\caption{disjoint intervals : phase-III}
		\label{fig:disj3}
	\end{figure}

	The last phase concerns small $A$ and large $B$ with a small separation between them. In this phase, the minimal EWCS is anchored on the RT surface connecting $b_1$ on either side as depicted in \cref{fig:disj3}. Once again, we consider two arbitrary points $R$ and $S$ parametrized by the angles $\phi_{R}$ and $\phi_{S}$, now on the smaller single crossing RT surface. The Poincar\'e coordinates of these points may be read off from \cref{parametrization-R,parametrization-S} with $b_4$ replaced by $b_1$. Utilizing \cref{Poincare-geod}, the total length of the candidate EWCS may be computed as follows
	\begin{align}
		d_{\text{III}}&=d_{PR}+d_{QS}\notag\\
		=&L_{\text{I}} \cosh^{-1}\left[\frac{\left(b_1\sin\phi_S-\left(\frac{b_2+b_3}{2}+\frac{b_3-b_2}{2}\sin\phi_{\text{I}}\right)\right)^2+\left(\frac{b_3-b_2}{2}\cos\phi_{\text{I}}\right)^2+\left(b _1 \cos\phi_S\right)^2}{2\left(\frac{b_3-b_2}{2}\cos \phi_{\text{I}}\right) b_1\cos\phi_S}\right]\notag\\
		&+L_{\text{II}}\cosh^{-1}\left[\frac{\left(b_1\sin\phi_R-\left(\frac{b_2+b_3}{2}+\frac{b_3-b_2}{2}\sin\phi_{\text{II}}\right)\right)^2+\left(\frac{b_3-b_2}{2}\cos\phi_{\text{II}}\right)^2+\left(b _1 \cos\phi_R\right)^2}{2\left(\frac{b _3-b _2}{2}\cos \phi_{\text{II}}\right) b_1\cos\phi_R}\right]\label{d-disj3}
	\end{align}
	Extremizing over the arbitrary angles $\phi_{\text{I}}\,,\,\phi_{\text{II}}\,,\,\phi_R\,,\,\phi_S$ we obtain
	\begin{align}
		\phi_\text{I}=\phi_\text{II}=\sin^{-1}\left[\frac{b_2^2-b_3^2}{b_2^2+b_3^2-2 b_1^2}\right]~~,~~\phi_R=\phi_S=\sin^{-1}\left[\frac{\left(b _2+b _3\right) b _1}{b _1^2+b _2 b _3}\right]
	\end{align}
	Substituting the above extremal values in \cref{d-disj3}, we obtain the minimal EWCS to be
	\begin{align}
		E_W(A:B)&=\frac{L_\text{I}+L_\text{II}}{4G_N} \cosh^{-1}\left[\frac{b_2b_3-b_1^2}{b_1 \left(b_3-b_2\right)}\right]\notag\\
		&=\frac{L_\text{I}+L_\text{II}}{4G_N}\log\left[\frac{b_2 b _3-b_1^2+\sqrt{\left(b_2^2-b_1^2\right) \left(b_3^2-b_1^2\right)}}{b_1 \left(b_3-b_2\right)}\right]\label{EW-dj-3}
	\end{align}
	
	\subsubsection{Double crossing configurations}\label{dj-double-crossing}
	Next, we consider configurations of the RT saddles corresponding to the entanglement entropy of $A\cup B$ such that one or more RT surfaces cross the brane twice. Within each such configuration, we will construct the bulk entanglement wedge dual to $\rho_{AB}$ and systematically investigate the phase transitions of the minimal EWCS for different subsystem sizes and the geometry. To proceed, we further divide the possible RT saddles into two sub-classes. First we consider the RT surfaces homologous to $C=[b_2,b_3]_\text{I}\cup [b_2,b_3]_\text{II}$ which do not cross the brane. In the second case we explore the possibility of $C$ owning an island by considering the RT surfaces homologous to $C$ which crosses the brane and comes back\footnote{\label{footnote11}There is yet another possibility where both the RT surfaces homologous to $C$ and $ABC$ has a double crossing topology. However, we have checked numerically that this situation fails to arise for a sufficiently large range of parameter values and hence in the following we shall drop this possibility from our discussion}.

	\subsubsection*{A. RT surfaces homologous to C which do not cross the brane}\label{section:double-crossing-C}
	We begin with the configuration where in the AdS$_3^\text{II}$ geometry, the geodesics homologous to the intervals $\left[{b }_1,{b }_2\right]_{\text{I}}\cup \left[b _3,b _4\right]_{\text{II}}$ in the CFT$_2^\text{II}$ have the following topology: 
	\begin{itemize}
		\item the geodesic connecting $b_1$ and $b_4$ crosses the EOW brane twice at the bulk points distant $\bar{y}^{*}_1$ and $\bar{y}^{*}_4$ along the brane from the interface. In other words, the geodesic is made up of three semi-circular segments, one of which resides entirely in the AdS$_3^\text{I}$ geometry. Note that the locations of the points $\bar{y}^{*}_1$ and $\bar{y}^{*}_4$ on the EOW brane should be determined by solving the extremization conditions\footnote{\label{footnote12}In this case, as is clear from the context, the parameters $\Theta_D$ and $k_D$ in \cref{doublecrossing-ext-k} should be replaced by $\Theta_{ABC}=\frac{b_4}{b_1}$ and $k_{ABC}=\sqrt{\frac{y_4}{\Theta_{ABC} y_1}}$ respectively.} given in \cref{doublecrossing-ext-k} and $\bar{y}_1=\frac{b_1}{k_{ABC}}$. We note that such configurations occur when $\Theta_{ABC}$ is larger than a critical value \cite{Anous:2022wqh}.
		\item the geodesic semi-circle connecting $b_2$ and $b_3$ never crosses the brane and has a dome like structure.
	\end{itemize} 
	
	On the other hand, the geodesics homologous to the subsystems in the CFT$_2^\text{I}$ consist of single semi-circles and have the topology of a dome. The schematics of this configuration is sketched in \cref{fig:disj-double1}. The bulk entanglement wedge is the region of the spacetime bounded by these geodesics and the corresponding subsystems as shown by the shaded regions in \cref{fig:disj-double1}. The entanglement entropy for $A\cup B$ in this phase is given by
	\begin{align}\label{EE_disj_2}
		S(A\cup B)=S_\text{double}\left([b_1,b_4]\right)+\frac{L_\text{I}+L_\text{II}}{2G_N}\log\left(\frac{b_3-b_2}{\epsilon}\right)+\frac{L_\text{I}}{2G_N}\log\left(\frac{b_4-b_1}{\epsilon}\right)\,,
	\end{align}
	where $S_\text{double}$ is defined in \cref{EE-double-cross}.
	
	The minimal EWCS for this configuration consists of two extremal curves, one of which resides entirely in the AdS$_3^\text{I}$ geometry and corresponds to the usual notion of EWCS in standard AdS$_3$/CFT$_2$ scenario. The minimal EWCS residing entirely in the AdS$_3^\text{I}$ region may be computed using the standard AdS$_3$/CFT$_2$ techniques \cite{Takayanagi:2017knl,Nguyen:2017yqw} and the result reads as
	\begin{align}
		E_W^{\,\text{I}}(A:B)=\frac{L_\text{I}}{4G_N}\log\left[1+2 \eta+2 \sqrt{\eta (\eta+1)}\right]\,,\label{standard-EW}
	\end{align}
	where the cross-ratio $\eta$ is given by
	\begin{align}
		\eta=\frac{\left(b _2-b _1\right) \left(b _4-b _3\right)}{\left(b _3-b _2\right) \left(b _4-b _1\right)}\,.\label{Dome-cross-ratio}
	\end{align}
	
	On the other hand, there are three possible choices for the other extremal curve which we shall consider below.
	\subsubsection*{Phase-I}
\begin{figure}[ht]
		\centering
		\includegraphics[scale=0.45]{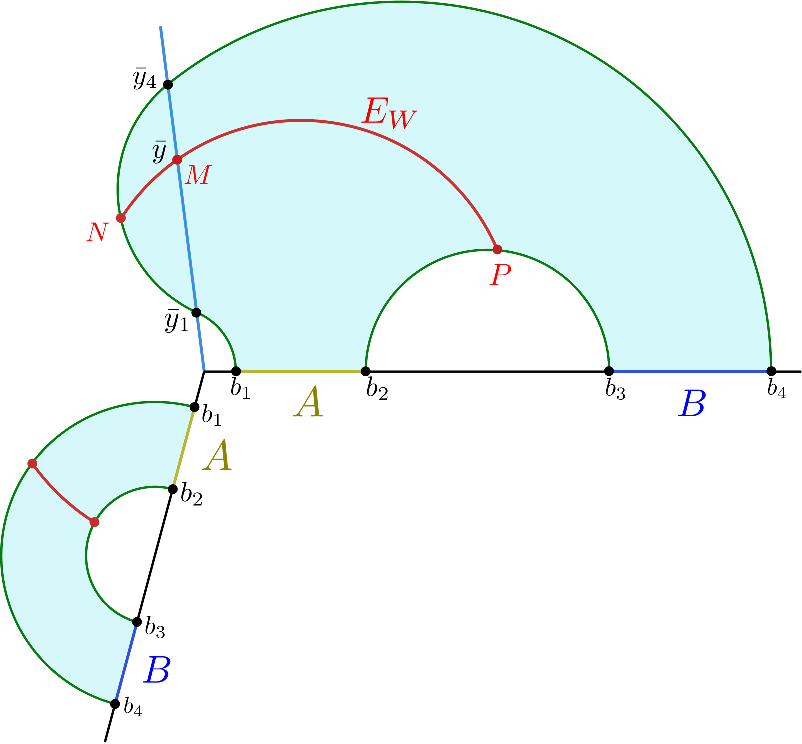}
		\caption{disjoint intervals : double-crossing phase-I}
		\label{fig:disj-double1}
	\end{figure}

\begin{figure}[ht]
		\centering
		\includegraphics[scale=0.65]{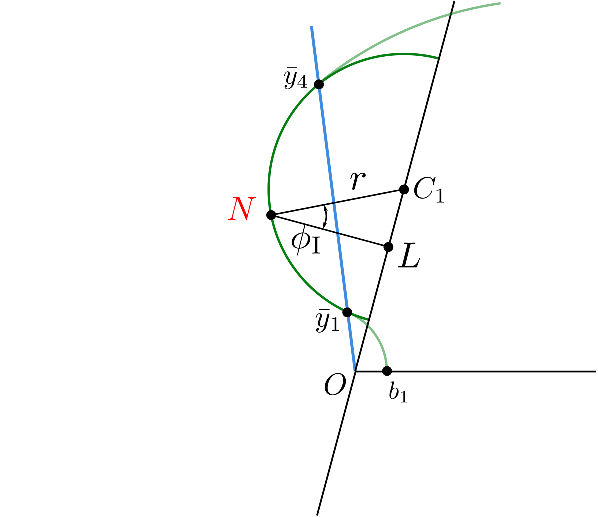}
		\caption{ Computation of the minimal EWCS for the disjoint intervals in double-crossing phase-I.}
		\label{fig:disj-double1-calc}
	\end{figure}

	In the first phase, we allow the candidate extremal curve for the minimal EWCS originating in the AdS$_3^\text{II}$ geometry to cross the brane and probe the geometry beyond the ``end of the world". This phase occurs when the sizes of the subsystems $A$ and $B$ are comparable. The schematics of this configuration is sketched in \cref{fig:disj-double1}. 
	To compute the length of this candidate extremal curve, note that it consists of two circular geodesic segments joined smoothly at the location of the brane. The segment $MP$ starts from the point P on the dome shaped RT surface connecting $b_2$ and $b_3$ and ends on the EOW brane at the point M on the EOW brane which is at a distance $\bar{y}$ from the interface O. The other circular arc $MN$ ends on the geodesic segment which connects the bulk points $\bar{y}_1$ and $\bar{y}_4$. Therefore, the total length of this surface is given by $d_{NP}=d_{MP}+d_{MN}$. The Poincar\'e coordinates of the points $P$ and $M$ may be read off from \cref{ccordinate-P,coordinate-R}. To obtain the coordinates of the point $N$, consider the diagram in \cref{fig:disj-double1-calc}, where $r$ and $x_0$ are the radius and center coordinates of the circular arc connecting $\bar{y}_1$ and $\bar{y}_4$, and the arbitrary angle $\phi_\text{I}$ parametrizes the position of $N$ on this circular arc.
	
	From \cref{fig:disj-double1-calc}, the Poincar\'e coordinates of $N$ may be read off as 
	\begin{align}
		N:(x_\text{I},z_\text{I})=\left(\overline{C_1L},\overline{NL}\right)=\left(\overline{OC_1}-\overline{OL},\overline{NL}\right)=\left(x_0-r\sin\phi_{\text{I}},r\cos\phi_{\text{I}}\right)\,,\label{N-parametrization}
	\end{align}
	where the center and the radius of the circular arc are given as
	\begin{align}
		r= \frac{\sqrt{\displaystyle\bar{y}_1^{*2}+\bar{y}_4^{*2}+2\bar{y}_1^*\bar{y}_4^* \cos\left(2 \psi _\text{I}\right)}}{2\sin\psi_\text{I}}~~ ,~~x_{0}= \frac{\bar{y}_1^*+\bar{y}_4^*}{2\sin\psi_\text{I}}\,.
	\end{align}
	Now utilizing the length formula in \cref{Poincare-geod}, we may obtain the length of the candidate EWCS as follows
	\begin{align}
		d_{NP}&=L_{\text{II}}\cosh^{-1}\left[\frac{\left(\frac{b_3+b_2}{2}+\frac{b_3-b_2}{2}\sin\phi_{\text{II}}+\bar{y}\sin\psi_{\text{II}}\right)^2+\left(\frac{b_3-b_2}{2}\cos\phi_{\text{II}}\right)^2+(\bar{y} \cos \psi_{\text{II}})^2}{2 \left(\frac{b_3-b_2}{2}\cos\phi_{\text{II}}\right) \bar{y}\cos\psi_{\text{II}}}\right]\notag\\
		&~~+L_\text{I}\cosh^{-1}\left[\frac{(x_0-r\sin\phi_\text{I}-\bar{y}\sin\psi_{\text{I}})^2+(\bar{y}\cos\psi_{\text{I}})^2+(r\cos\phi_{\text{I}})^2}{2\left(\bar{y}\cos\psi_{\text{I}}\right)\left(r\cos\phi_{\text{I}}\right)}\right]
	\end{align}
	Extremizing the above length over $\phi_{\text{I}}$ and $\phi_{\text{II}}$, we obtain the following extremal values
	\begin{align}
		&\phi_{\text{I}}=\sin ^{-1}\left[\frac{2 r (x_0-\bar{y}\sin\psi_\text{I})}{r^2+x_0^2-2 x_0 \bar{y} \sin\psi_\text{I}+\bar{y}^2}\right]\notag\\
		&\phi_{\text{II}}=\sin^{-1}\left[\frac{(b_2-b_3) (b_2+b_3+2 \bar{y} \sin\psi_\text{II})}{b_2^2+b_3^2+2 \bar{y}^2+2 \bar{y} (b_2+b_3) \sin\psi_\text{II}}\right]
	\end{align}
	Substituting these and subsequently extremizing over the remaining parameter $\bar{y}$, we obtain
	\begin{align}
		&\frac{L_\text{I} \left(r^2-x_0^2+\bar{y}^2\right)}{\bar{y}\sqrt{\left(r^2-2 x_0 \bar{y} \sin\psi_{\text{I}}+x_0^2+\bar{y}^2\right){}^2-4 r^2 \left(x_0-\bar{y} \sin\psi_{\text{I}}\right){}^2}}\notag\\
		&\qquad\qquad\qquad\qquad\qquad=\frac{L_{\text{II}} \left(\bar{y}^2-b _2 b _3\right)}{\bar{y}\sqrt{\left(2 b _2 \bar{y} \sin\psi _{\text{II}}+b _2^2+\bar{y}^2\right) \left(2 b _3 \bar{y} \sin \psi _{\text{II}}+b _3^2+\bar{y}^2\right)}}\label{extr-disj4}
	\end{align}
	The algebraic equation in \cref{extr-disj4} may now be solved for $\bar{y}$ and the corresponding extremal value $\bar{y}=\bar{y}^*$ determines the minimal EWCS to be
	\begin{align}
		E_W(A:B)=&\frac{L_{\text{II}}}{4G_N}\cosh^{-1}\left[\frac{\sqrt{\left(\bar{y}^*{}^2+b _2^2+2b_2 \bar{y}^*\sin\psi_{\text{II}}\right) \left(\bar{y}^*{}^2+b _3^2+2b_3 \bar{y}^*\sin\psi_{\text{II}}\right)}}{\bar{y}^*\left(b_3-b_2\right)\cos\psi_{\text{II}}}\right]\notag\\
		&+\frac{L_\text{I}}{4G_N} \cosh ^{-1}\left[\frac{\sqrt{\left(r^2-2 x_0 \bar{y}^* \sin\psi_{\text{I}}+x_0^2+\bar{y}^*{}^2\right){}^2-4 r^2 \left(x_0-\bar{y}^* \sin\psi_{\text{I}}\right){}^2}}{2\bar{y}^* r  \cos\psi_\text{I}}\right]\notag\\
		&+\frac{L_\text{I}}{4G_N}\log\left[1+2 \eta+2 \sqrt{\eta (\eta+1)}\right]\,,
	\end{align}
	where we have included the contribution from the left geometry in the final expression. In the large tension limit $\delta\to 0$, the extremization condition in \cref{extr-disj4} reduces to
	\begin{align}
		\frac{L_\text{II} \left(\bar{y}^2-b _2 b _3\right)}{\left(b _2+\bar{y}\right) \left(b _3+\bar{y}\right)}+L_\text{I} \left(\frac{\bar{y}}{\bar{y}-\bar{y}_4^*}+\frac{\bar{y}_1^*}{\bar{y}-\bar{y}_1^*}\right)=0\,.\label{Ext-EW-disj-zeroT-doubleI}
	\end{align}
	Substituting the extremal value $\bar{y}=\bar{y}^*$, the EWCS may now be obtained in the large tension limit to be
	\begin{align}
		E_W(A:B)=&\frac{L_\text{I}}{4G_N}\log\left[\frac{\left(\bar{y}^*_4-\bar{y}^*\right)\left(\bar{y}^*-\bar{y}^*_1\right)}{\bar{y}^*\left(\bar{y}^*_4-\bar{y}^*_1\right)}\right]+\frac{L_\text{II}}{4G_N}\log\left[\frac{\left(\bar{y}^*+b_2\right)\left(\bar{y}^*+b_3\right)}{\bar{y}^*\left(b_3-b_2\right)}\right]+S_\text{int}^{(\delta)}\notag\\
		&+\frac{L_\text{I}}{4G_N}\log\left[1+2 \eta+2 \sqrt{\eta (\eta+1)}\right]\,,\label{EW-disj-zeroT-doubleI}
	\end{align}
	where $S_\text{int}^{(\delta)}$ is the $\delta\to 0$ limit of the interface entropy, defined in \cref{S-int}.

	\subsubsection*{Phase-II}
\begin{figure}[H]
		\centering
		\includegraphics[scale=0.45]{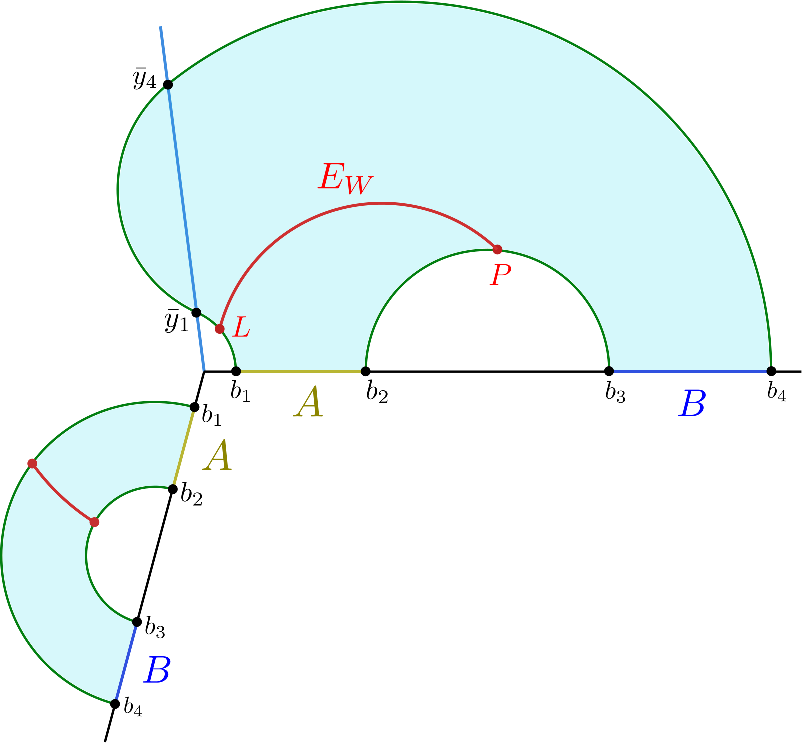}
		\caption{disjoint intervals : double-crossing phase-II}
		\label{fig:disj-double2}
	\end{figure}

	In the next phase, when the size of the subsystem $A$ is small compared to that of $B$, the minimal EWCS ends on the smaller segment of the double crossing RT surface anchored on $b_1$, as depicted in \cref{fig:disj-double2}.
	
\begin{figure}[ht]
		\centering
		\includegraphics[scale=1]{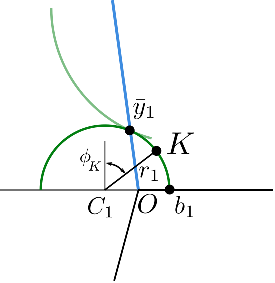}
		\caption{Computation of the minimal EWCS for two disjoint subsystems in double-crossing phase-II}
		\label{fig:disj-double2-calc}
	\end{figure}

	To compute the length of the minimal EWCS we consider a candidate surface which ends on an arbitrary point $K$ parametrized by an angle $\phi_K$, on the segment of the RT surface anchored on $b_1$. From \cref{fig:disj-double2-calc}, the Poincar\'e coordinates of the point $K$ may be read off as follows
	\begin{align}
		K:(x_\text{II},z_\text{II})=\left(\overline{OT},\overline{TK}\right)=\left(\overline{C_1T}-\overline{OC_1},\overline{TK}\right)=\left(r_1\sin\phi_K-x_{C_1},r_1\cos\phi_K\right)\,,
	\end{align}
	where the radius $r_1$ and the center coordinate $|x_{C_1}|$ of the circular geodesic connecting $b_1$ and $\bar{y}_1^*$ are given by
	\begin{align}
		r_1=b _1+x_{C_1}~~,~~x_{C_1}=-\frac{\bar{y}_1^{*2}-b _1^2}{2 \left(b _1+\bar{y}_1^* \sin \left(\psi _{\text{II}}\right)\right)}\,.
	\end{align}
	The other endpoint of the candidate EWCS may be parametrized by another arbitrary angle $\phi_\text{II}$ similar to \cref{ccordinate-P}. Now, utilizing, the general formula for the geodesic length in \cref{Poincare-geod}, we may obtain the length of the candidate surface as follows
	\begin{align}
		d_{PK}=L_\text{II} \cosh ^{-1}\left[\frac{\left(x_{C_1}-r_1\sin\phi_K+\frac{b_2+b_3}{2}+\frac{b_3-b_2}{2}\sin\phi_{\text{II}}\right){}^2+\left(\frac{b_3-b_2}{2}\cos\phi_{\text{II}}\right){}^2+\left(r_1\cos\phi_K\right){}^2}{2\left(\frac{b _3-b _2}{2}\cos \phi_{\text{II}}\right)r_1\cos\phi _K }\right]\label{candidate-EW-double-2}
	\end{align}
	Extremizing the above length over the arbitrary angles $\phi_\text{II}$ and $\phi_K$, we obtain the extremal solutions to be
	\begin{align}
		&\phi _K=\sin ^{-1}\left[\frac{r_1 \left(b _2+b _3+2 x_{C_1}\right)}{r_1^2+\left(b _2+x_{C_1}\right) \left(b _3+x_{C_1}\right)}\right]\notag\\
		&\phi _{\text{II}}= \sin ^{-1}\left[\frac{\left(b _2-b _3\right) \left(b _2+b _3+2 x_{C_1}\right)}{-2 r_1^2+b _2^2+b _3^2+2 \left(b _2+b _3\right) x_{C_1}+2 x_{C_1}^2}\right]
	\end{align}
	Substituting these in \cref{candidate-EW-double-2} the minimal EWCS is obtained as follows
	\begin{align}
		&E_W(A:B)\notag\\
		&=\frac{L_\text{I}}{4G_N}\log\left[1+2 \eta+2 \sqrt{\eta (\eta+1)}\right]+\frac{L_\text{II}}{4G_N}\cosh^{-1}\left[\frac{r_1^2-\left(x_{C_1}+b _2\right) \left(x_{C_1}+b _3\right)}{r_1 \left(b _3-b _2\right)}\right]\notag\\
		&=\frac{L_\text{I}}{4G_N}\log\left[1+2 \eta+2 \sqrt{\eta (\eta+1)}\right]\notag\\
		&+\frac{L_\text{II}}{4G_N}\cosh^{-1}\left[\frac{2 \left(b _1^2-b _2 b _3\right) \bar{y}_1^* \sin\psi _{\text{II}}+b _1 \left(b _1 \left(b _2+b _3\right)-2 b _2 b _3\right)+\left(2 b _1-b _2-b _3\right) \bar{y}_1^{*2}}{	\left(b _3-b _2\right) \left(2 b _1 \bar{y}_1^* \sin\psi _{\text{II}}+b _1^2+\bar{y}_1^{*2}\right)}\right]\,.
	\end{align}
	In the above expression, we have included the contribution from the left geometry as well.
	In the $\delta\to 0$ limit, the minimal EWCS reduces to
	\begin{align}
		E_W(A:B)=\frac{L_\text{I}}{4G_N}\log\left[1+2 \eta+2 \sqrt{\eta (\eta+1)}\right]+\frac{L_\text{II}}{4G_N}\cosh^{-1}\left[1+2\frac{\left(b _1-b _2\right) \left(b _3+\bar{y}_1^*\right)}{\left(b _2-b _3\right) \left(b _1+\bar{y}_1^*\right)}\right]\label{EW-dj-double-II}
	\end{align}

	\subsubsection*{Phase-III}
\begin{figure}[ht]
		\centering
		\includegraphics[scale=0.45]{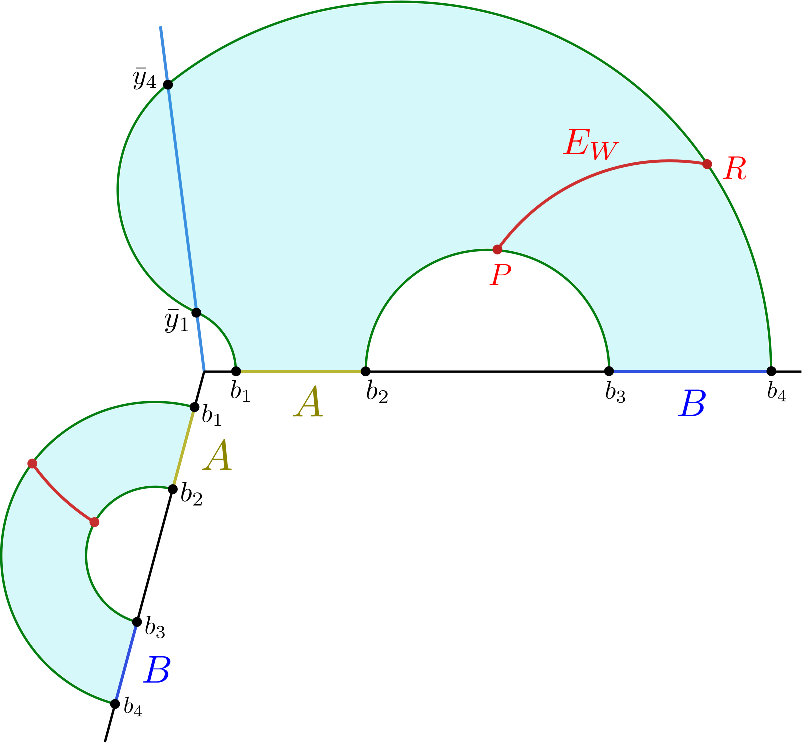}
		\caption{disjoint intervals : double-crossing phase-III}
		\label{fig:disj-double3}
	\end{figure}

	The final phase of the EWCS considering the present structure of the entanglement entropy of $A\cup B$ concerns a geodesic in the AdS$^\text{II}_3$ geometry, emanating from the dome connecting $b_2$ and $b_3$ and ending on the larger segment of the double crossing RT surface anchored on $b_4$. The schematics of the configuration is depicted in \cref{fig:disj-double3}. 
	
	The radius $r_2$ and the coordinate of the center of the circular geodesic segment connecting $b_4$ and $\bar{y}^{*}_4$ are given by
	\begin{align}
		r_2=b_4+x_{C_2}~~,~~x_{C_2}=\frac{b_4^2-\bar{y}_4^{*2}}{2 \left(\bar{y}_4^* \sin \left(\psi _{\text{II}}\right)+b _4\right)}\,.
	\end{align}
	The computation of the length of the minimal EWCS follows very closely the analysis in the previous subsection and hence we skip the details here. The minimal EWCS, including the contribution from the left geometry, is then given by 
	\begin{align}
		E_W(A:B)=\frac{L_\text{I}}{4G_N}\cosh^{-1}\left[1+2 \eta\right]+\frac{L_\text{II}}{4G_N}\cosh^{-1}\left[\frac{r_2^2-\left(x_{C_2}-b_2\right) \left(x_{C_2}-b_3\right)}{r_2 \left(b_3-b_2\right)}\right]
	\end{align}
	In the large tension limit, the above expression simplifies to
	\begin{align}
		E_W(A:B)=\frac{L_\text{I}}{4G_N}\log\left[1+2 \eta+2 \sqrt{\eta (\eta+1)}\right]+\frac{L_\text{II}}{4G_N}\cosh^{-1}\left[1+2\frac{ \left(b _3-b _4\right) \left(b _2+\bar{y}_4^*\right)}{\left(b _2-b _3\right) \left(b _4+\bar{y}_4^*\right)}\right]\label{EW-dj-double-III}
	\end{align}
	
	\subsection*{B. Double crossing RT surface for subsystem $C$}
\begin{figure}[h!]
		\centering
		\includegraphics[scale=0.4]{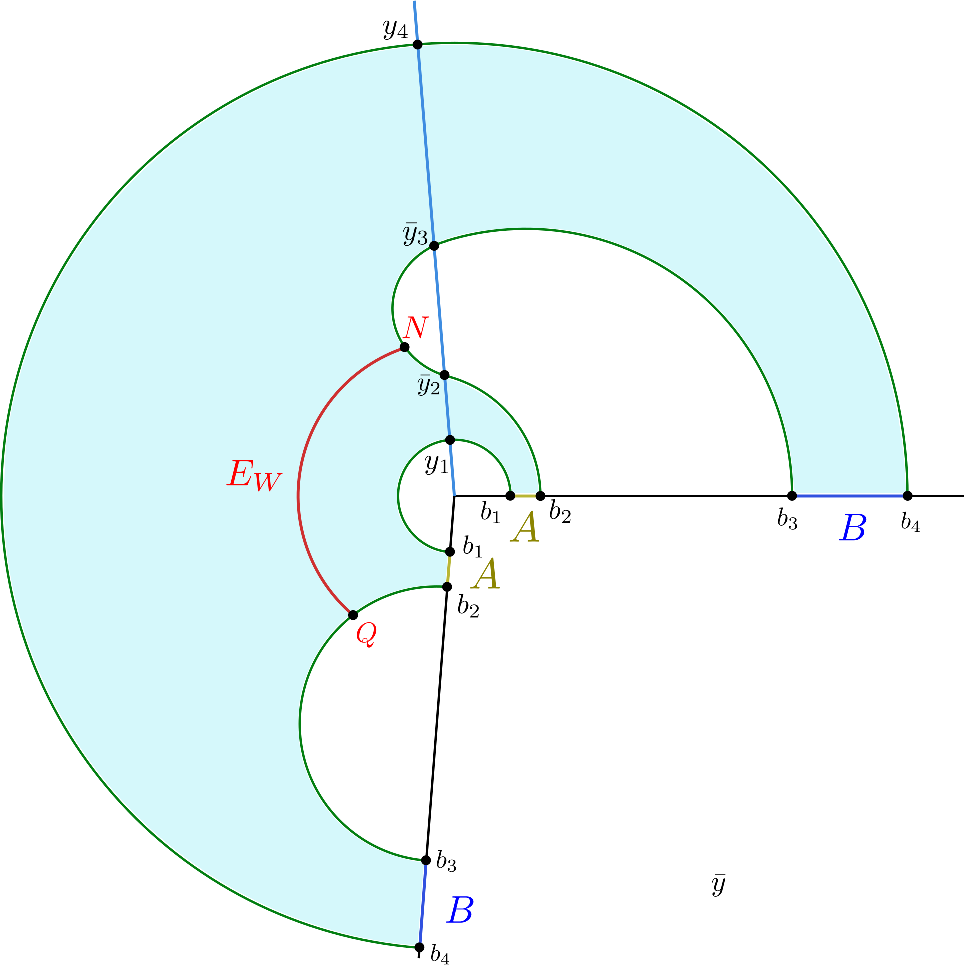}
		\caption{disjoint intervals : double-crossing phase-IV}
		\label{fig:disj-newdouble1}
	\end{figure}

	Next we consider another phase for the RT saddles corresponding to the entanglement entropy of $A\cup B$ sketched in \cref{fig:disj-newdouble1}. In this phase, there exists a double crossing RT surface homologous to the interval $[b_2,b_3]_\text{II}$ on the CFT$^\text{II}_2$, which crosses the EOW brane twice at the bulk points $\bar{y}_2^*$ and $\bar{y}_3^*$ respectively. This phase becomes dominant when $\Theta_C=\frac{b_3}{b_2}$ is greater than its critical value. The locations of these bulk points are determined by solving the extremization condition in \cref{doublecrossing-ext-k} together with $\bar{y}_2=\frac{b_2}{k_C}$, where $k_C^2=\frac{\bar{y}_3}{\Theta_C \bar{y}_2}$.
	
	This configuration may be understood from the single-crossing one in \cref{fig:SingleEEdisj} as the dome shaped geodesic connecting $b_2$ and $b_3$ undergoes a phase transition to a double-crossing one. Recall that, as determined earlier, the single crossing geodesics in \cref{fig:SingleEEdisj} cross the EOW brane at the bulk points $y^*_1=b_1$ and $y^*_4=b_4$ respectively. Therefore, in this phase, the entanglement entropy of $A\cup B$ is given by
	\begin{align}\label{EE_disj_3}
		S(A\cup B)=S_\text{double}\left([b_2,b_3]\right)+\frac{L_\text{I}+L_\text{II}}{4G_N}\left[\log\left(\frac{2b_1}{\epsilon}\right)+\log\left(\frac{2b_4}{\epsilon}\right)\right] +\frac{\rho_{\text{I}}^*+\rho_{\text{II}}^*}{2 G_N}
	\end{align}
	
	The entanglement wedge dual to the density matrix $\rho_{AB}$ is depicted by the shaded region in \cref{fig:disj-newdouble1}. Within this phase of $S(A\cup B)$, the minimal EWCS can undergo phase transitions depending upon the subsystem sizes and their relative distances from the interface. Note that, in principle, segments of the candidate curves for the EWCS may penetrate into the AdS$^{\text{II}}_3$ geometry. However, similar to \cite{Anous:2022wqh}, we may argue that passing through the geometry which is less curved (recall that $L_\text{II}>L_\text{I}$) increases the total length. Hence, we may conclude that the minimal curves reside within the AdS$^{\text{I}}_3$ geometry and never probe the AdS$^{\text{II}}_3$ as depicted in \cref{fig:disj-newdouble1,fig:disj-newdouble2,fig:disj-newdouble3}. 
	
	\subsubsection*{Phase-IV}
	In the first phase sketched in \cref{fig:disj-newdouble1}, the EWCS is given by the minimal geodesic between the double crossing RT surface emanating from the right asymptotic boundary and the dome-shaped RT surface connecting $b_2$ and $b_3$ on the left asymptotic boundary. We may parametrize an arbitrary point $N$ on the double-crossing geodesic by the angle $\phi_N$, similar to \cref{N-parametrization}, with the radius and center coordinate of the semi-circular arc in AdS$^\text{I}_3$ given as follows
	\begin{align}
		r=\frac{\sqrt{\displaystyle \bar{y}_2^{*2}+\bar{y}_3^{*2}+2 \bar{y}_3^* \bar{y}_2^* \cos \left(2 \psi _1\right)}}{2 \sin \left(\psi _1\right)}~~,~~x_0=\frac{\bar{y}_2^*+\bar{y}_3^*}{2 \sin \left(\psi _1\right)}\,.
	\end{align}
	On the other hand the Poincar\'e coordinates of the arbitrary point $Q$ on the dome-shaped RT in the left geometry are given in \cref{ccordinate-P} with $\phi_\text{II}$ replaced by $\phi_Q$. Therefore, utilizing the formula in \cref{Poincare-geod}, the length of this candidate surface may be computed as follows
	\begin{align}
		d_{NQ}=L_\text{I}\cosh^{-1}\left[\frac{\left(\frac{b_3+b_2}{2}+\frac{b_3-b_2}{2}\sin\phi_{Q}+x_0-r \sin\phi _N\right){}^2+\left(\frac{b_3-b_2}{2}\cos\phi_{Q}\right){}^2+\left(r\cos\phi _N\right){}^2}{2\left(\frac{b_3-b_2}{2}\cos\phi_{Q}\right)r\cos\phi _N}\right]
	\end{align}
	Extremizing over the arbitrary angles $\phi_N$ and $\phi_{Q}$, we obtain the extremal values to be
	\begin{align}
		&\phi _N=\sin ^{-1}\left[\frac{r \left(b_2+b_3+2 x_0\right)}{\left(b_2+x_0\right) \left(b_3+x_0\right)+r^2}\right]\notag\\
		&\phi _Q=\sin ^{-1}\left[\frac{\left(b_2-b_3\right) \left(b_2+b_3+2 x_0\right)}{2 \left(b_2+b_3\right) x_0+b_2^2+b_3^2-2 r^2+2 x_0^2}\right]
	\end{align}
	Substituting these in the expression for the length, we obtain the minimal EWCS to be
	\begin{align}
		E_W(A:B)&=\frac{L_\text{I}}{4G_N}\cosh^{-1}\left[\frac{\left(b_2+x_0\right) \left(b_3+x_0\right)-r^2}{\left(b_3-b_2\right) r}\right]\notag\\
		&=\frac{L_\text{I}}{4G_N}\cosh^{-1}\left[\frac{2 \left(b_2 b_3+\bar{y}_2^* \bar{y}_3^*\right) \sin \left(\psi _1\right)+\left(b_2+b_3\right) \left(\bar{y}_2^*+\bar{y}_3^*\right)}{\left(b_3-b_2\right) \sqrt{\bar{y}_2^{*2}+\bar{y}_3^{*2}+2 \bar{y}_3^* \bar{y}_2^* \cos \left(2 \psi _1\right)}}\right]
	\end{align}
	In the limit of large brane tension $\delta\to 0$, the above expression reduces to
	\begin{align}\label{djdc1}
		E_W(A:B)=\frac{L_\text{I}}{4G_N}\log\left[1+2\zeta+2\sqrt{\zeta(\zeta+1)}\right]\,,
	\end{align}
	where the cross-ratio $\zeta$ is given by
	\begin{align}
		\zeta=\frac{\left(b_3+\bar{y}_2^*\right) \left(b_2+\bar{y}_3^*\right)}{\left(b_3-b_2\right) \left(\bar{y}_2^*-\bar{y}_3^*\right)}\,.
	\end{align}
	
	\subsubsection*{Phase-V}
\begin{figure}[ht]
		\centering
		\includegraphics[scale=0.4]{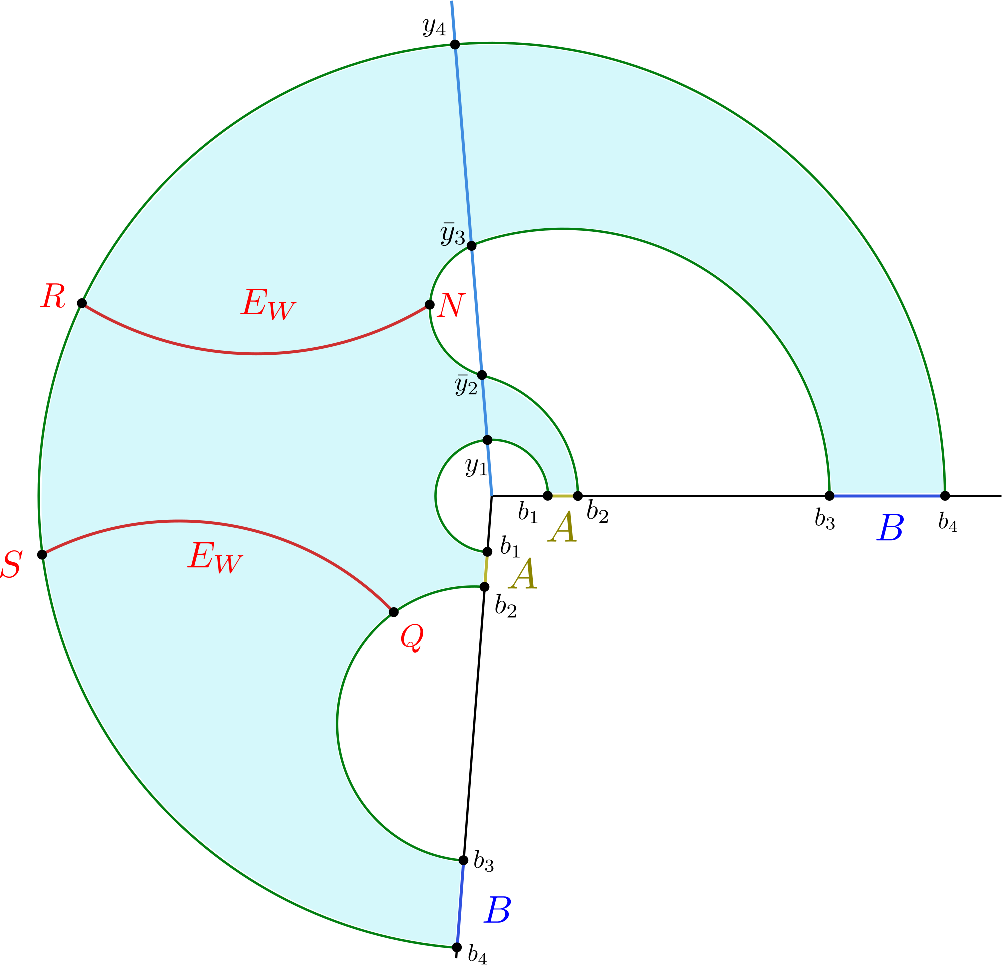}
		\caption{disjoint intervals : double-crossing phase-V}
		\label{fig:disj-newdouble2}
	\end{figure}

	Next, we consider the configuration where the candidate EWCS comprises of two disconnected geodesic segments one of which connects the double crossing RT surface and the bigger single crossing one connecting $b_4$ on either side. On the other hand, the second segment connects the dome shaped RT surface and the bigger single crossing one. The schematics of the configuration is depicted in \cref{fig:disj-newdouble2}.
	
	As described earlier, we may parametrize the endpoints of these geodesic segments on the double crossing and dome-shaped RT surfaces similar to \cref{N-parametrization,ccordinate-P}. Furthermore, recall that the single crossing RT surface cross the EOW brane at $y_4^*=b_4$ and hence the endpoints of the geodesic segments on this surface may be parametrized as in \cref{parametrization-S}. Therefore, utilizing \cref{Poincare-geod}, the total length of the two geodesics segments in \cref{fig:disj-newdouble2} may be computed as follows
	\begin{align}
		d&=d_{NR}+d_{QS}\notag\\
		&=L_\text{I}\cosh^{-1}\left[\frac{\left(x_0-r\sin\phi_N+b_4\sin\phi_R\right){}^2+\left(r\cos\phi_N\right){}^2+\left(b_4\cos\phi_R\right){}^2}{2 rb_4\cos\phi _N\cos\phi_R}\right]\notag\\
		&~~+L_{\text{I}} \cosh^{-1}\left[\frac{\left(b_4\sin\phi_S-\left(\frac{b_2+b_3}{2}+\frac{b_3-b_2}{2}\sin\phi_Q\right)\right)^2+\left(\frac{b_3-b_2}{2}\cos\phi_Q\right)^2+\left(b _4 \cos\phi_S\right)^2}{2\left(\frac{b_3-b_2}{2}\cos \phi_Q\right) b_4\cos\phi_S}\right]
	\end{align}
	Extremizing over the arbitrary angles $\phi_N$, $\phi_{Q}$, $\phi_{R}$ and $\phi_{S}$, we obtain
	\begin{align}
		&\phi_R=-\sin ^{-1}\left[\frac{2 b _4 x_0}{-r^2+b _4^2+x_0^2}\right]~~,~~\phi_N=\sin ^{-1}\left[\frac{2 r x_0}{r^2-b _4^2+x_0^2}\right]\notag\\
		&\phi_Q=\sin ^{-1}\left[\frac{b _2^2-b _3^2}{b _2^2+b _3^2-2 b _4^2}\right]~~,~~\phi_S=\sin ^{-1}\left[\frac{\left(b _2+b _3\right) b _4}{b _4^2+b _2 b _3}\right]
	\end{align}
	Substituting these extremal values, we obtain the minimal EWCS to be
	\begin{align}
		E_W(A:B)&=\frac{L_\text{I}}{4G_N}\left(\cosh ^{-1}\left[\frac{b _4^2-b _2 b _3}{\left(b _3-b _2\right) b _4}\right]+\cosh ^{-1}\left[\frac{2 r b _4}{r^2+b _4^2-x_0^2}\right]\right)\notag\\
		&=\frac{L_\text{I}}{4G_N}\left(\cosh ^{-1}\left[\frac{b _4^2-b _2 b _3}{\left(b _3-b _2\right) b _4}\right]+\cosh ^{-1}\left[\frac{\sin\psi_{\text{I}} \left(b _4^2-\bar{y}_2^* \bar{y}_3^*\right)}{b _4 \sqrt{2 \bar{y}_3^* \bar{y}_2^* \cos \left(2 \psi_\text{I}\right)+\bar{y}_2^{*2}+\bar{y}_3^{*2}}}\right]\right)\label{disj-newdouble2}
	\end{align}
	In the $\delta\to 0$ limit, the above expression reduces to
	\begin{align}\label{djdc2}
		E_W(A:B)&=\frac{L_\text{I}}{4G_N}\left(\cosh ^{-1}\left[\frac{b _4^2-b _2 b _3}{\left(b _3-b _2\right) b _4}\right]+\log\left[1+2\xi+2\sqrt{\xi(\xi+1)}\right]\right)
	\end{align}
	where the cross ratio $\xi$ is given by
	\begin{align}
		\xi=\frac{\left(b _4-\bar{y}_2^*\right) \left(b _4+\bar{y}_3^*\right)}{2 b _4 \left(\bar{y}_2^*-\bar{y}_3^*\right)}.
	\end{align}
	
	\subsubsection*{Phase-VI}
\begin{figure}[H]
		\centering
		\includegraphics[scale=0.4]{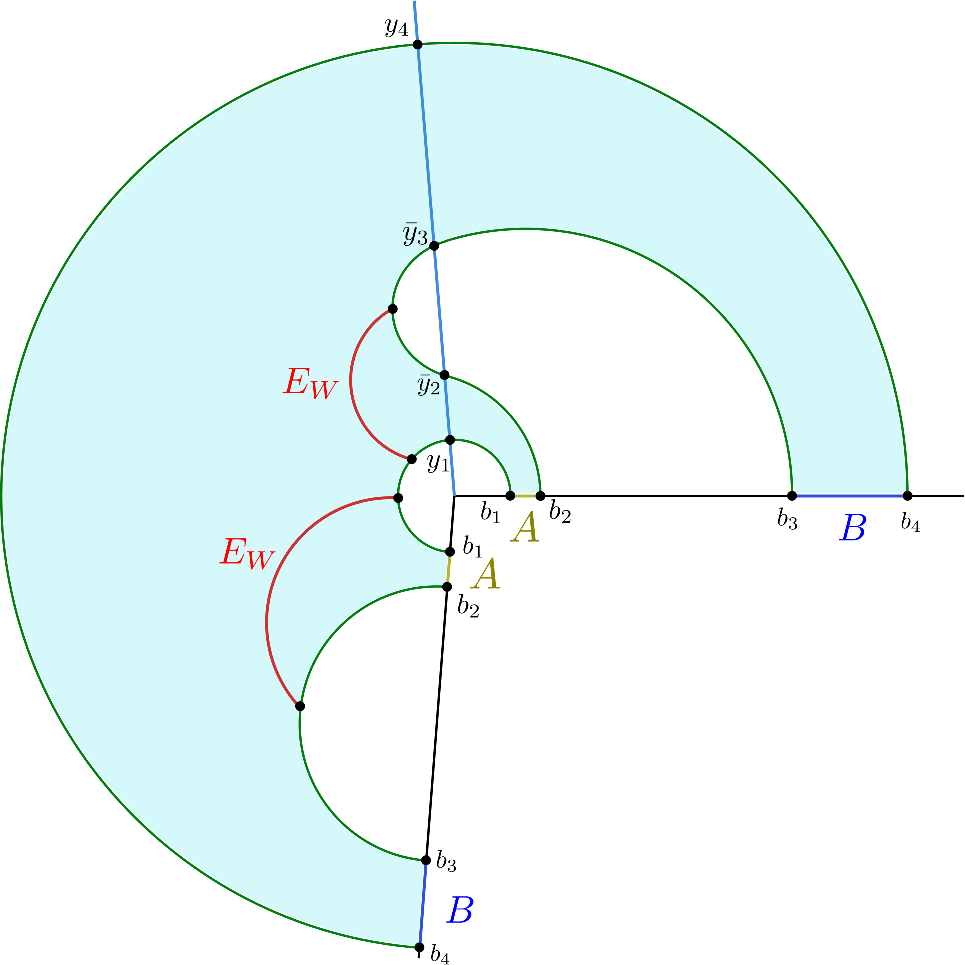}
		\caption{disjoint intervals : double-crossing phase-VI}
		\label{fig:disj-newdouble3}
	\end{figure}

	There is one more possibility for the EWCS where two disconnected geodesic segments land on the smaller single crossing RT surface, as depicted in \cref{fig:disj-newdouble3}. The computation of the lengths are similar to that in the previous subsection and we may obtain the expression from \cref{disj-newdouble2} via the replacement $b_4\rightarrow b_1$ as follows
	\begin{align}
		E_W(A:B)=\frac{L_\text{I}}{4G_N}\left(\cosh ^{-1}\left[\frac{b _2 b _3-b _1^2}{\left(b _3-b _2\right) b _1}\right]+\cosh ^{-1}\left[\frac{\sin\psi_{\text{I}} \left(b _1^2-\bar{y}_2^* \bar{y}_3^*\right)}{b _1 \sqrt{2 \bar{y}_3^* \bar{y}_2^* \cos \left(2 \psi_\text{I}\right)+\bar{y}_2^{*2}+\bar{y}_3^{*2}}}\right]\right)
	\end{align}
	In the large tension limit ($\delta\to 0$), the above expression reduces to
	\begin{align}\label{djdc3}
		E_W(A:B)&=\frac{L_\text{I}}{4G_N}\left(\cosh ^{-1}\left[\frac{b _2 b _3-b _1^2}{\left(b _3-b _2\right) b _1}\right]+\cosh^{-1}\left[\frac{b _1^2-\bar{y}_2^* \bar{y}_3^*}{b _1 \left(\bar{y}_2^*-\bar{y}_3^*\right)}\right]\right).
	\end{align}

	\subsection*{RT saddles with no brane crossing}
	\begin{figure}[ht]
		\centering
		\includegraphics[scale=0.45]{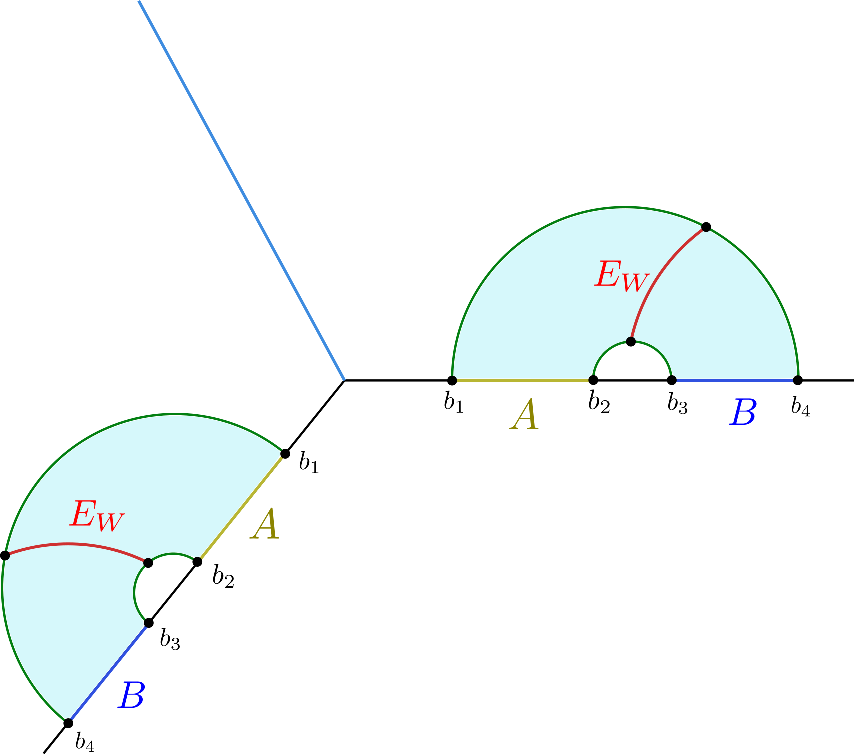}
		\caption{disjoint intervals : EWCS without brane-crossing}
		\label{fig:Disj-2domes}
	\end{figure} 
	
	Finally, we consider the simplest RT saddle homologous to $A\cup B$ which never cross the EOW brane and has dome-shaped structures in each AdS$_3$ geometry as sketched in \cref{fig:Disj-2domes}. Once again, we only consider the configuration with a connected (though disjoint into two parts in each spacetime) entanglement wedge. The corresponding entanglement wedge cross-section may be computed utilizing standard AdS$_3$/CFT$_2$ techniques as follows \cite{Takayanagi:2017knl,Nguyen:2017yqw}
	\begin{align}\label{djnc}
		E_W(A:B)=\frac{L_\text{I}+L_\text{II}}{4G_N}\log\left[1+2 \eta+2 \sqrt{\eta (\eta+1)}\right]\,,
	\end{align}
	where the cross-ratio $\eta$ is given in \cref{Dome-cross-ratio}.
	
	\section{Reflected entropy from island prescription: vacuum state}\label{sec:SRIs0}
	
	In this section, we will discuss the effective lower dimensional perspective of the setup where the gravitational theory on the brane is coupled to two non-gravitating bath CFT$_2$s. As described in \cite{Anous:2022wqh}, the gravitational theory on the brane in the effective intermediate picture is obtained by integrating out the bulk AdS$_3$ degrees of freedom on either side of the brane. 
	
\begin{figure}[ht]
		\centering
		\includegraphics[scale=0.5]{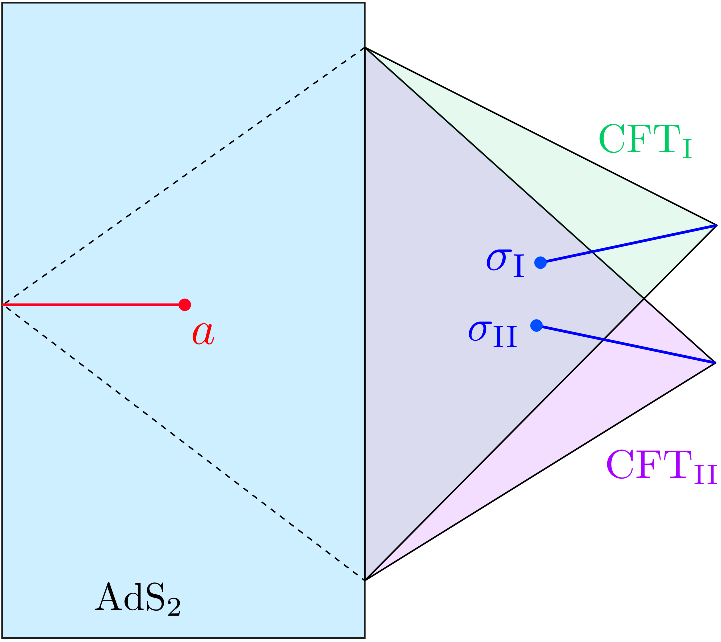}
		\caption{Two-dimensional effective model obtained from integrating out the three-dimensional bulk geometry in the AdS/ICFT model. Figure modified from \cite{Anous:2022wqh}. }
		\label{fig:Island-ICFT}
	\end{figure}

	In the large tension limit $T\to T_\text{max}$, the theory on the brane is given by two CFT$_2$s coupled to the weakly fluctuating (AdS$_2$) metric. The nature of the CFTs on the brane also follows from the dimensional reduction of the bulk geometry. In the large tension regime, we obtain a non-local action \cite{Chen:2020uac,Fallows:2021sge} which may be rewritten in terms of the Polyakov action by introducing two auxiliary fields $\varphi_k~(k=\text{I,II})$ as follows \cite{Anous:2022wqh,Afrasiar:2023nir}:
	\begin{align}
		I=\sum_{k=\text{I,II}}\frac{L_k}{32 \pi G_N}\int_\Sigma \text{d}^2y\sqrt{-h}\left[-\frac{1}{2}h^{ab}\nabla_a\varphi_k\nabla_b\varphi_k+\varphi_k R^{(2)}-\frac{2}{L_k}e^{-\varphi_k}\right]\,,\label{Polyakov}
	\end{align}
	where $h_{ab}$ is the induced metric on the brane and $R^{(2)}$ is the corresponding Ricci scalar. The above Polyakov action may be interpreted as two CFT$_2$s with central charges $c_k=\frac{3L_k}{4G_N}$ located on the AdS$_2$ brane $\Sigma$. Hence, as advocated in \cite{Anous:2022wqh}, we have two CFT$_2$s on the whole real line interacting through the common metric on the AdS$_2$ brane and decoupled on the other halves as depicted in \cref{fig:Island-ICFT}. This constitutes the setup of a QFT coupled to gravity on a hybrid manifold, usual in the island paradigm \cite{Almheiri:2019hni,Almheiri:2019qdq,Almheiri:2019yqk,Penington:2019kki}.
	
	From the Polyakov action \cref{Polyakov}, the transverse area term of a co-dimension two surface $\chi$ appearing in the island formula may be obtained as follows \cite{Fallows:2021sge,Afrasiar:2023nir}
	\begin{align}
		\frac{\text{Area}(\chi)}{4G_N^{(2)}}&=\frac{1}{8G_N}\sum_{k=\text{I,II}}L_k\,\varphi_k(\chi)=\frac{1}{8G_N}\sum_{k=\text{I,II}}L_k\log\left[-\frac{2}{L_k^2\,R^{(2)}}\right]\notag\\
		&=\frac{c_\text{I}}{6}\log\left(\frac{1}{\cos\psi_{\text{I}}}\right)+\frac{c_\text{II}}{6}\log\left(\frac{1}{\cos\psi_{\text{II}}}\right)\equiv \Phi_0\,,\label{area-term}
	\end{align}
	where in the last equality, we have used the Brown-Henneaux relations as well as the fact that the Ricci scalar on the brane $\Sigma$ is given by \cite{Anous:2022wqh,Afrasiar:2023nir}
	\begin{align}
		R^{(2)}=-\frac{2}{\ell_\text{eff}^2}\equiv-\frac{2}{L_k^2}\cos^2\psi_k~~,~~k=\text{I,II}\,.
	\end{align}
	
	Now we discuss the computation of entanglement entropy of subsystems in the bath CFT$_2$s utilizing the concept of generalized entropy and the island formalism. The generalized R\'enyi entropy for subsystems in the baths is computed through an Euclidean path integral on the replica manifold obtained by sewing $n$ copies of the original manifold along branch cuts present on the subsystems under consideration \cite{Almheiri:2019qdq,Dong:2020uxp}. As the bath CFT$_2$s couple to the gravitational theory on the brane, in certain saddles to the gravitational part of the path integral, additional smooth branch points may emerge at the replica fixed points on copies of the brane theory. These are the endpoints of the so called \textit{island} region corresponding to the bath subsystems.
	
	However, unlike the usual scenario with a single bath coupled to gravity \cite{Almheiri:2019hni,Almheiri:2019qdq,Almheiri:2019yqk,Penington:2019kki}, in the present case, the existence of an additional bath leads to novel saddle points to the gravitational path integral. In \cite{Afrasiar:2023nir}, these novel island saddles were termed as \textit{induced islands}. In the presence of two baths, consider the entanglement entropy of the union of two subsystems on either bath. In the usual scenario, both these subsystems are responsible for the appearance of additional branch points on the brane. However, when the central charge of one of the CFT$_2$s is larger than the other, branch points in the gravitating manifold may emerge solely due to the subsystem in the CFT with the larger central charge. Since the CFTs interact on the gravitating manifold, the other CFT also realizes the same branch points and perceives an induced island. Note that, as the island region is induced from the CFT with larger central charge, it bears no signature of the subsystem in the other CFT\footnote{See \cite{Afrasiar:2023nir} for more details and the corresponding generalized island formulae.}. In the following we will assume $c_\text{II}>c_\text{I}$ without loss of generality,  and hence the induced islands will only appear under the influence of the subsystem in CFT$_2^\text{II}$.
	
	The origin of the conventional and induced islands may also be understood from the doubly holographic (three dimensional bulk) perspective. The conventional island appears in the effective intermediate picture when the RT saddle homologous to the subsystems crosses the brane only once. On the other hand for a double-crossing RT saddle homologous to the subsystem in CFT$_2^\text{II}$, we obtain an induced island in the lower dimensional perspective. In both the cases, the island region is bounded by the crossing points on the brane. 
	
	The generalized entanglement entropy for $A\cup B$ in the presence of an island $\text{Is}(AB)=[-y_3,-y_1]$ may be expressed as\footnote{Note that the correlation functions of the twist operators, $\tilde{G}^{n}_{\mathrm{CFT}_\text{I}}$ and $G^{n}_{\mathrm{CFT}_{\text{II}}}$, generically need not have the same structure due to the presence of induced islands.} \cite{Almheiri:2019qdq,Almheiri:2019yqk}
	\begin{align}
		S_{\text{gen}}(AB)&=\frac{\mathcal{A}\left(\partial\, \mathrm{Is}(AB)\right)}{4G_N}+\lim_{n\to 1}	S^{(n)}_\mathrm{eff}\left(AB \cup \mathrm{Is}(AB)\right)\\
		&= \Phi_0+\lim_{n\to 1}\frac{1}{1-n}\log\bigg[\prod_{i=1,3}\Omega_\text{I}(y_i)^{\Delta^{(\text{I})}}\Omega_\text{II}(y_i)^{\Delta^{(\text{II})}}\, \tilde{G}^{n}_{\mathrm{CFT}_\text{I}}G^{n}_{\mathrm{CFT}_{\text{II}}} \bigg],\label{Adj-EE}
	\end{align}
	where  $\Omega_\text{I,II}(y_i)=|y_i|$ is the Weyl factor corresponding to the point $y_i$ on the brane, $\text{Is}(X)$ denotes the island region contributing to the entanglement entropy of subsystem $X$ and $S^{(n)}_\mathrm{eff}$ is the effective Renyi entanglement entropy of quantum matter fields on the fixed background. In \cref{Adj-EE}, the conformal dimensions of the twist operators $\Delta^{(\text{I,II})}$ are given by \cite{Calabrese:2004eu}
	\begin{align}
		\Delta^{(k)}=\frac{c_{k}}{24}\left(n-\frac{1}{n}\right)~~,~~k=\text{I,II}.
	\end{align}
	The island prescription now dictates that the entanglement entropy is obtained by extremizing the generalized entropy over all possible island configurations as follows \cite{Almheiri:2019qdq,Almheiri:2019yqk}
	\begin{align}
		S(AB)=\underset{\text{Is}(AB)}{\textrm{Min~Ext}} \bigg[S_{\text{gen}}(AB)\bigg].
	\end{align}
	
	Once the entanglement entropy island $\text{Is}(AB)$ for $A\cup B$ is determined, the reflected entropy in the effective intermediate perspective is obtained by splitting $\text{Is}(AB)$ into the respective reflected entropy islands $\text{Is}_R(A)$ and $\text{Is}_R(B)$ at the \textit{island cross section} $Q=\partial \text{Is}_R(A)\cap \partial \text{Is}_R(B)$ as follows \cite{Chandrasekaran:2020qtn,Li:2020ceg}
	\begin{align}\label{adj-reflected-formula}
		S_R(A:B)=\underset{Q}{\textrm{Ext}} \bigg[\frac{\mathrm{Area}(Q)}{2 G_N^{(2)}}+\lim_{n\to 1}\lim_{m_e\to 1}S_{R}^{\,\mathrm{eff}}\left(A \cup \mathrm{Is}_{R}(A): B \cup \mathrm{Is}_{R}(B)\right)\bigg]\,.
	\end{align}
	In \cref{adj-reflected-formula}, the effective reflected entropy from its R\'enyi generalization may be computed through the (normalized) partition function $Z_{n, m}$ on a $m\times n$ sheeted replica manifold as follows \cite{Dutta:2019gen}
	\begin{equation}\label{SRGmn0}
		\begin{aligned}
			S_{R}^{\left(m,n\right),\mathrm{eff}}\left(A \cup \mathrm{Is}_{R}(A): B \cup \mathrm{Is}_{R}(B)\right)&=\frac{1}{1-n}\log\bigg[\frac{Z_{n, m}}{\left(Z_{1, m}\right)^{n}}\bigg]\\
			&= \frac{1}{1-n}\log\bigg[\frac{\prod_{i}\Omega(y_i)^{\Delta_i} \, \tilde{G}^{m,n}_{\mathrm{CFT}_{\text{I}}}G^{m,n}_{\mathrm{CFT}_{\text{II}}}}{ (\tilde{G}^{m}_{\mathrm{CFT}_{\text{I}}}G^{m}_{\mathrm{CFT}_{\text{II}}})^n}\bigg].
		\end{aligned}
	\end{equation}
	In the above expression, $\Omega(y_i)$ corresponds to the Weyl factor corresponding to the point $y_i$ on the AdS$_2$ brane, $\tilde{G}^{m,n}_{\mathrm{CFT}_\text{I}}$ and $G^{m,n}_{\mathrm{CFT}_\text{II}}$ are appropriate correlation functions of twist operators inserted at the endpoint of the subsystems and their corresponding reflected entropy islands on the replica manifold. 
	
	\subsection {Adjacent Subsystems}
	Here we compute the entanglement entropy and  reflected entropy for the configuration of adjacent subsystems $A=[\tilde{b}_1,\tilde{b}_2]_\text{I}\cup[b_1,b_2]_{\text{II}}$ and $B=[\tilde{b}_2,\tilde{b}_3]_\text{I}\cup[b_2,b_3]_{\text{II}}$ in the lower dimensional effective perspective described above, by employing the replica technique developed in \cite{Dutta:2019gen}.  To this end we first consider different saddles for the entanglement island for $A\cup B$ and subsequently discuss the phase transitions of reflected entropy between $A$ and $B$ within each phase of the entanglement entropy.
	\subsubsection{Conventional Island}
	We begin by considering the case of conventional islands, where the entanglement entropy island $\mathrm{Is}(AB)=[-y_3,-y_1]$ conceived on the brane depends on the degrees of freedom of the subsystems in both CFT$_2$s. In this case, the correlation functions $\tilde{G}^{n}_{\mathrm{CFT}_\text{I}}$ and $G^{n}_{\mathrm{CFT}_{\text{II}}}$ have the same large-$c$ structure\footnote{Note that on the right hand side, we have suppressed the subscripts $\mathrm{CFT}_\mathrm{I}^{\otimes n}$ for compactness of the expressions. In the following, unless specified explicitly, we will continue to adopt this simplification of notations.}:
	\begin{align}
		\tilde{G}^{n}_{\mathrm{CFT}_\text{I}}&=\langle\sigma_{g_n}(\tilde{b}_1) \sigma_{g_n^{-1}}(\tilde{b}_3)\sigma_{g_n}(-y_3)\sigma_{g_n^{-1}}(-y_1)\rangle\nonumber\\&\approx\langle\sigma_{g_n}(\tilde{b}_1) \sigma_{g_n^{-1}}(-y_1)\rangle
		\langle\sigma_{g_n^{-1}}(\tilde{b}_3)\sigma_{g_n}(-y_3)\rangle,
	\end{align}
	with a similar factorization for $G^{n}_{\mathrm{CFT}_{\text{II}}}=\langle\sigma_{g_n}(b_1) \sigma_{g_n^{-1}}(b_3)\sigma_{g_n}(-y_3)\sigma_{g_n^{-1}}(-y_1)\rangle$. We are going to follow this convention for the rest of the article.
	Now from \cref{Adj-EE,area-term}, we obtain
	\begin{align}\label{Adj-EE-I}
		S_\text{gen}(y_1,y_3)=\Phi_0+\frac{c_\text{\,I}}{6} \log\bigg[\frac{(y_1+\tilde{b}_1)^2 (y_3+\tilde{b}_3)^2}{\epsilon^2 y_1   y_3 }\bigg]+\frac{c_{\text{II}}}{6} \log\bigg[\frac{(y_1+b_1)^2 (y_3+b_3)^2}{\epsilon^2 y_1   y_3}\bigg]\,.
	\end{align}
	where the constant area contribution denoted as $\Phi_0$ is defined in \cref{area-term}.
	On extremizing the above equation with respect to $y_1$ and $y_3$, the positions of the endpoints of the island are given by
	\begin{equation}\label{Adj-island-position-I}
		\begin{aligned}
			y_i^*=\frac{(c_{\text{II}}-c_\text{I}) (b _i-\tilde{b }_i)+\sqrt{4 b_i \tilde{b}_i(c_\text{I}+c_{\text{II}})^2 +(c_\text{I}-c_{\text{II}})^2 (b_i-\tilde{b}_i)^2}}{2 (c_\text{I}+c_{\text{II}})}~~,~~(i=1,3).
		\end{aligned}
	\end{equation}
	The entanglement entropy for the adjacent subsystems in the effective intermediate perspective may be obtained by substituting \cref{Adj-island-position-I} in \cref{Adj-EE-I}. Utilizing \cref{T-max,angle-small}, in the $\delta\to 0$ limit, the above expression is seen to match identically with the large tension limit of \cref{y*-entropy}. Incidentally, in the $\delta\to 0$ limit one obtains
	\begin{align}
		\Phi_0^{(\delta)}=\frac{c_\text{I}}{6}\log\left[\frac{(L_\text{I}+L_\text{II})}{L_\text{I}\,\delta}\right]+\frac{c_\text{II}}{6}\log\left[\frac{(L_\text{I}+L_\text{II})}{L_\text{II}\,\delta}\right]+\mathcal{O}\left(\delta\right)\equiv S_\text{int}^{(\delta)}-\frac{c_\text{I}+c_\text{II}}{6}\log 2\,,\label{Phi0-delta}
	\end{align}
	and hence the large tension limit of the entanglement entropy obtained from \cref{Adj-bulk-EE-single} also matches with \cref{Adj-EE-I}.
	
	We now compute the island contributions to the reflected entropy for the configuration of two adjacent subsystems when the entanglement entropy island is conventional. We divide the entropy island Is(A$\cup$ B) into the respective reflected islands as follows: $\text{Is}_R(A)=[-y,-y_1^*]$ and $\text{Is}_R(B)=[-y_3^*,-y]$ at the island cross-section $Q=y$ such that Is$_\text{R}$(A) $\cup$ Is$_\text{R}$(B)=Is(A$\cup$ B) \cite{Chandrasekaran:2020qtn}. The twist correlation function computing the effective reflected entropy between $A \cup \mathrm{Is}_{R}(A)$ and $B \cup \mathrm{Is}_{R}(B)$ is generically obtained through the six point function which is given as
	\begin{align}
		G^{m,n}_{\mathrm{CFT}_{\text{II}}}&=\langle\sigma_{g_A}(b_1) \sigma_{g_B g_A^{-1}}(b_2) \sigma_{g_B^{-1}}(b_3)\sigma_{g_B}(y_3^*)\sigma_{g_Ag_B^{-1}}(y)\sigma_{g_A^{-1}}(y_1^*)\rangle\,.
	\end{align}
	The expression for $\tilde{G}$ is of the same form as above with coordinates $b$ replaced by $\tilde{b}$ while points on the brane remain the same. The correlation function $G^{m}_{\mathrm{CFT}_{\text{II}}}$ on the $m$-sheeted Riemann surface for the configuration of adjacent subsystems may be expressed as
	\begin{align}\label{Gm}
		G^{m}_{\mathrm{CFT}_{\text{II}}}=\langle\sigma_{g_m}(b_1) \sigma_{g_m^{-1}}(b_3) \sigma_{g_m}(y_3)\sigma_{g_m^{-1}}(y_1^*)\rangle.
	\end{align}
	It has the same form for the CFT$_\text{I}$ with $b$ replaced by the $\tilde{b}$ coordinates.
	The scaling dimensions of the relevant twist operators are given as follows ($k=\text{I,II}$) \cite{Dutta:2019gen}
	\begin{align}
		&\Delta^{(k)}_{\sigma_{g_A}}=\Delta^{(k)}_{\sigma_{g_A^{-1}}}=\Delta^{(k)}_{\sigma_{g_B}}=\Delta^{(k)}_{\sigma_{g_B^{-1}}}=\frac{n\,c_{k}}{12}\left(m-\frac{1}{m}\right)=n \Delta_m^{(k)} \notag\\
		&\Delta^{(k)}_{\sigma_{g_A^{-1} g_B}}=\Delta^{(k)}_{\sigma_{g_B^{-1} g_A}}=\frac{c_{k}}{12}\left(n-\frac{1}{n}\right)=2 \Delta^{(k)}_n\label{Scaling-dimensions-SR}.
	\end{align}
	The form of the six point function in a CFT is not known explicitly, however it can be determined in the large central charge limit leading to various phases which we discuss in the following subsections.

	\subsubsection*{Phase-I}
\begin{figure}[ht]
		\centering
		\includegraphics[scale=0.7]{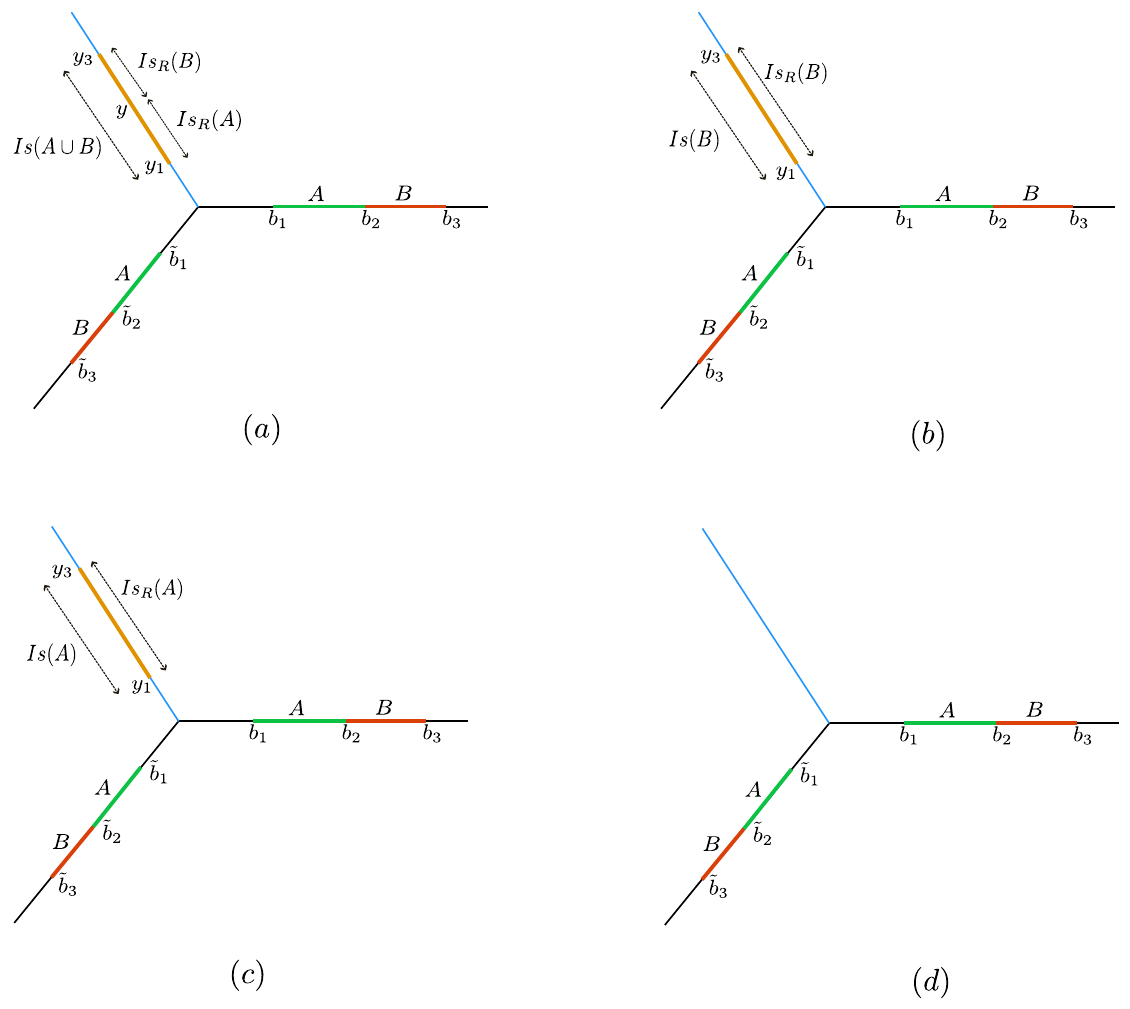}
		\caption{Various phases depicting the island contribution to the reflected entropy of
adjacent subsystems. We replace $y$ by $\bar{y}$ for the case of induced islands. }
		\label{Adjacent-no-crossing-ft}
	\end{figure}

	We choose the size of the subsystems $A$ and $B$ such that both subsystems admit their own islands on the brane region. In this case, the six point twist correlator factorizes into three two point functions (cf. \cref{Adjacent-no-crossing-ft}\textcolor{blue}{(a)}) as
	\begin{equation}\label{G-1}
		\begin{aligned}
			G^{m,n}_{\mathrm{CFT}_{\text{II}}}&=\langle\sigma_{g_A}(b_1) \sigma_{g_B g_A^{-1}}(b_2) \sigma_{g_B^{-1}}(b_3)\sigma_{g_B}(y_3^*)\sigma_{g_Ag_B^{-1}}(y)\sigma_{g_A^{-1}}(y_1^*)\rangle\\
			&=\langle\sigma_{g_A}(b_1)\sigma_{g_A^{-1}}(y_1^*)\rangle\langle\sigma_{g_B g_A^{-1}}(b_2)\sigma_{g_Ag_B^{-1}}(y)\rangle\langle\sigma_{g_B^{-1}}(b_3)\sigma_{g_B}(y_3^*))\rangle.
		\end{aligned}
	\end{equation}
	The correlation function $G^m$ in this phase is given by
	\begin{equation}\label{Gm-1}
		G^{m}_{\mathrm{CFT}_{\text{II}}}=\langle\sigma_{g_m}(b_1)\sigma_{g_m^{-1}}(y_1^*) \rangle\langle\sigma_{g_m}(b_3) \sigma_{g_m}(y_3^*)\rangle.
	\end{equation}
	Similar factorizations occur for $\tilde{G}$ and $\tilde{G}^m$. We now utilize \cref{G-1,Gm-1} in \cref{SRGmn0} to obtain the generalized reflected entropy in the replica limit as
	\begin{equation}
		\begin{aligned}
			S_R^{\text{gen}}(y)=2\Phi_0+\frac{c_{\text{I}}}{3} \log\bigg[\frac{(y+\tilde{b}_2)^2}{\epsilon\, y}\bigg]+\frac{c_{\text{II}}}{3} \log\bigg[\frac{(y+b_2)^2}{\epsilon\, y}\bigg],
		\end{aligned}
	\end{equation}
	where the area term $\Phi_0$ is defined in \cref{area-term}. After extremizing the above expression with respect to $y$, the location of the island cross-section is given by \cref{Adj-island-position-I} with $i=2$.
	The reflected entropy for this phase may now be obtained in the limit $\delta \to 0$ as 
	
	\begin{equation}\label{Adj-SR-I}
		\begin{aligned}
			S_R(A:B)=\frac{c_{\text{I}}}{3} \log \left[ \frac{(y^* + \tilde{b}_2)^2}{2y^* \, \epsilon}\right]+ \frac{c_{ \text{II}}}{3} \log \left[ \frac{(y^* + b_2)^2}{2y^* \, \epsilon} \right]+S^{(\delta)}_\text{int} \ .
		\end{aligned}
	\end{equation}
	where $S^{(\delta)}_\text{int}$ is the interface entropy in the $\delta\to 0$ limit, defined in \cref{S-int}. In this limit, we observe that the location of the brane crossing point $y^*$ in the $3d$ bulk picture given in \cref{Adj-y*-I-EW} matches with the island cross-section obtained in the effective lower dimensional scenario. Furthermore, the large tension limit of the EWCS given in \cref{Adj-I-EW} matches identically with half of the reflected entropy obtained in \cref{Adj-SR-I}.

	\subsubsection*{Phase-II}
	In phase-II, the size of the subsystem $B$ is much larger than $A$ such that the entire island belongs to $B$ and $A$ does not have a corresponding island. This configuration is depicted in \cref{Adjacent-no-crossing-ft}\textcolor{blue}{(b)}. Note that, there is no non-trivial island cross-section for this phase and hence no extremization is involved. The corresponding twist correlation function computing the effective reflected entropy in this phase is given by
	\begin{align}
		G^{m,n}_{\mathrm{CFT}_{\text{II}}}&=\langle\sigma_{g_A}(b_1) \sigma_{g_B g_A^{-1}}(b_2) \sigma_{g_B^{-1}}(b_3)\sigma_{g_B}(y_3^*)\sigma_{g_B^{-1}}(y_1^*)\rangle\notag\\
		&\approx\langle\sigma_{g_A}(b_1) \sigma_{g_B g_A^{-1}}(b_2)\sigma_{g_B^{-1}}(y_1^*)\rangle\langle \sigma_{g_B^{-1}}(b_3)\sigma_{g_B}(y_3^*)\rangle.
	\end{align}
	As earlier, the correlation function on the $m$-sheeted surface factorizes following \cref{Gm-1}. Similar factorizations occur for the CFT$_\mathrm{I}$ correlators and the two point functions cancel from the numerator and denominator. Furthermore, note that $y_1$ and $y_3$ are fixed to the extremal values $y_1^*$ and $y_3^*$ respectively, by the entanglement island corresponding to the subsystem $A\cup B$. The three point function is fixed by the conformal symmetry up to an OPE coefficient which for the present case is given by \cite{Dutta:2019gen}
	\begin{align}
		C^{(k)}_{n,m}=(2m)^{2\Delta^{(k)}_n}\,.\label{OPE-coeff}
	\end{align}
	Therefore, the reflected entropy may be obtained in this phase by utilizing \cref{Scaling-dimensions-SR,OPE-coeff} followed by taking the replica limit as
	\begin{equation}
		S_R(A:B)=\frac{c_{\text{I}}}{3} \log \left[\frac{2 (\tilde{b}_2-\tilde{b}_1) (\tilde{b}_2+y_1^*)}{\epsilon_1 (\tilde{b}_1+y_1^*)}\right]+\frac{c_{\text{II}}}{3} \log \left[\frac{ 2(b_2-b_1) (b_2+y_1^*)}{\epsilon_2 (b_1+y_1^*)}\right]\,,
	\end{equation}
	where $y_1^*$ is given in \cref{Adj-island-position-I}.
	We observe that the above result matches exactly with twice the large tension limit of the EWCS in the bulk perspective, obtained in \cref{Adj-sing-II-EW}.
	\subsubsection*{Phase-III}
	As opposed to the previous case, the size of the subsystem $A$ is much larger than that of $B$ in this phase. Hence the entire entanglement entropy island now belongs to $A$ and $B$ does not posses an island as shown in \cref{Adjacent-no-crossing-ft}\textcolor{blue}{(c)}. Similar to the previous phase, the correlation function in this phase factorizes as
	\begin{align}
		G^{m,n}_{\mathrm{CFT}_{\text{II}}}&=\langle\sigma_{g_B g_A^{-1}}(b_2)\sigma_{g_B^{-1}}(b_3)\sigma_{g_A}(y_3^*)\rangle\langle \sigma_{g_A}(b_1) \sigma_{g_A^{-1}}(y_1^*)\rangle\,,\notag\\
		G^{m}_{\mathrm{CFT}_{\text{II}}}&=\langle\sigma_{g_m^{-1}}(b_3)\sigma_{g_m}(y_3^*)\rangle\langle \sigma_{g_m}(b_1) \sigma_{g_m^{-1}}(y_1^*))\rangle\,.
	\end{align}
	The reflected entropy in phase III may now be obtained in a similar manner to the previous phase as follows
	\begin{equation}
		S_R(A:B)=\frac{c_{\text{I}}}{3} \log \left[\frac{2 (\tilde{b}_2-\tilde{b}_3) (\tilde{b}_2+y_3^*)}{\epsilon_1 (\tilde{b}_3+y_3^*)}\right]+\frac{c_{\text{II}}}{3} \log \left[\frac{2(b_2-b_3) (b_2+y_3^*)}{\epsilon_2 (b_3+y_3^*)}\right]\,.
	\end{equation}
	Once again, since the $\delta\to 0$ limit of \cref{y*-entropy} is identical to \cref{Adj-island-position-I}, it is easy to verify that the reflected entropy obtained above matches identically with twice the corresponding large tension expression for the EWCS in the doubly holographic framework, as given in \cref{Adj-sing-III-EW}.
	\subsubsection{Induced Island}
	Next we consider the induced island $\mathrm{\overline{Is}}(AB)=[-\bar{y}_3,-\bar{y}_1]$ where the island region for the subsystem in CFT$_2^\text{I}$ is induced by the subsystem in CFT$_2^\text{II}$. As a result, although the CFT$_2^\text{II}$ correlator has the same structure as earlier, the CFT$_2^\text{I}$ correlator $\tilde{G}^{n}_{\mathrm{CFT}_\text{I}}$ has a different factorization in the large central charge limit (the island region is independent of the degrees of freedom on the CFT$_\text{I}$ subsystem)
	\begin{align}
		\tilde{G}^{n}_{\mathrm{CFT}_\text{I}}&=\langle\sigma_{g_n}(\tilde{b}_1) \sigma_{g_n^{-1}}(\tilde{b}_3)\sigma_{g_n}(-\bar{y}_3)\sigma_{g_n^{-1}}(-\bar{y}_1)\rangle\notag\\
		&\approx\langle\sigma_{g_n}(\tilde{b}_1)  \sigma_{g_n^{-1}}(\tilde{b}_3)\rangle\langle\sigma_{g_n}(-\bar{y}_3)\sigma_{g_n^{-1}}(-\bar{y}_1)\rangle.
	\end{align}
	The generalized entropy may now be obtained using \cref{Adj-EE,area-term} as follows
	\begin{align}\label{Adj-EE-II}
		S_{\text{gen}}(\bar{y}_1,\bar{y}_3)=2\Phi_0+&\frac{c_\text{I}}{3}\log\bigg[\frac{\tilde{b}_3-\tilde{b}_1}{\epsilon}\bigg]+\frac{c_\text{I}}{6} \log \left[\frac{\left(\bar{y}_1-\bar{y}_3\right){}^2}{\bar{y}_1 \bar{y}_3}\right]\nonumber\\+&
		\frac{c_{\text{II}}}{6} \log\left[\frac{\left(\bar{y}_1+b_1\right){}^2\left(\bar{y}_3+b_3\right){}^2}{\epsilon^2 \bar{y}_1 \bar{y}_3}\right]\,.
	\end{align}
	Extremizing over the locations of the quantum extremal surfaces $\bar{y}^*_1$ and $\bar{y}^*_3$, we obtain
	\begin{align}
		&c_{\text{II}} \left(\bar{y}^*_1-\bar{y}^*_3\right) \left(\bar{y}^*_1-b_k\right)+c_\text{I} \left(\bar{y}^*_1+\bar{y}^*_3\right) \left(b_k+\bar{y}^*_1\right)=0~~,~~(k=1,3).\label{islands-double-crossing}
	\end{align}
	The entanglement entropy may be obtained upon substituting the physical solution to the above equations in \cref{Adj-EE-II} and subsequently choosing the minimal saddle. Using the parametrization given in \cref{new-variables}, it is now straightforward to verify that \cref{Double-crossing-EE-ext} together with the solution $\bar{y}_i^*=\frac{b_i}{k_D^*}$, conforms to the locations of the quantum extremal surfaces in the $2d$ effective theory as obtained from \cref{islands-double-crossing}.
	
	In the following, we compute the induced island contributions to the reflected entropy between the adjacent subsystems $A$ and $B$. Once again, we divide the induced entanglement island $\overline{\text{Is}}(A\cup B)$ into the respective reflected islands $\overline{\text{Is}}_R(A)=[-\bar{y},-\bar{y}_1^*]$ and $\overline{\text{Is}}_R(B)=[-\bar{y}_3^*,-\bar{y}]$ such that $\overline{\text{Is}}_\text{R}$(A) $\cup$ $\overline{\text{Is}}_\text{R}(B)=\overline{\text{Is}}(A\cup B)$. Note that, similar to the entanglement island, the reflected entropy islands for the CFT$^\text{I}$ degrees of freedom appearing on the AdS$_2$ brane is induced by the subsystem in CFT$^\text{II}$. As earlier, the twist correlators computing the effective reflected entropy between $A\cup \overline{\text{Is}}_R(A)$ and $B\cup \overline{\text{Is}}_R(B)$ are generically given by the six point function
	\begin{align}
		G^{m,n}_{\mathrm{CFT}_{\text{II}}}&=\langle\sigma_{g_A}(b_1) \sigma_{g_B g_A^{-1}}(b_2) \sigma_{g_B^{-1}}(b_3)\sigma_{g_B}(\bar{y}_3^*)\sigma_{g_Ag_B^{-1}}(\bar{y})\sigma_{g_A^{-1}}(\bar{y}_1^*)\rangle\,,
	\end{align}
	in CFT$^\text{II}$, with a similar expression holding in CFT$^\text{I}$. Unlike the earlier phases, these correlators factorize differently in CFT$_{\text{I}}$ and CFT$_{\text{II}}$s as discussed in the following subsections.
	\subsubsection*{Phase-I}
	In the first phase, the portions of the subsystems $A$ and $B$ residing in CFT$^\text{II}$ admit their own islands and correspondingly induce islands for their counterparts in CFT$^\text{I}$. 
	The correlation function in the CFT$_{\text{I}}$ (cf. \cref{Adjacent-no-crossing-ft}\textcolor{blue}{(a)}) factorizes as
	\begin{equation}\label{G-5-I}
		\begin{aligned}
			\tilde{G}^{m,n}_{\mathrm{CFT}_{\text{I}}}
			&=\langle\sigma_{g_A}(\tilde{b}_1) \sigma_{g_B g_A^{-1}}(\tilde{b}_2) \sigma_{g_B^{-1}}(\tilde{b}_3)\rangle\langle\sigma_{g_B}(\bar{y}_3)\sigma_{g_Ag_B^{-1}}(\bar{y})\sigma_{g_A^{-1}}(\bar{y}_1^*)\rangle,
			\\
			\tilde{G}^{m}_{\mathrm{CFT}_{\text{I}}}&=\langle\sigma_{g_m}(\tilde{b}_1)\sigma_{g_m^{-1}}(\tilde{b}_3)\rangle_{\mathrm{CFT}_\text{I}^{ \otimes m }}\langle \sigma_{g_m}(\bar{y}_3^*) \sigma_{g_m^{-1}}(\bar{y}_1^*))\rangle_{\mathrm{CFT}_{\text{I}}^{ \otimes m }},
		\end{aligned}
	\end{equation}
	while the correlator in the CFT$_{\text{II}}$ factorizes into two point twist correlators as
	\begin{equation}\label{G-5-II}
		\begin{aligned}
			G^{m,n}_{\mathrm{CFT}_{\text{II}}}&=\langle\sigma_{g_A}(b_1)\sigma_{g_A^{-1}}(\bar{y}_1^*)\rangle\langle\sigma_{g_B g_A^{-1}}(b_2)\sigma_{g_Ag_B^{-1}}(\bar{y})\rangle\langle\sigma_{g_B^{-1}}(b_3)\sigma_{g_B}(\bar{y}_3^*))\rangle\\
			G^{m}_{\mathrm{CFT}_{\text{II}}}&=\langle\sigma_{g_m}(b_1)\sigma_{g_m^{-1}}(\bar{y}_1^*)\rangle\langle \sigma_{g_m^{-1}}(b_3) \sigma_{g_m}(\bar{y}_3^*))\rangle.
		\end{aligned}
	\end{equation}
	Now, the generalized reflected entropy in this phase may be obtained using \cref{G-5-I,G-5-II,OPE-coeff} in \cref{SRGmn0} in the replica limit as follows
	\begin{equation}
		\begin{aligned}
			S_R^\text{gen}(A:B)=2\Phi_0&+\frac{c_{\text{I}}}{3}\log\left[\frac{2(\tilde{b}_2-\tilde{b}_1)(\tilde{b}_3-\tilde{b}_2)}{\epsilon(\tilde{b}_3-\tilde{b}_1)}\right]+\frac{c_{\text{I}}}{3} \log \left[\frac{2 (\bar{y}-\bar{y}_1^*) (\bar{y}-\bar{y}_3^*)}{\bar{y}(\bar{y}_1^*-\bar{y}_3^*)  }\right]\\&+\frac{c_{\text{II}}}{3} \log \left[\frac{(b_2+\bar{y})^2}{\epsilon \,\bar{y}}\right].
		\end{aligned}
	\end{equation}
	Extremization of the above expression with respect to the (induced) island cross-section $\bar{y}$ leads precisely to \cref{Adj-double-I-delta-extr}, where $\bar{y}_1^*$ and $\bar{y}_3^*$ are fixed according to the solution of \cref{islands-double-crossing}. Utilizing \cref{Phi0-delta}, the reflected entropy in the effective lower dimensional perspective matches identically with twice the large tension limit of the corresponding EWCS obtained in \cref{EW-double1} in the doubly holographic perspective. 
	\subsubsection*{Phase-II}
	In the next phase, we consider the subsystem $A$ to be much larger than $B$ as described in \cref{Adjacent-no-crossing-ft}\textcolor{blue}{(c)}, so that the entire (induced) island belongs to $A$. In this case, there is no non-trivial island cross section on the AdS$_2$ region and the following factorization occurs 
	\begin{equation}\label{G-6-I}
		\begin{aligned}
			\tilde{G}^{m,n}_{\mathrm{CFT}_{\text{I}}}&=\langle\sigma_{g_A}(\tilde{b}_1) \sigma_{g_B g_A^{-1}}(\tilde{b}_2) \sigma_{g_B^{-1}}(\tilde{b}_3)\rangle\langle\sigma_{g_B}(\bar{y}_3^*)\sigma_{g_A^{-1}}(\bar{y}_1^*)\rangle,
			\\
			\tilde{G}^{m}_{\mathrm{CFT}_{\text{I}}}&=\langle\sigma_{g_A}(\tilde{b}_1) \sigma_{g_B^{-1}}(\tilde{b}_3)\rangle\langle \sigma_{g_m}(\bar{y}_3^*) \sigma_{g_m^{-1}}(\bar{y}_1^*))\rangle.
		\end{aligned}
	\end{equation}
	The correlation function in the CFT$_{\text{II}}$ factorizes in the following way
	\begin{equation}\label{G-6-II}
		\begin{aligned}
			G^{m,n}_{\mathrm{CFT}_{\text{II}}}&=\langle\sigma_{g_B g_A^{-1}}(b_2)\sigma_{g_B^{-1}}(b_3)\sigma_{g_A}(\bar{y}_3^*)\rangle\langle \sigma_{g_A}(b_1) \sigma_{g_A^{-1}}(\bar{y}_1^*)\rangle,\\
			G^{m}_{\mathrm{CFT}_{\text{II}}}&=\langle\sigma_{g_m^{-1}}(b_3)\sigma_{g_m}(\bar{y}_3^*)\rangle\langle \sigma_{g_m}(b_1) \sigma_{g_m^{-1}}(\bar{y}_1^*))\rangle.
		\end{aligned}
	\end{equation}
	The reflected entropy for this phase may now be determined as follows
	\begin{equation}
		S_R(A:B)=\frac{c_{\text{I}}}{3}\log\left[\frac{2(\tilde{b}_2-\tilde{b}_1)(\tilde{b}_3-\tilde{b}_2)}{\epsilon(\tilde{b}_3-\tilde{b}_1)}\right]+\frac{c_{\text{II}}}{3}\log \left[\frac{2(b_3-b_2)(b_2+\bar{y}_3^*)}{\epsilon (b_3+\bar{y}_3^*)}\right],
	\end{equation}
	where $\bar{y}_3^*$ is fixed by the entanglement island of $A\cup B$, as given in \cref{islands-double-crossing}. This matches identically with the large tension limit of the EWCS in the doubly holographic picture, given in \cref{adj-EW-doubleII}.
	\subsubsection*{Phase-III}
	In this phase, the subsystem $B$ is much larger than the subsystem $A$ as shown in \cref{Adjacent-no-crossing-ft}\textcolor{blue}{(b)}. Hence the factorization of correlator remains same as in the previous case for CFT$_\mathrm{I}$ while for CFT$_{\text{II}}$ we have
	\begin{equation}\label{G-7-II}
		\begin{aligned}
			G^{m,n}_{\mathrm{CFT}_{\text{II}}}&=\langle\sigma_{g_A}(b_1) \sigma_{g_B g_A^{-1}}(b_2)\sigma_{g_B^{-1}}(\bar{y}_1^*)\rangle\langle \sigma_{g_B^{-1}}(b_3)\sigma_{g_B}(\bar{y}_3)\rangle.
		\end{aligned}
	\end{equation}
	The reflected entropy for this phase may be obtained in a similar manner to the previous phase as follows
	\begin{equation}
		S_R(A:B)=\frac{c_{\text{I}}}{3}\log\left[\frac{2(\tilde{b}_2-\tilde{b}_1)(\tilde{b}_3-\tilde{b}_2)}{\epsilon(\tilde{b}_3-\tilde{b}_1)}\right]+\frac{c_{\text{II}}}{3}\log \left[\frac{2(b_2-b_1)(b_2+\bar{y}_1^*)}{\epsilon (b_1+\bar{y}_1^*)}\right],
	\end{equation}
	where $\bar{y}_1^*$ is given by \cref{islands-double-crossing} and the corresponding minimal EWCS obtained in \cref{EW-adj-double-III} from the double holographic perspective, matches with the above expression in the limit of large brane tension.
	\subsubsection{No Island saddle}
	When the sizes the subsystems $A$ and $B$ are small enough such that they do not posses any  entanglement entropy islands as shown in \cref{Adjacent-no-crossing-ft}\textcolor{blue}{(d)}, the corresponding entanglement entropies are computed via the usual two-point functions in either CFT \cite{Calabrese:2004eu}. The correlation function computing the reflected entropy between $A$ and $B$ may be written as a three point function in either CFT:
	\begin{equation}
		\begin{aligned}
			G^{m,n}_{\mathrm{CFT}_{\text{II}}}&=\langle\sigma_{g_A}(b_1) \sigma_{g_B g_A^{-1}}(b_2) \sigma_{g_B^{-1}}(b_3)\rangle\,.
		\end{aligned}
	\end{equation}
	The $\mathrm{CFT}_{\text{II}}$ correlator $\tilde{G}^{m,n}$ is given by a similar two point function with $b$ replaced by $\tilde{b}$.
	Therefore, the reflected entropy may be obtained in a straightforward manner as follows
	\begin{equation}
		S_R(A:B)=\frac{c_{\text{I}}}{3}\log\left[\frac{2(\tilde{b}_2-\tilde{b}_1)(\tilde{b}_3-\tilde{b}_2)}{\epsilon(\tilde{b}_3-\tilde{b}_1)}\right]
		+\frac{c_{\text{II}}}{3}\log\left[\frac{2(b_2-b_1)(b_3-b_2)}{\epsilon(b_3-b_1)}\right]\,,
	\end{equation}
	which matches identically with the corresponding EWCS in the $3d$ bulk perspective, given in \cref{EW-adj-no-crossing}.

	\subsection {Disjoint subsystems}
	
	\begin{figure}[h!]
		\centering
		\begin{subfigure}[b]{0.45\textwidth}
			\centering
			\includegraphics[width=6cm,height=5cm]{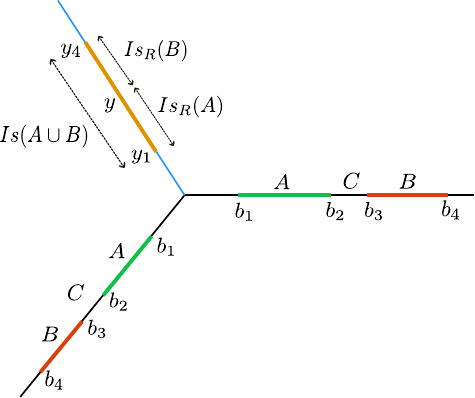}
			\caption{}
			\label{SRDJ0CFT1}
		\end{subfigure}
		\hspace{.5cm}
		\begin{subfigure}[b]{0.45\textwidth}
			\centering
			\includegraphics[width=6cm,height=5cm]{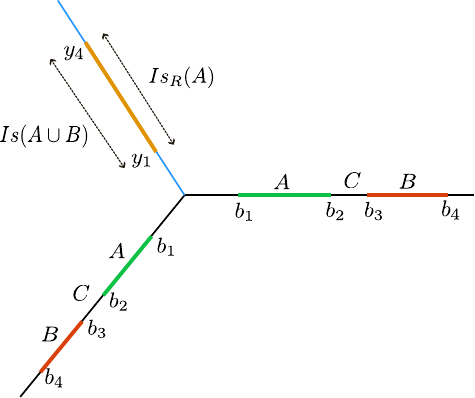}
			\caption{}
			\label{SRDJ0CFT2}
		\end{subfigure}
		\hspace{.5cm}
		\begin{subfigure}[b]{0.45\textwidth}
			\centering
			\includegraphics[width=6cm,height=5cm]{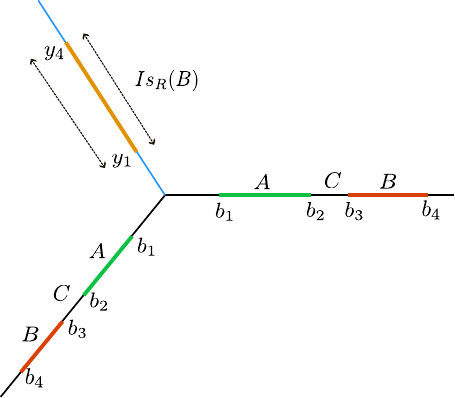}
			\caption{}
			\label{SRDJ0CFT3}
		\end{subfigure}
		\vspace{.5cm}
		\begin{subfigure}[b]{0.45\textwidth}
			\centering
			\includegraphics[width=6cm,height=5cm]{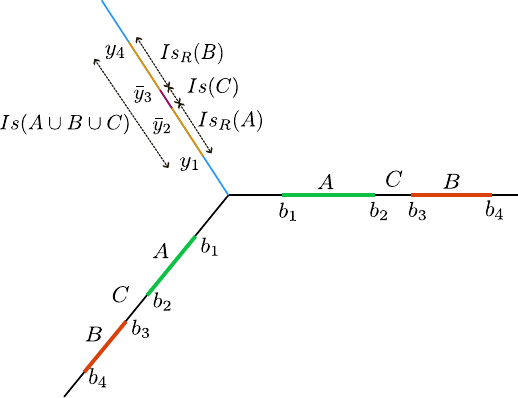}
			\caption{}
			\label{SRDJ0CFT4}
		\end{subfigure}
		\caption{Various phases depicting the island contributions to the reflected entropy of disjoint intervals in a zero temperature holographic  ICFT$_2$ . }\label{SRDJ0CFT}
	\end{figure}
	
In this section we determine  the island contributions to the reflected entropy for two disjoint subsystems A$=[b_1,b_2]_\mathrm{I}\cup [b_1,b_2]_\mathrm{II}$ and B$=[b_3,b_4]_\mathrm{I}\cup [b_3,b_4]_\mathrm{II}$ in the lower dimensional effective theory described by dynamical gravity on the AdS$_2$ manifold coupled to two flat Minkowski baths., utilizing the replica technique \cite{Dutta:2019gen,Chandrasekaran:2020qtn}.
	
Similar to the earlier investigation with adjacent subsystems, in the following, we will discover both conventional and induced island regions conceived in the gravitating manifold depending on the locations and (relative) sizes of the subsystems.

\subsubsection{Conventional Island}
We begin by considering the case of the conventional entanglement island for $A\cup B$, denoted as $\text{Is}(AB)=[-y_1,-y_4]$. Recall that a conventional island on the gravitational manifold depends on the degrees of freedom from the subsystems on both baths. Hence, the twist correlators computing the effective R\'enyi entropy corresponding to $A\cup B$ have the same kind of factorization in the large-c limit. For the present configuration of two disjoint subsystems with the corresponding conventional island $\text{Is}(AB)$, the effective R\'enyi entropy is computed through six point correlation functions of twist operators as follows 
	\begin{align}\label{G12EESym}
		G^{n}_{\mathrm{CFT_{I/II}}}&=\langle\sigma_{g_n}(b_1) \sigma_{g_n^{-1}}(b_2) \sigma_{g_n}(b_3) \sigma_{g_n^{-1}}(b_4)\sigma_{g_n}(-y_4)\sigma_{g_n^{-1}}(-y_1)\rangle\nonumber
	\end{align}
In the large central charge limit, the above twist correlator may be factorized as follows \cite{Hartman:2013mia,Almheiri:2019yqk,Almheiri:2019qdq}
	\begin{align}
		G^{n}_{\mathrm{CFT_{I/II}}}&=\langle \sigma_{g_n^{-1}}(b_2) \sigma_{g_n}(b_3)\rangle \langle\sigma_{g_n}(b_1)\sigma_{g_n^{-1}}(-y_1)\rangle \langle \sigma_{g_n^{-1}}(b_4)\sigma_{g_n}(-y_4) \rangle
	\end{align}
Substituting the above correlation in \cref{Adj-EE} and accounting for the appropriate Weyl factors for the points $y_i$ on the AdS$_2$ region given as $\Omega(y_i)=|y_i|$, the generalized entanglement entropy may be obtained as follows
	\begin{align}\label{ciee}
		S_{gen}(AB)=\Phi_0 &+\frac{\mathrm{c_{I}+c_{II}}}{6}\left(2\log\left(\frac{b_3-b_2}{\epsilon}\right)+\log\left[\frac{(b_4+y_4)^2}{\epsilon\, y_1}\right]+\log\left[\frac{(b_4+y_4)^2}{\epsilon\, y_4}\right]\right)
	\end{align}
Extremizing the above expression with respect to $y_i$ we obtain $y_i^*=b_i$ leading to the following expression for the entanglement entropy
	\begin{align}
		S(AB)=\Phi_0+\frac{\mathrm{c_{I}+c_{II}}}{6}\log 2+\frac{\mathrm{c_{I}+c_{II}}}{6}\left[2\log\left(\frac{b_3-b_2}{\epsilon}\right)+\log\left(\frac{2b_1}{\epsilon}\right)+\log\left(\frac{2b_4}{\epsilon}\right)\right]
	\end{align}
In the large tension regime, utilizing \cref{Phi0-delta}, the above result is seen to be an exact match of the corresponding expression obtained from the bulk geometry given in \cref{EE_disj}.

Having obtained the entanglement entropy we now compute the reflected entropy of two disjoint subsystems for phases involving the conventional islands utilizing the replica technique developed in \cite{Dutta:2019gen}
\subsubsection*{Phase-I}
We begin by considering the configuration described by figure \ref{SRDJ0CFT1}. The twist correlation function characterizing the reflected entropy of $AB$ in this phase is given by $G^{m,n}_{\mathrm{CFT_{I/II}}}$ which corresponds to the following seven point correlation function
	\begin{align}
		G^{m,n}_{\mathrm{CFT_{I/II}}}&=\langle\sigma_{g_A}(b_1) \sigma_{g_A^{-1}}(b_2) \sigma_{g_B}(b_3) \sigma_{g_B^{-1}}(b_4)\sigma_{g_B}(-b_4)\sigma_{g_Ag_B^{-1}}(-y)\sigma_{g_A^{-1}}(-b_1)\rangle\nonumber
	\end{align}
Note that the two correlations $G^{m,n}_{\mathrm{CFT_{I}}}$ and $G^{m,n}_{\mathrm{CFT_{II}}}$ are identical in this case because of the symmetry of chosen configuration. Since the seven point function is hard to determine analytically even in the large-$c$ limit, we take $b_4,b_1$ away from $b_2,b_3$ such that the following factorization occurs
	\begin{align}\label{Gmn}
		G^{m,n}_{\mathrm{CFT_{I/II}}}&=\langle\sigma_{g_A}(b_1) \sigma_{g_A^{-1}}(-b_1)\rangle \sigma_{g_B^{-1}}(b_4)\sigma_{g_B}(-b_4)\rangle\langle\sigma_{g_A^{-1}}(b_2) \sigma_{g_B}(b_3)\sigma_{g_Ag_B^{-1}}(-y)\rangle\nonumber
	\end{align}
Note that $_{\mathrm{CFT_{I/II}}}$ on LHS indicates that the structure is same for both CFT$_{\text{II}}$ and CFT$_{\text{I}}$. The corresponding correlation functions of the $m$-sheeted Riemann surface which for this phase are given as
	\begin{align}
		G^{m}_{\mathrm{CFT_{I/II}}}&=\langle\sigma_{g_m}(b_1) \sigma_{g_m^{-1}}(b_2) \sigma_{g_m}(b_3) \sigma_{g_m^{-1}}(b_4)\sigma_{g_m}(-b_4)\sigma_{g_m^{-1}}(-b_1)\rangle\\
		&=\langle\sigma_{g_m}(b_1) \sigma_{g_m^{-1}}(-b_1)\rangle\langle \sigma_{g_m^{-1}}(b_4)\sigma_{g_m}(-b_4)\rangle\langle\sigma_{g_m^{-1}}(b_2) \sigma_{g_m}(b_3)\rangle\label{Gm0}
	\end{align}
where the second equality comes from the factorization specific to this phase. 
Note that the two point functions in the numerator and denominators exactly cancel. Also the Weyl factors  in the numerator  and denominators cancel for all the operators except  $\sigma_{g_Ag_B^{-1}}(y)$. We may now obtain the generalized Renyi reflected entropy by substituting the expressions for the correlators in \cref{Gmn,Gm0} in (\ref{SRGmn0}) as follows
	\begin{align}
		S_{R}&^{\left(m,n,\mathrm{eff}\right)}\left(A \cup \mathrm{Is}_{R}(A): B \cup \mathrm{Is}_{R}(B)\right)\nonumber\\
		&=\frac{1}{1-n}\log\bigg[\frac{\Omega_\text{I}(-y)^{\Delta^{(\text{I})}_{g_Ag_B^{-1}}} \langle\sigma_{g_A^{-1}}(b_2) \sigma_{g_B}(b_3)\sigma_{g_Ag_B^{-1}}(-y)\rangle_{\text{CFT}^{\otimes mn}_\text{I}}}{\left(\langle\sigma_{g_m^{-1}}(b_2) \sigma_{g_m}(b_3)\rangle_{\text{CFT}^{\otimes m}_\text{I}}\right)^n}\bigg]+(\text{I}\leftrightarrow\text{II})
	\end{align}
The three point function is fixed by the conformal symmetry up to the OPE constant given in \cref{OPE-coeff}. This leads to the following expression for the generalized reflected entropy in the replica limit
	\begin{align}
		S_R^\text{gen}(b_0)&=2\Phi_0+\frac{\mathrm{c_{I}+c_{II}}}{3} \log\bigg[\frac{2(y+b_2)(y+b_3)}{y(b_3-b_2)}\bigg]
	\end{align}
Extremizing the above expression over the island cross-section $Q=-y$ leads to $y=\sqrt{b_2 b_3}$ and hence we may obtain the following expression for the reflected entropy in this phase
	\begin{align}
		S_R(A:B)=2\left(\Phi_0+\frac{\mathrm{c_{I}+c_{II}}}{6} \log 2\right)+\frac{\mathrm{c_{I}+c_{II}}}{3} \log\Bigg(\frac{\sqrt{b_2}+\sqrt{b_3}}{\sqrt{b_3}-\sqrt{b_2}}\Bigg)\,.
	\end{align}
Utilizing \cref{Phi0-delta,S-int}, it is straightforward to verify that the above expression matches identically with twice the large tension limit of the EWCS given in \cref{EW-phase1}.
	\subsubsection*{Phase-II}
	
	We now consider the reflected entropy for phase-II of the disjoint interval configuration which is described in figure \ref{SRDJ0CFT2}. In this phase-II the subsystem-$B$ does not posses an island and hence the entire island belongs to $A$.    The corresponding correlation function may be factorized in the corresponding OPE channel as follows
	\begin{align}
		G^{m,n}_{\mathrm{CFT_{I/II}}}&=\langle\sigma_{g_A}(b_1) \sigma_{g_A^{-1}}(b_2)  \sigma_{g_B}(b_3)\sigma_{g_B^{-1}}(b_4)\sigma_{g_A}(-b_4)\sigma_{g_A^{-1}}(-b_1)\rangle\nonumber\\
        &=\langle\sigma_{g_A}(b_1) \sigma_{g_A^{-1}}(-b_1)\rangle\langle\sigma_{g_A}(-b_4)\sigma_{g_A^{-1}}(b_2) \sigma_{g_B}(b_3)\sigma_{g_B^{-1}}(b_4)\rangle
	\end{align}
A similar factorization holds for the correlation function on $m$-sheeted surface. As earlier, the two point functions cancel from the numerator and the denominator which leads to the following expression for the reflected entropy
	\begin{align}
		S_{R}^{(m,n,\mathrm{eff})}(&A \cup \mathrm{Is}_{R}(A): B \cup \mathrm{Is}_{R}(B)\nonumber\\
		&=\frac{1}{1-n}\log\bigg[\frac{ \langle\sigma_{g_A}(-b_4)\sigma_{g_A^{-1}}(b_2) \sigma_{g_B}(b_3)\sigma_{g_B^{-1}}(b_4)\rangle_{\text{CFT}^{\otimes mn}_\text{I}}}{\left(\langle\sigma_{g_m}(-b_4)\sigma_{g_m^{-1}}(b_2) \sigma_{g_m}(b_3)\sigma_{g_m^{-1}}(b_4)\rangle_{\text{CFT}^{\otimes m}_\text{I}}\right)^n}\bigg]+\left(\text{I}\leftrightarrow\text{II}\right)
	\end{align}
Note that since the subsystem $B$ does not posses any island there is no island cross-section and hence no extremization involved in this phase. The conformal block that gives dominant contribution to the above four point function(s) is well known in the large central charge limit \cite{Dutta:2019gen} to be of the following form
	\begin{align}
		\log\mathcal{F}_{(k)}(n\Delta_m,2\Delta_n|x)=-2n\Delta^{(k)}_m\log x-2\Delta^{(k)}_{n}\log\bigg(\frac{1+\sqrt{x}}{1-\sqrt{x}}\bigg)\label{Large-c-block}
	\end{align}
	where $x=\frac{b_{12}b_{34}}{b_{13}b_{24}}$ is the cross-ratio. Hence, in the replica limit the above expression directly leads to the reflected entropy as follows
	\begin{align}
		S_R(A:B)=\frac{\mathrm{c_{I}}+\mathrm{c_{II}}}{3} \log\Bigg[\frac{b_4^2-b_2b_3+\sqrt{\displaystyle(b_3^2-b_4^2)(b_2^2-b_4^2)}}{b_4(b_3-b_2)}\Bigg]\,,\label{SR-dj-2}
	\end{align}
which matches identically with  half of the corresponding EWCS in the $3d$ bulk description, given in \cref{EW-disj-II}.
	\subsubsection*{Phase-III}

	Phase-III of the disjoint interval configuration with the conventional island saddle for entanglement entropy is depicted in figure.\ref{SRDJ0CFT3}. In this phase the subsystem $A$ does not possess any reflected entropy island whereas $B$ does. The computation of the generalized Renyi reflected entropy proceeds similar to the previous phase and we may as well replace $b_4$ by $b_1$ in \cref{SR-dj-2} for the present case, to obtain 
	\begin{align}
		S_R(A:B)&=\frac{\mathrm{c_{I}}+\mathrm{c_{II}}}{3} \log\bigg[\frac{b_1^2-b_2b_3+\sqrt{\displaystyle(b_3^2-b_1^2)(b_2^2-b_1^2)}}{b_1(b_3-b_2)}\bigg]\,.
	\end{align}
	The above expression is trivially seen to match with the corresponding EWCS in \cref{EW-dj-3}.

	\subsubsection{Induced Islands}
	Next we consider situation involving induced islands for various subsystems under consideration. This can be further subdivided into phases based on whether the subsystem $C$ sandwiched between $A$ and $B$ in either baths claims an induced island as follows:
	\begin{itemize}
		\item The subsystem $A\cup B\cup C$ is large enough to possess an induced island. This situation arises when $\Theta_{ABC}=\frac{b_4}{b_1}$ exceeds its critical value (cf. the discussion in \cref{dj-double-crossing}). We simultaneously require the subsystem $C$ to be small enough to be lacking any induced island. 
		\item In the second case, $A\cup B\cup C$ possesses the conventional island while $\Theta_{C}=\frac{b_3}{b_2}$ exceeds its critical value giving access to the induced island for subsystem $C$.
		\item Both $C$ and $A\cup B\cup C$ possess their induced islands. However, as discussed in \cref{footnote11}, we do not encounter this scenario for a large range of parameter values and hence will be omitted in the following discussion.
	\end{itemize}
In the following, we will investigate each of these situations individually and discuss the phase transitions for the reflected entropy within each scenario.	 
\subsubsection*{A. Subsystem $C$ lacking an island}
We begin with the case where $C$ does not have an island which results in the following expression for the correlation functions computing the R\'enyi entropy for $A\cup B$,
	\begin{align}
		G^{n}_{\mathrm{CFT_{I/II}}}&=\langle\sigma_{g_n}(b_1) \sigma_{g_n^{-1}}(b_2) \sigma_{g_n}(b_3) \sigma_{g_n^{-1}}(b_4)\rangle\langle\sigma_{g_n}(-\bar{y}_4)\sigma_{g_n^{-1}}(-\bar{y}_1)\rangle\nonumber
	\end{align}
	As mentioned above in the large-$c$ limit the above correlators factorize differently in the two CFTs as follows
	\begin{align}
		G^{n}_{\mathrm{CFT_{I}}}
		&=\langle\sigma_{g_n}(b_1) \sigma_{g_n^{-1}}(b_4)\rangle \langle \sigma_{g_n^{-1}}(b_2) \sigma_{g_n}(b_3) \rangle\langle\sigma_{g_n}(-\bar{y}_1)\sigma_{g_n^{-1}}(-\bar{y}_4)\rangle\nonumber\\
		G^{n}_{\mathrm{CFT_{II}}}&=\langle\sigma_{g_n}(b_1)\sigma_{g_n^{-1}}(-\bar{y}_1)\rangle \langle \sigma_{g_n^{-1}}(b_2) \sigma_{g_n}(b_3)\rangle\langle\sigma_{g_n}(-\bar{y}_4) \sigma_{g_n^{-1}}(b_4)\rangle\label{GMCFTDCEE}
	\end{align}
	Since the above correlation functions are expressed in terms of the two point functions completely fixed by conformal symmetry, we may readily obtain the generalized entanglement entropy from \cref{Adj-EE} as follows
	\begin{align}
		S_\text{gen}(AB)=\Phi_0&+\frac{\mathrm{c_{I}}}{6}\log\left[\frac{(b_2-b_3)^2(b_1-b_4)^2(\bar{y}_1-\bar{y}_4)^2}{\epsilon^4\bar{y}_1\bar{y}_4}\right]\nonumber\\&+\frac{\mathrm{c_{II}}}{6}\log\left[\frac{(b_2-b_3)^2(b_1+\bar{y}_1)^2(b_4+\bar{y}_4)^2}{\epsilon^4\bar{y}_1\bar{y}_4}\right]\label{Sgen-dj-double1}
	\end{align}
	Extremizing the above equation w.r.t $y_1$ and $y_4$ we get
	\begin{align}\label{EEdisjextF}
			&c_{\text{II}} \left(\bar{y}_1-\bar{y}_4\right) \left(\bar{y}_1-b_i\right)+c_\text{I} \left(\bar{y}_1+\bar{y}_4\right) \left(b_i+\bar{y}_1\right)=0~~,~~(i=1,4).
	\end{align}
	The above result exactly matches with the corresponding expression obtained from the bulk geometry given in \cref{Double-crossing-EE-ext}, together with the solution $\bar{y}_1^*=\frac{b_1}{k^*_{ABC}}$ in the large tension limit\footnote{Note that in the present context,  $k_D\equiv k_{ABC}$ and $\Theta_D\equiv\Theta_{ABC}=\frac{b_4}{b_1}$ in the parametrization given in \cref{new-variables} (cf. \cref{footnote12}) with $i=1,4$.}.
	Finally, the entanglement entropy for the present configuration may be obtained by substituting the solutions $\bar{y}_1^*$ and $\bar{y}_4^*$ to the above extremization equations in the expression for the generalized entropy in \cref{Sgen-dj-double1}.
	
	We now proceed to compute the islands contributions to the reflected entropy for phases involving induced islands for $A\cup B\cup C$. As earlier, we divide the induced entanglement island $\bar{\text{Is}}(AB)=[-\bar{y}_4^*,-\bar{y}_1^*]$ into the corresponding reflected entropy islands $\bar{\text{Is}}_R(A)=[-\bar{y},-\bar{y}_1^*]$ and $\bar{\text{Is}}_R(B)=[-\bar{y}_4^*,-\bar{y}]$ by placing the island cross-section at $Q=-\bar{y}$. The twist correlator computing the effective R\'enyi reflected entropy is then given by a seven point function 
	\begin{align}
		G^{m,n}_{\mathrm{CFT_{I/II}}}&=\langle\sigma_{g_A}(b_1) \sigma_{g_A^{-1}}(b_2) \sigma_{g_B}(b_3) \sigma_{g_B^{-1}}(b_4)\sigma_{g_B}(-\bar{y}_4^*)\sigma_{g_Ag_B^{-1}}(-\bar{y})\sigma_{g_A^{-1}}(-\bar{y}_1^*)\rangle\nonumber\,.
	\end{align}
	 In this case, the corresponding correlation functions of both CFT$_{\text{I}}$ and CFT$_{\text{II}}$ factorize differently. These phases  correspond to the double crossing geodesics in the dual bulk geometry. Note that the diagrams depicting induced island phases remain same as  \cref{SRDJ0CFT}. The difference however is in the way correlators factorize.
	\subsubsection*{Phase-I}
	We now compute the reflected entropy for the disjoint subsystems when both $A$ and $B$ admit their reflected entropy islands.   In this phase depicted in  \cref{SRDJ0CFT1} (replace $y$ with $\bar{y}$ and $y_i$ with $\bar{y}_i$)  $G^{m,n}_{\mathrm{CFT_{I}}},\, \tilde{G}^{m,n}_{\mathrm{CFT_{II}}}$ corresponds to the seven point correlation functions which factorize in the large-$c$ limit as follows
	\begin{align}
		G^{m,n}_{\mathrm{CFT_{I}}}&=\langle\sigma_{g_A}(b_1) \sigma_{g_A^{-1}}(b_2) \sigma_{g_B}(b_3) \sigma_{g_B^{-1}}(b_4)\sigma_{g_B}(-\bar{y}_1^*)\sigma_{g_Ag_B^{-1}}(-\bar{y})\sigma_{g_A^{-1}}(-\bar{y}_4^*)\rangle\nonumber\\
		&=\langle\sigma_{g_A}(b_1) \sigma_{g_A^{-1}}(b_2) \sigma_{g_B}(b_3) \sigma_{g_B^{-1}}(b_4)\rangle\langle\sigma_{g_B}(-\bar{y}_1^*)\sigma_{g_Ag_B^{-1}}(-\bar{y})\sigma_{g_A^{-1}}(-\bar{y}_4^*)\rangle\nonumber\\
		G^{m,n}_{\mathrm{CFT_{II}}}&=\langle\sigma_{g_A}(b_1) \sigma_{g_A^{-1}}(b_2) \sigma_{g_B}(b_3) \sigma_{g_B^{-1}}(b_4)\sigma_{g_B}(-\bar{y}_1^*)\sigma_{g_Ag_B^{-1}}(-\bar{y})\sigma_{g_A^{-1}}(-\bar{y}_4^*)\rangle\nonumber\\
		&=\langle\sigma_{g_A}(b_1)\sigma_{g_A^{-1}}(-\bar{y}_1^*)\rangle \langle \sigma_{g_A^{-1}}(b_2) \sigma_{g_B}(b_3)\sigma_{g_Ag_B^{-1}}(-\bar{y})\rangle\langle\sigma_{g_B}(-\bar{y}_4^*) \sigma_{g_B^{-1}}(b_4)\rangle\label{GMNCFTDC}
	\end{align}
	
	Note that $[-\bar{y}_1^*,-\bar{y}_4^*]$ corresponds to the entanglement entropy island of $AB$ which were obtained in \cref{EEdisjextF}. Observe that in the second line we have specific factorization of the correlation function in the large-$c$ limit in this phase. Also a similar factorization exists for the corresponding correlation functions on the $m$-sheeted Riemann surface which are expressed as follows
	\begin{align}
		G^{m}_{\mathrm{CFT_{I}}}&=\langle\sigma_{g_m}(b_1) \sigma_{g_m^{-1}}(b_2) \sigma_{g_m}(b_3) \sigma_{g_m^{-1}}(b_4)\sigma_{g_m}(-\bar{y}_1^*)\sigma_{g_m^{-1}}(-\bar{y}_4^*)\rangle\nonumber\\
		&=\langle\sigma_{g_m}(b_1) \sigma_{g_m^{-1}}(b_2) \sigma_{g_m}(b_3) \sigma_{g_m^{-1}}(b_4)\rangle\langle\sigma_{g_m}(-\bar{y}_1^*)\sigma_{g_m^{-1}}(-\bar{y}_4^*)\rangle\nonumber\\
		\tilde{G}^{m}_{\mathrm{CFT_{II}}}&=\langle\sigma_{g_m}(b_1) \sigma_{g_m^{-1}}(b_2) \sigma_{g_m}(b_3) \sigma_{g_m^{-1}}(b_4)\sigma_{g_m}(-\bar{y}_1^*)\sigma_{g_m^{-1}}(-\bar{y}_4^*)\rangle\nonumber\\
		&=\langle\sigma_{g_m}(b_1)\sigma_{g_m^{-1}}(-\bar{y}_1^*)\rangle \langle \sigma_{g_m^{-1}}(b_2) \sigma_{g_m}(b_3)\rangle\langle\sigma_{g_m}(-\bar{y}_4^*) \sigma_{g_m^{-1}}(b_4)\rangle\label{GMCFTDC}
	\end{align}
	Utilizing the above expressions given by \cref{GMNCFTDC}, \cref{GMCFTDC} in \cref{SRGmn0} we determine the following result for the generalized or effective reflected entropy in the replica limit
	\begin{align}
		S_R(A:B)=2\left(\Phi_0+\frac{c_\text{I}+c_\text{II}}{6}\log 2\right)+\frac{c_\text{I}}{3} \log\bigg[\frac{1+\sqrt{x}}{1-\sqrt{x}}\bigg]&+\frac{c_\text{I}}{3} \log\bigg[\frac{(\bar{y}-\bar{y}_1^*)(\bar{y}_4^*-\bar{y})}{\bar{y}(\bar{y}_4^*-\bar{y}_1^*)}\bigg]\notag\\&+\frac{\mathrm{c_{II}}}{3} \log\bigg[\frac{(\bar{y}+b_2)(\bar{y}+b_3)}{\bar{y}(b_3-b_2)}\bigg]
	\end{align}
	where  $x=\frac{b_{12}b_{34}}{b_{13}b_{24}}$ is the cross-ratio. Extremizing the above expression w.r.t $\bar{y}$ we obtain the following relation
	\begin{align}
		\mathrm{c_{I}}(\bar{y}^2-\bar{y}_1^*\bar{y}_4^*)(\bar{y}+b_2)(\bar{y}+b_3)+\mathrm{c_{II}}(\bar{y}^2-b_2b_3)(\bar{y}-\bar{y}_1^*)(\bar{y}-\bar{y}_4^*)=0
	\end{align}
	which is identical to its doubly holographic counterpart in \cref{Ext-EW-disj-zeroT-doubleI} in the large tension limit. Consequently upon utilizing \cref{Phi0-delta}, the large tension limit of the reflected entropy matches identically with the corresponding large tension value of the EWCS given in \cref{EW-disj-zeroT-doubleI}.
	
	\subsubsection*{Phase-II}
	
	In this phase only the reflected entropy island for $A$ receives the  entire island contribution as depicted in \cref{SRDJ0CFT2} (with $y_i\rightarrow\bar{y}_i$). Hence, the required correlation function factorize in this phase as follows
	\begin{align}\label{Gmnp12dc}
		G^{m,n}_{\mathrm{CFT_{I}}}&=\langle\sigma_{g_A}(b_1) \sigma_{g_A^{-1}}(b_2) \sigma_{g_B}(b_3) \sigma_{g_B^{-1}}(b_4)\rangle\langle \sigma_{g_A}(-\bar{y}_1^*)\sigma_{g_A^{-1}}(-\bar{y}_4^*)\rangle\nonumber\\
		G^{m,n}_{\mathrm{CFT_{II}}}&=\langle\sigma_{g_A}(b_1) \sigma_{g_A^{-1}}(-\bar{y}_1^*)\rangle\langle\sigma_{g_A}(-\bar{y}_4^*)\sigma_{g_A^{-1}}(b_2) \sigma_{g_B}(b_3)\sigma_{g_B^{-1}}(b_4)\rangle
	\end{align}
	A similar factorization holds for the correlation function on $m$-sheeted surface.
Once again the two point functions cancel from the numerator and the denominator leading to the following expression for the reflected entropy
	\begin{align}
		&S_{R}^{\left(m,n,\mathrm{eff}\right)}\left(A \cup \mathrm{Is}_{R}(A): B \cup \mathrm{Is}_{R}(B)\right)\nonumber\\&=\frac{1}{1-n}\log\bigg[\frac{ \langle\sigma_{g_A}(b_1)\sigma_{g_A^{-1}}(b_2) \sigma_{g_B}(b_3)\sigma_{g_B^{-1}}(b_4)\rangle_{\text{CFT}^{\otimes mn}_\text{I}}}{\left(\langle\sigma_{g_m}(b_1)\sigma_{g_m^{-1}}(b_2) \sigma_{g_m}(b_3)\sigma_{g_m^{-1}}(b_4)\rangle_{\text{CFT}^{\otimes m}_\text{I}}\right)^n}\bigg]\notag\\&
		+\frac{1}{1-n}\log\bigg[\frac{ \langle\sigma_{g_A}(-\bar{y}_4^*)\sigma_{g_A^{-1}}(b_2) \sigma_{g_B}(b_3)\sigma_{g_B^{-1}}(b_4)\rangle_{\text{CFT}^{\otimes mn}_\text{II}}}{\left(\langle\sigma_{g_m}(-\bar{y}_4^*)\sigma_{g_m^{-1}}(b_2) \sigma_{g_m}(b_3)\sigma_{g_m^{-1}}(b_4)\rangle_{\text{CFT}^{\otimes m}_\text{II}}\right)^n}\bigg]
	\end{align}
Note that since the subsystem-$B$ does not posses any island there is no island cross-section and hence no extremization involved in this phase. Now utilizing \cref{Large-c-block}, the replica limit of the above expression directly leads to the reflected entropy as follows
	\begin{align}
		S_R(A:B)=\frac{c_{I}}{3} \log\bigg[\frac{1+\sqrt{x}}{1-\sqrt{x}}\bigg]+\frac{\mathrm{c_{II}}}{3} \log\bigg[\frac{1+\sqrt{x_0}}{1-\sqrt{x_0}}\bigg]\,,
	\end{align}
	where the cross ratios are given by $x=\frac{b_{12}b_{34}}{b_{13}b_{24}}$ and $x_0=\frac{(\bar{y}_4^*+b_2)(b_{3}-b_4)}{(\bar{y}_4^*+b_3)(b_{2}-b_4)}$. Once again, the reflected entropy obtained above matches identically with twice the large tension limit of the bulk EWCS in \cref{EW-dj-double-II}, obtained in the doubly holographic framework.
	\subsubsection*{Phase-III}
	
	The computation of the reflected entropy in this phase proceeds exactly as the previous phase except that the subsystem-$B$ receives the reflected island contribution whereas $A$ does not. This phase may be described by replacing $y_i$ with $\bar{y}_i$ in \cref{SRDJ0CFT3}. As earlier, the correlation functions $G^{m,n}_{\mathrm{CFT_{I,II}}}$ factorize as follows
	\begin{align}\label{Gmnp3dc}
		G^{m,n}_{\mathrm{CFT_{I}}}&=\langle\sigma_{g_A}(b_1) \sigma_{g_A^{-1}}(b_2) \sigma_{g_B}(b_3) \sigma_{g_B^{-1}}(b_4)\rangle\langle \sigma_{g_B}(-\bar{y}_4^*)\sigma_{g_B^{-1}}(-\bar{y}_1^*)\rangle\,,\nonumber\\
		G^{m,n}_{\mathrm{CFT_{II}}}&=\langle\sigma_{g_B}(-\bar{y}_4^*) \sigma_{g_B^{-1}}(b_4)\rangle\langle\sigma_{g_A}(b_1)\sigma_{g_A^{-1}}(b_2) \sigma_{g_B}(b_3)\sigma_{g_B^{-1}}(-\bar{y}_1^*)\,,\rangle
	\end{align}
leading to the following expression for the reflected entropy in the replica limit
	\begin{align}
		S_R(A:B)=\frac{c_\text{I}}{3} \log\bigg[\frac{1+\sqrt{x}}{1-\sqrt{x}}\bigg]+\frac{\mathrm{c_{II}}}{3} \log\bigg[\frac{1+\sqrt{x_1}}{1-\sqrt{x_1}}\bigg]\,,
	\end{align}
	where $x=\frac{b_{12}b_{34}}{b_{13}b_{24}}$ and $x_1=\frac{(b_{1}-b_2)(b_3+\bar{y}_1^*)}{(b_{1}-b_3)(b_{2}+\bar{y}_1^*)}$ are the corresponding cross ratios. The above expression matches with twice the EWCS obtained in \cref{EW-dj-double-III} in the doubly holographic perspective.
	
	\subsubsection*{B. Subsystem C with an induced island}
	
	Next we will focus on the computations of the reflected entropy for specific configurations where $A\cup B\cup C$ possesses a conventional island $\text{Is}(A\cup B\cup C)=[-y_4^*,-y_1^*]$ and the subsystem $C$ claims an induced island $\bar{\text{Is}}(C)=[-\bar{y_3}^*,-\bar{y_2}^*]$ (\cref{SRDJ0CFT4}). Following the extremization of \cref{ciee}, we have $y_4^*=b_4$ and $y_1^*=b_1$ respectively. However for $\bar{y}_2^*$ and $\bar{y}_3^*$ we need to extremize the expression for the entanglement entropy of $C$ which leads to a similar set of equations given in \cref{islands-double-crossing} or \cref{EEdisjextF}. Note that when $C$ has an induced island of its own, the corresponding induced reflected entropy islands for $A$ and $B$ are disconnected as depicted in \cref{SRDJ0CFT4}. This is a novel aspect of the induced islands absent in earlier investigations of reflected entropy involving islands. The relevant twist correlators for this case are given as
	\begin{align}
		G^{m,n}_{\mathrm{CFT_{I,II}}}&=\langle\sigma_{g_A}(b_1) \sigma_{g_A^{-1}}(b_2) \sigma_{g_B}(b_3)\sigma_{g_B^{-1}}(b_4)\sigma_{g_A^{-1}}(-b_1)\sigma_{g_A}(-\bar{y}_2^*)\sigma_{g_B^{-1}}(-\bar{y}_3^*)\sigma_{g_B}(-b_4)\rangle,\label{Gmnall}
	\end{align} 
	There are three different phases I,II and III for this configuration depending on how the above correlators factorize.

	\subsubsection*{Phase-IV}
	In this phase the subsystems $A$ and $B$ are comparable in size where the correlators in \cref{Gmnall} factorize as follows
	\begin{align}\label{Gmnp12dcp9}
		G^{m,n}_{\mathrm{CFT_{I}}}=&\langle\sigma_{g_A}(-\bar{y}_2^*) \sigma_{g_A^{-1}}(b_2)  \sigma_{g_B}(b_3)\sigma_{g_B^{-1}}(-\bar{y}_3^*)\rangle\langle \sigma_{g_A}(b_1)\sigma_{g_A^{-1}}(-b_1)\rangle\langle \sigma_{g_B^{-1}}(b_4)\sigma_{g_B}(-b_4)\rangle,\nonumber\\
		G^{m,n}_{\mathrm{CFT_{II}}}=&\langle\sigma_{g_A}(b_1) \sigma_{g_A^{-1}}(-b_1)\rangle\langle\sigma_{g_A}(-\bar{y}_2^*)\sigma_{g_A^{-1}}(b_2)\rangle\langle \sigma_{g_B}(b_3)\sigma_{g_B^{-1}}(-\bar{y}_3^*)\rangle \langle \sigma_{g_B}(-b_4)\sigma_{g_B^{-1}}(b_4)\rangle.
	\end{align}
	A similar factorization holds for the correlation function on the $m$-sheeted Riemann surface.
	Substituting the above expressions in \cref{SRGmn0} and utilizing \cref{Large-c-block} we determine the reflected entropy to be as follows
	\begin{align}
		S_R(A:B)=\frac{c_\text{I}}{3} \log\bigg[\frac{1+\sqrt{w}}{1-\sqrt{w}}\bigg],
	\end{align}
	where $w=\frac{(\bar{y}_2+b_2)(b_3+\bar{y}_3)}{(\bar{y}_2+b_3)(b_2+\bar{y}_3)}$ is the cross ratio. This matches identically with the large tension limit of twice the EWCS in the doubly holographic picture, given in \cref{djdc1}.
	
	\subsubsection*{Phase-V}
	In phase-V, the size of the subsystem $A$ is sufficiently large compared to the subsystem $B$. Here the eight point correlation function $G^{m,n}_{\mathrm{CFT_{I}}}$ in \cref{Gmnall} factorizes into a six point function and a two point function as follows
	\begin{align}
		G^{m,n}_{\mathrm{CFT_{I}}}=&\langle\sigma_{g_A}(b_1) \sigma_{g_A^{-1}}(-b_1) \rangle \langle  \sigma_{g_A^{-1}}(b_2)  \sigma_{g_B}(b_3)\sigma_{g_B^{-1}}(b_4)\sigma_{g_A}(-\bar{y}_2^*)\sigma_{g_B^{-1}}(-\bar{y}_3^*)\sigma_{g_B}(-b_4)\rangle.
	\end{align}		
	The six point function on the RHS of the above equation further factorizes into a product of two four point function in the large-$c$ limit \footnote{As demonstrated in \cite{Banerjee:2016qca}, in the OPE channel corresponding to the present configuration the six point conformal black factorizes into two four point conformal blacks in the large central charge limit. Assuming the dominance of the $|\sigma_{g_A^{-1} g_B}\rangle$ block, the corresponding six point correlator may in turn be factorized into two four point correlators. Note that a similar factorization has been demonstrated in \cite{Basu:2023jtf}.}
	\begin{align}
		G^{m,n}_{\mathrm{CFT_{I}}}=\langle\sigma_{g_A}(b_1) \sigma_{g_A^{-1}}(-b_1) \rangle &\langle  \sigma_{g_A^{-1}}(b_2)  \sigma_{g_B}(b_3)\sigma_{g_A}(-\bar{y}_2^*)\sigma_{g_B^{-1}}(-\bar{y}_3^*)\rangle\nonumber\\
		&\times \langle  \sigma_{g_A^{-1}}(b_2)  \sigma_{g_B}(b_3) \sigma_{g_B^{-1}}(b_4)\sigma_{g_B}(-b_4)\rangle.
	\end{align}		
	The CFT$_{\text{II}}$ correlation function in \cref{Gmnall} on the other hand factorizes into product of two point function as follows
	\begin{align}
		G^{m,n}_{\mathrm{CFT_{II}}}=&\langle\sigma_{g_A}(b_1) \sigma_{g_A^{-1}}(-b_1)\rangle\langle\sigma_{g_A}(-\bar{y}_2^*)\sigma_{g_A^{-1}}(b_2)\rangle\langle \sigma_{g_B}(b_3)\sigma_{g_B^{-1}}(-\bar{y}_3^*)\rangle \langle \sigma_{g_B}(-b_4)\sigma_{g_B^{-1}}(b_4)\rangle.\label{factorization-dj-double-II}
	\end{align}
	Substituting the above expressions in \cref{SRGmn0} and utilizing \cref{Large-c-block} we determine the reflected entropy to be as follows
	\begin{align}\label{djdc2cft}
		S_R(A:B)=\frac{c_\text{I}}{3} \log\bigg[\frac{1+\sqrt{w_0}}{1-\sqrt{w_0}}\bigg]+\frac{c_\text{I}}{3} \log\bigg[\frac{1+\sqrt{w_1}}{1-\sqrt{w_1}}\bigg],
	\end{align}
	with cross ratios $w_0=\frac{(b_4-\bar{y}_3^*)({b}_4+\bar{y}_2^*)}{(b_4+\bar{y}_3^*)({b}_4-\bar{y}_2^*)}$ and $w_1=\frac{(b_3-b_4)(b_2+{b}_4)}{(b_2-b_4)(b_3+{b}_4)}$. Note that the reflected entropy computed in the effective lower dimensional perspective in \cref{djdc2cft} is exactly the twice of EWCS in the large tension limit evaluated in \cref{djdc2} in the context of double holography.
	
	\subsubsection*{Phase-VI}
	In this phase, we observe the opposite situation compared to the previous phase, the subsystem $B$ is larger than $A$. As a result, the correlator $G^{m,n}_{\mathrm{CFT_{I}}}$ in \cref{Gmnall} again factorizes into a six point function and a two point function. However, the factorization is different from the earlier case,
	\begin{align}
		G^{m,n}_{\mathrm{CFT_{I}}}=&\langle\sigma_{g_B}(-b_4) \sigma_{g_B^{-1}}(b_4) \rangle \langle\sigma_{g_A}(b_1) \sigma_{g_A^{-1}}(b_2) \sigma_{g_B}(b_3)\sigma_{g_A^{-1}}(-b_1)\sigma_{g_A}(-\bar{y}_2^*)\sigma_{g_B^{-1}}(-\bar{y}_3^*)\rangle.
	\end{align}		
	As earlier the six point function on the RHS of the above equation once again factorizes into a product of two four point function in the large-$c$ limit
	\begin{align}
		G^{m,n}_{\mathrm{CFT_{I}}}=&\langle\sigma_{g_B}(-b_4) \sigma_{g_B^{-1}}(b_4) \rangle\langle\sigma_{g_A}(b_1) \sigma_{g_A^{-1}}(b_2) \sigma_{g_B}(b_3)\sigma_{g_A^{-1}}(-b_1)\rangle\nonumber\\
		& \langle\sigma_{g_A}(b_1)  \sigma_{g_A^{-1}}(-b_1)\sigma_{g_A}(-\bar{y}_2^*)\sigma_{g_B^{-1}}(-\bar{y}_3^*)\rangle.
	\end{align}		
	The CFT$_{\text{II}}$ correlation function in \cref{Gmnall} on the other hand factorizes into product of two point functions as in \cref{factorization-dj-double-II}.
	Substituting the above expressions in \cref{SRGmn0} and utilizing \cref{Large-c-block} we may determine the reflected entropy as
	\begin{align}
		S_R(A:B)=\frac{c_\text{I}}{3} \log\bigg[\frac{1+\sqrt{w_2}}{1-\sqrt{w_2}}\bigg]+\frac{c_\text{I}}{3} \log\bigg[\frac{1+\sqrt{w_3}}{1-\sqrt{w_3}}\bigg],
	\end{align}
	where $w_2=\frac{(b_3+{b}_1)({b}_1-b_2)}{(b_1-b_3)({b}_1+b_2)}$ and $w_3=\frac{(\bar{y}_3^*+b_1)({b}_1-\bar{y}_2^*)}{(b_1+\bar{y}_2^*)({b}_1-\bar{y}_3^*)}$ are the cross ratios. Similar to the earlier phases, the reflected entropy calculated here matches with the twice of EWCS computed in \cref{djdc3}.

\subsubsection{No Island saddle}

In this phase all the subsystems $A,B$ and $C$ are very small in size such that neither of $A,B,C$ and $A\cup B\cup C$ has any island. The effective Renyi reflected entropy is therefore given by
\begin{align}
	S_{R}^{(m,n,\mathrm{eff})}(&A \cup \mathrm{Is}_{R}(A): B \cup \mathrm{Is}_{R}(B))\nonumber\\&=\frac{1}{1-n}\log\bigg[\frac{ \langle\sigma_{g_A}(b_1)\sigma_{g_A^{-1}}(b_2) \sigma_{g_B}(b_3)\sigma_{g_B^{-1}}(b_4)\rangle_{\text{CFT}^{\otimes mn}_\text{I}}}{\left(\langle\sigma_{g_m}(b_1)\sigma_{g_m^{-1}}(b_2) \sigma_{g_m}(b_3)\sigma_{g_m^{-1}}(b_4)\rangle_{\text{CFT}^{\otimes m}_\text{I}}\right)^n}\bigg]+\left(\text{I}\leftrightarrow\text{II}\right).
\end{align}
Considering the contributions from both the CFTs and taking the replica limit we obtain
\begin{align}
	S_R(A:B)=\frac{\mathrm{c_{I}}+\mathrm{c_{II}}}{3} \log\bigg[\frac{1+\sqrt{x}}{1-\sqrt{x}}\bigg],
\end{align}
where $x=\frac{b_{12}b_{34}}{b_{13}b_{24}}$ is the cross-ratio. We observe an exact agreement between the above expression and the corresponding EWCS given in \cref{djnc}.

	\section{Black Hole Evaporation : Time Evolution of EWCS}\label{sec:EWCST}

	In this section, we study the time evolution of the EWCS dual to the reflected entropy  for bipartite mixed states in the AdS/ICFT setup, in the context of the black hole information loss paradox. We will consider bipartite mixed state configurations in the thermal baths located outside an eternal black hole, where these baths collect Hawking radiation emitted from the black hole. The holographic dual of the eternal black hole in AdS$_3$ is the thermofield double (TFD) state \cite{Maldacena:2001kr}, which may be obtained from the vacuum state of the ICFT$_2$ via a conformal map. As described in \cite{Anous:2022wqh}, the TFD state in the ICFT$_2$ may be prepared from a path integral on half of an infinite cylinder with a circular interface between the two CFT$_2$s dividing the cylinder into two distinct parts. Furthermore, as explained in \cite{Anous:2022wqh}, the TFD state on the cylinder may be obtained via a series of conformal transformations from the vacuum state of the ICFT on the complex plane described by $\zeta=x+i t_E$. The planar interface $x_i=0$ between the two CFT$_2$s may be mapped into a circle of length $\ell$, $\tilde{x}^2+\tilde{t}^2=\ell^2$ by utilizing the transformation
	\begin{align}\label{planeCircle}
		p=\frac{4\ell^2}{2\ell-\zeta}-\ell\,,
	\end{align}
	where $p=\tilde{x}+i\tilde{t}_E$ is the complex coordinate on the plane where the CFT$_\text{I/II}$s are defined respectively on the interior and exterior of the circle of length $\ell$ centered at the origin. Subsequently, the conformal transformation
	\begin{align}
		p=\ell\, e^{\frac{2\pi q}{\beta}}\,,\label{thermal-map}
	\end{align}
	maps the planar geometry into the infinite cylinder with the interface mapped to a disk $\mathfrak{Re}(q)=0$ with $q=u+i v_E$ denoting the complex coordinate on the thermal cylinder on which the TFD state is defined.
	
	The bulk dual geometry corresponding to the ICFT defined on the plane with a circular interface will be important for computational purposes in the following. Note that the transformation in \cref{planeCircle} is nothing but a $SL(2,R)$ transformation\footnote{In particular, it is given by a special conformal transformation followed by a translation.} for which the bulk dual geometry may be easily found by looking for the AdS$_3$ isometry which maps straight lines into circles\footnote{Alternatively, one may use the Banados formalism \cite{Banados:1998gg,Shimaji:2018czt}.} \cite{Takayanagi:2011zk}. The corresponding coordinate transformations are given as follows
	\begin{align}
		&\tilde{x}_i=\frac{x_i -\frac{x^2 _i+z^2 _i-t^2 _i}{2\ell}}{1-\frac{x_i}{\ell}+\frac{x^2 _i+z^2 _i-t^2 _i}{4\ell^2}}+l\,~~,~~ \tilde{z}_i= \frac{z_i}{1-\frac{x_i}{\ell} +\frac{x^2 _i+z^2 _i-t^2 _i}{4\ell^2}}\,~~,~~\tilde{t}_i= \frac{t_i}{1-\frac{x_i}{\ell} +\frac{x^2 _i+z^2 _i-t^2 _i}{4\ell^2}}\,,~~~~\left(i=\text{I,II}\right)\label{tilde-coordinates}
	\end{align}
	where we have Wick rotated the time coordinate to $t=-i t_E$. Being an isometry, the above bulk transformations do not change the metric and hence we have Poincar\'e AdS$_3$ on either side of the spherical EOW brane with profile
	\begin{align}
		&\tilde{x}_\text{I}^2-\tilde{t}_\text{I}^2+(\tilde{z}_\text{I}-\ell\tan\psi_\text{I})^2=\ell^2\sec^2\psi_\text{I}\,,\notag\\
		&\tilde{x}_\text{II}^2-\tilde{t}_\text{II}^2+(\tilde{z}_\text{II}+\ell\tan\psi_\text{II})^2=\ell^2\sec^2\psi_\text{II}\,.
	\end{align}
	
	On the cylinder, we will consider Lorentzian time evolution utilizing the Wick rotation $v=-i v_E$. Having described the required notations and conventions we will now compute the EWCS dual to the reflected entropy of the mixed state configuration involving two adjacent subsystems of an ICFT$_{\text{2}}$ in a TFD state. Note that unlike the vacuum case, here we will restrict ourselves to the computation for the adjacent intervals as we do not expect that the disjoint interval case will reveal any novel physical aspects that are qualitatively different from those observed in the adjacent interval scenario.

	Consider the mixed state configuration of adjacent subsystems $A\cup B$ where $A=A_L\cup A_R$ and $B=B_L\cup B_R$ with
	\begin{align}\label{ABTFD}
	A_L&=[(u_0,v),(u_1,v)]_{\text{I}}\cup[(-u_0,v),(-u_1,v)]_{\text{II}}\,,\nonumber\\ B_L&=[(u_1,v),(\infty,v)]_{\text{I}}\cup[(-u_1,v),(-\infty,v)]_{\text{II}}\,,\nonumber\\
	A_R&=\left[\left(u_0,-v+\frac{i\beta}{2}\right),\left(u_1,-v+\frac{i\beta}{2}\right)\right]_{\text{I}}\cup\left[\left(-u_0,-v+\frac{i\beta}{2}\right),\left(-u_1,-v+\frac{i\beta}{2}\right)\right]_{\text{II}}\,,\nonumber\\ B_R&=\left[\left(u_1,-v+\frac{i\beta}{2}\right),\left(\infty,-v+\frac{i\beta}{2}\right)\right]_{\text{I}}\cup\left[\left(-u_1,-v+\frac{i\beta}{2}\right),\left(-\infty,-v+\frac{i\beta}{2}\right)\right]_{\text{II}}\,.
		\end{align}
	The entanglement entropy for $A\cup B$ has been investigated in \cite{Anous:2022wqh}, where it was observed that in the doubly holographic picture the entanglement entropy is computed through the lengths of two competing sets of RT surfaces: the Hartman-Maldacena (HM) surfaces \cite{Maldacena:2001kr} and a pair of geodesics which cross the EOW brane. The Page time at which these two saddles change dominance, was found to be \cite{Anous:2022wqh}
	\begin{align}
		v^{}_P=\frac{\beta}{2\pi}\text{cosh}^{-1}\left[\sinh\left(\frac{2\pi u_0}{\beta}\right)\frac{6 S_{\text{int}}}{c_\text{I}+c_\text{II}}\right]\,,\label{Page-time}
	\end{align}
	where $S_\text{int}$ denotes the interface entropy. Within each phase of the entanglement entropy of $A\cup B$, the entanglement wedge cross section corresponding to the mixed state $\rho_{AB}$ undergoes various phase transitions with time $v$. In the following, we will systematically investigate all these phases and compute the EWCS in each case.
	
	\subsection{Before Page Time}
\begin{figure}
		\centering
		\begin{subfigure}[b]{0.45\textwidth}
			\centering
			\includegraphics[width=0.7\textwidth]{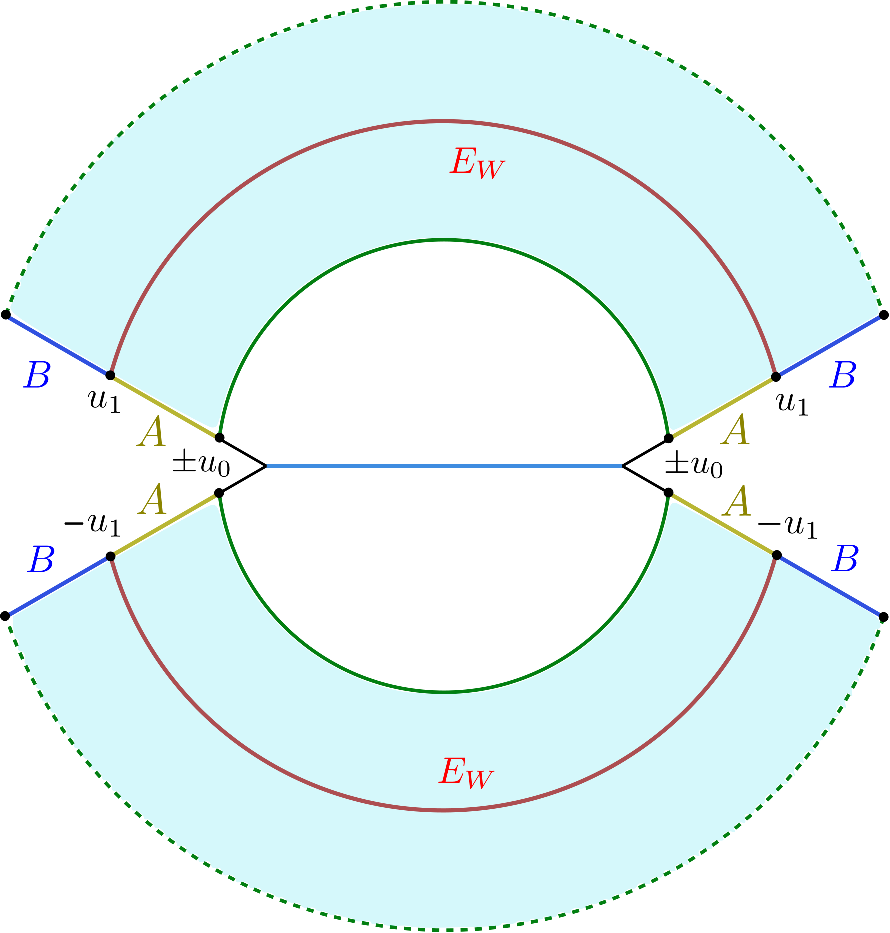}
			\caption{}
			\label{FT1}
		\end{subfigure}
		\hfill
		\begin{subfigure}[b]{0.45\textwidth}
			\centering
			\includegraphics[width=0.7\textwidth]{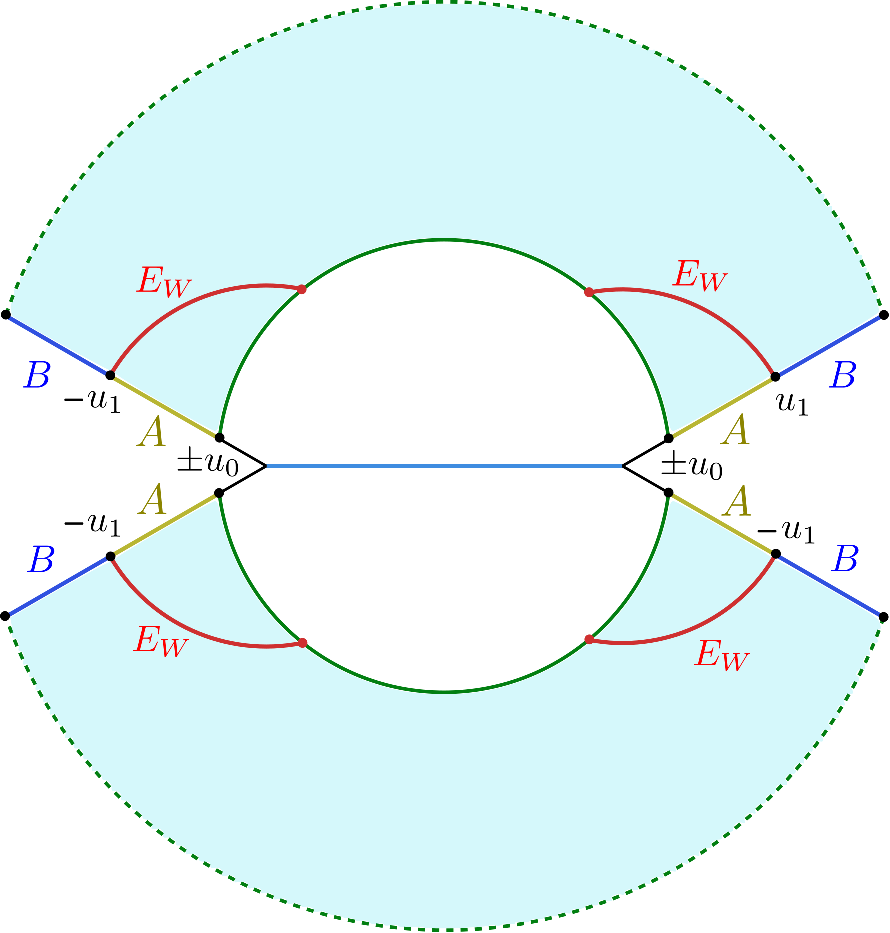}
			\caption{}
			\label{FT2}
		\end{subfigure}
		\caption{Phase transitions of the EWCS corresponding to two adjacent subsystems in the radiation bath, before the Page time: (a) The EWCS does not terminate on the RT of $A\cup B$, (b) EWCS terminates on the smaller HM surface.}
		\label{fig:finiteT-before-vP}
	\end{figure}

	Before the Page time, the RT saddle contributing to the entanglement entropy of $A\cup B$ consists of two HM surfaces which connect the endpoints of the subsystems from both copies. In the effective intermediate picture, this corresponds to the non-island phase. In this phase, the EWCS between $A$ and $B$ is either HM surface connecting the shared boundary of $A$ and $B$ on each copy or terminates on the smaller HM surface, as sketched in \cref{fig:finiteT-before-vP}.
	
	\subsection*{Phase-I}
	To compute the length of the EWCS, we employ the trick utilized in \cite{Anous:2022wqh}, namely perform the computations in the planar tilde coordinates in \cref{tilde-coordinates} and finally transform back to the cylinder. From \cref{thermal-map} the Lorentzian trajectories of the endpoints of $A$ and $B$ in the left copy of the ICFT are given by
	\begin{align}
		&\tilde{x}_\text{I}(u_k,v)=\ell e^{-\frac{2\pi u_k}{\beta}}\cosh\left(\frac{2\pi v}{\beta}\right)~~,~~\tilde{t}_\text{I}(u_k,v)=\ell e^{-\frac{2\pi u_k}{\beta}}\sinh\left(\frac{2\pi v}{\beta}\right)\notag\\
		&\tilde{x}_\text{II}(u_k,v)=\ell e^{\frac{2\pi u_k}{\beta}}\cosh\left(\frac{2\pi v}{\beta}\right)~~,~~\tilde{t}_\text{II}(u_k,v)=\ell e^{\frac{2\pi u_k}{\beta}}\sinh\left(\frac{2\pi v}{\beta}\right)\,,\label{Lorezian-trajectory}
	\end{align}
	while similar expressions hold for the TFD copy with $\tilde{x}$ replaced by $-\tilde{x}$. For the present phase, as described earlier, the EWCS is a HM surface whose length has already been computed in \cite{Anous:2022wqh}. Following this, we may write down the EWCS as follows
	\begin{align}\label{FTP1}
		E_W(\rho_{AB})&=\frac{L_\text{I}}{2G_N}\log\left[\frac{2\tilde{x}_\text{I}(u_1,v)}{\tilde{\epsilon}_\text{I}(u_1,v)}\right]+\frac{L_\text{II}}{2G_N}\log\left[\frac{2\tilde{x}_\text{II}(u_1,v)}{\tilde{\epsilon}_\text{II}(u_1,v)}\right]\notag\\
		&=\frac{L_\text{I}+L_\text{II}}{2G_N}\log\left[\frac{\beta}{\pi\epsilon}\cosh\left(\frac{2\pi v}{\beta}\right)\right]\,,
	\end{align}
	where we have utilized the fact that the cut-off in tilde coordinates is related to the cylinder cut-off as follows (cf. \cref{thermal-map})
	\begin{align}
		\tilde{\epsilon}=\left(\frac{\beta}{2\pi\ell}e^{-\frac{\pi(q+\bar{q})}{\beta}}\right)\epsilon.\label{cut-off-FT}
	\end{align}
	\subsection*{Phase-II}
	Next, we compute the lengths of the EWCS landing on the HM surfaces, utilizing the Poincar\'e AdS$_3$ geometry described by the tilde coordinates.

	In the following, we will suppress the subscripts $\text{I,II}$ to keep the notation simple. Due to the symmetry of the setup, the computation reduces to finding the length of a minimal surface from $(\tilde{x}_1,\tilde{t}_1)$ to the RT surface described by equation $\tilde{x}^2+\tilde{z}^2=\tilde{x}_0^2$ at $\tilde{t}=\tilde{t}_0$. Parametrizing a point $P:(\tilde{x},\tilde{t},\tilde{z})=(\tilde{x}_0\sin\theta,\tilde{t}_0,\tilde{x}_0\cos\theta)$ on the RT surface, we may obtain the length of the surface from $Q=(\tilde{x}_1,\tilde{t}_1)$ as follows
	\begin{align}
		d_{PQ}=L\,\text{cosh}^{-1}\left[\frac{(\tilde{x}_0\sin\theta-\tilde{x}_1)^2-(\tilde{t}_0-\tilde{t}_1)^2+(\tilde{x}_0\cos\theta)^2}{2\tilde{\epsilon}_1\tilde{x}_0\cos\theta}\right]\,.\label{d-FiniteII}
	\end{align}
	Extremization with respect to the arbitrary angle $\theta$ leads to
	\begin{align}
		\partial_\theta d_{PQ}=0 \Rightarrow \theta=\sin^{-1}\left[\frac{2\tilde{x}_0\tilde{x}_1}{\tilde{x}_0^2+\tilde{x}_1^2-(\tilde{t}_0-\tilde{t}_1)^2}\right]\,.
	\end{align}
	Substituting this back into \cref{d-FiniteII}, we obtain the minimal length to be 
	\begin{align}
		d^\text{min}_{PQ}=L\log\left[\frac{\sqrt{\left(\tilde{x}_0^2+\tilde{x}_1^2-(\tilde{t}_0-\tilde{t}_1)^2\right)^2-4\tilde{x}_0^2\tilde{x}_1^2}}{\tilde{\epsilon}_1\tilde{x}_0}\right].\label{d-FT2}
	\end{align}
	Now restoring the subscripts $\text{I,II}$, utilizing \cref{Lorezian-trajectory,cut-off-FT} and accounting for the pair of geodesics for both TFD copies of the setup, we obtain the minimal EWCS for this phase to be
	\begin{align}\label{FTP2}
		E_W(\rho_{AB})=\frac{L_\text{I}+L_\text{II}}{2G_N}\log\left[\frac{\beta\left(e^{\frac{2\pi u_1}{\beta}}-e^{\frac{2\pi u_0}{\beta}}\right)\sqrt{e^{\frac{4\pi u_0}{\beta}}+e^{\frac{4\pi u_1}{\beta}}+2e^{\frac{2\pi (u_0+u_1)}{\beta}}\cosh\left(\frac{4\pi v}{\beta}\right)}}{2\pi\epsilon\, e^{\frac{2\pi (u_0+u_1)}{\beta}}\cosh\left(\frac{2\pi v}{\beta}\right)}\right].
	\end{align}
	\subsection{After Page time}
\begin{figure}[h!]
		\centering
		\begin{subfigure}[b]{0.3\textwidth}
			\centering
			\includegraphics[width=\textwidth]{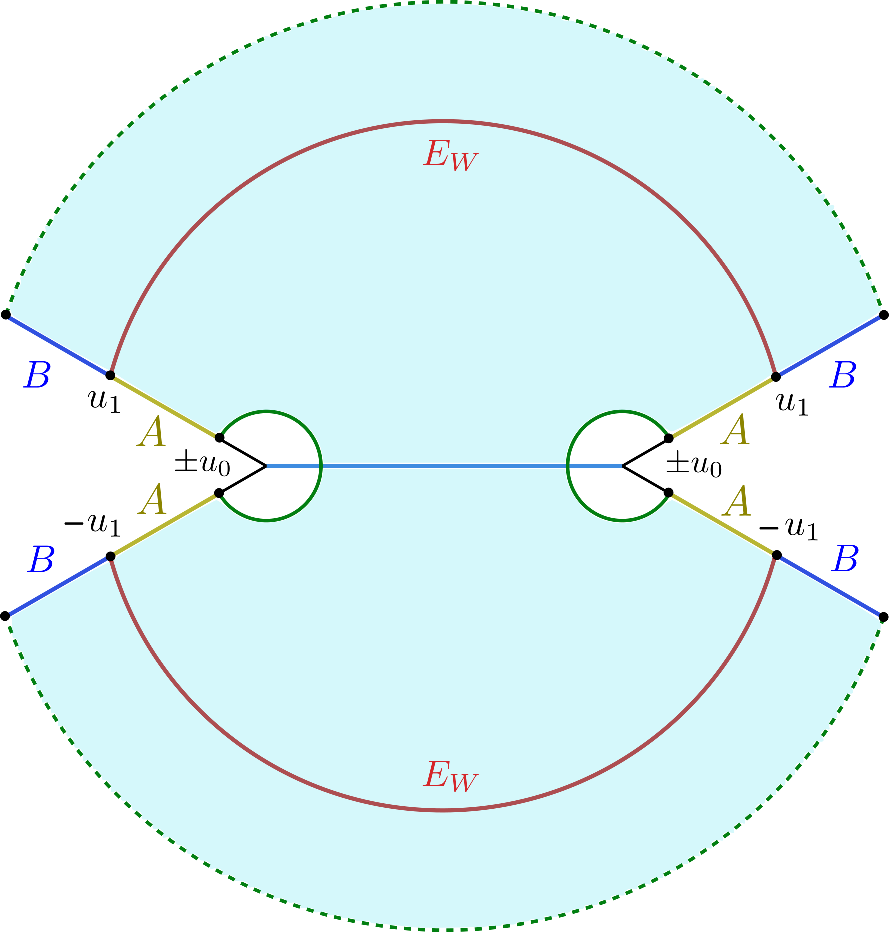}
			\caption{}
			\label{FT3}
		\end{subfigure}
		\hfill
		\begin{subfigure}[b]{0.3\textwidth}
			\centering
			\includegraphics[width=\textwidth]{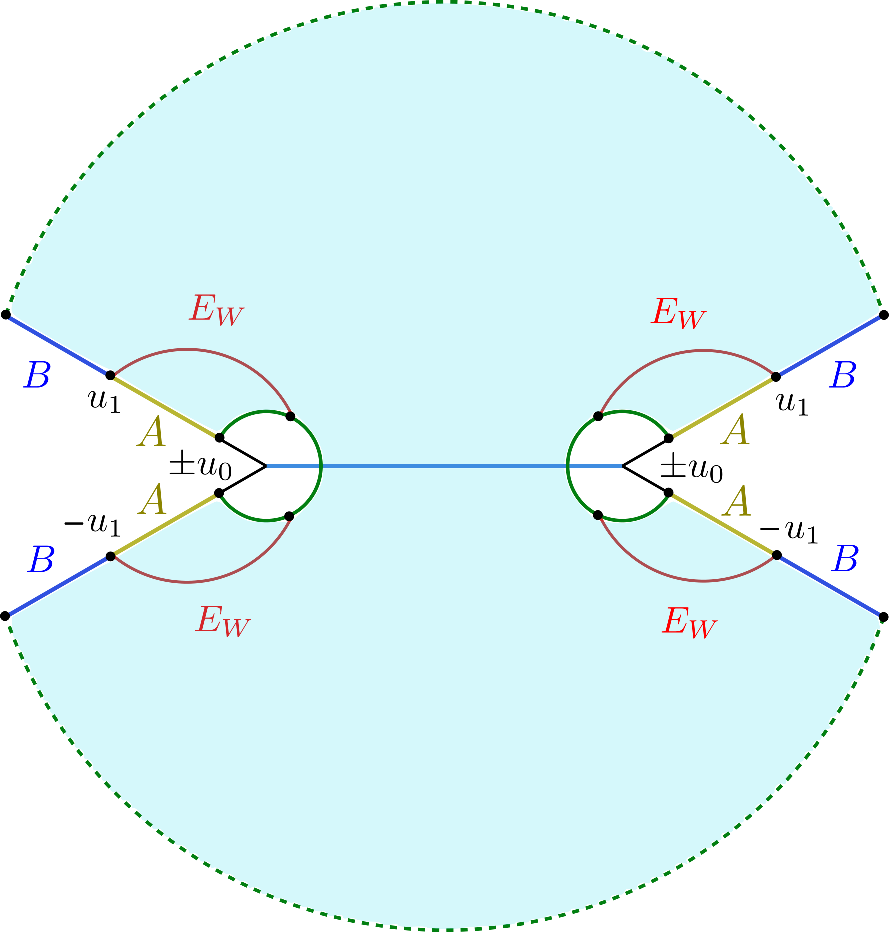}
			\caption{}
			\label{FT}
		\end{subfigure}
		\hfill
		\begin{subfigure}[b]{0.3\textwidth}
			\centering
			\includegraphics[width=\textwidth]{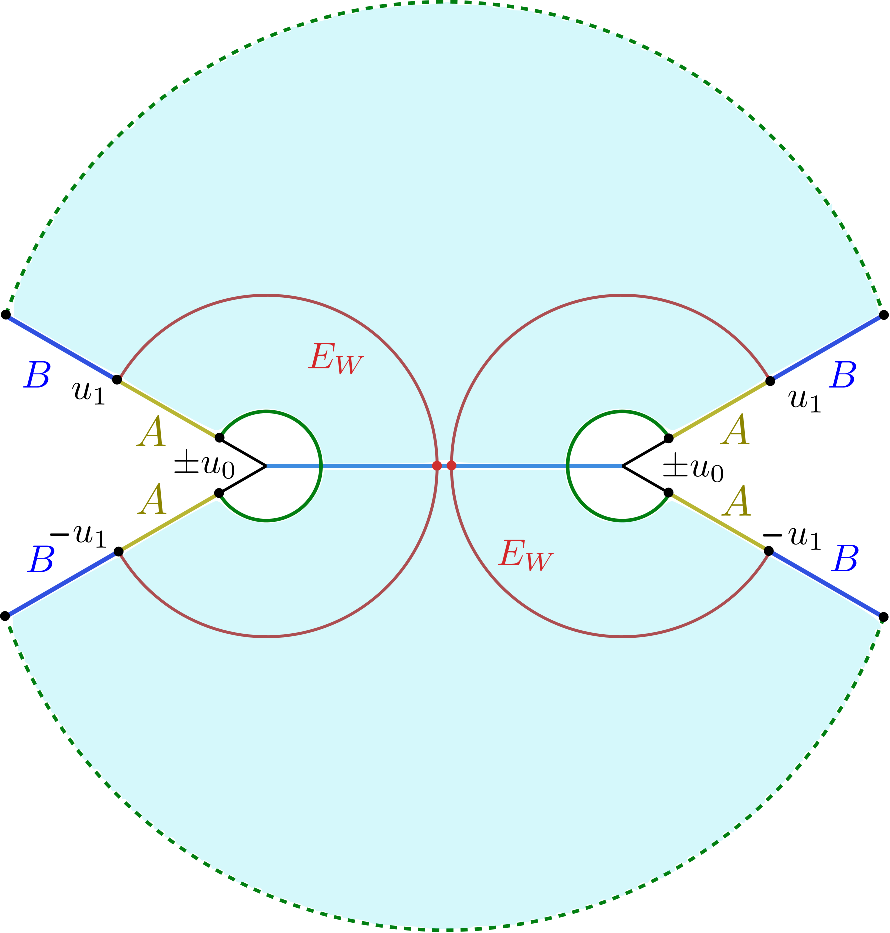}
			\caption{}
			\label{FT5}
		\end{subfigure}
		\caption{Phase transitions of the EWCS corresponding to two adjacent subsystems in the radiation bath, after the Page time: (a) The EWCS is a HM surface connecting the shared boundary of $A$ and $B$ on both copies, (b) EWCS terminates on the RT surface of $A\cup B$, (c) The EWCS crosses the brane. In the intermediate braneworld perspective, non-trivial island cross-sections for the reflected entropy appear on the brane.}
		\label{fig:finiteT-after-vP}
	\end{figure}

	After the Page time $v^{}_P$, the RT saddle for $A\cup B$ crosses the EOW brane which corresponds to the appearance of an island in the intermediate braneworld picture. In this case, the minimal EWCS has three possible phases which are depicted in \cref{fig:finiteT-after-vP}.
	
	\subsection*{Phase-III}
	As shown in \cref{fig:finiteT-after-vP}$\textcolor{blue}{(a)}$, this phase corresponds to a HM surface connecting the shared boundary of $A$ and $B$ on both copies, which is identical to phase-I. Hence, the EWCS is given by
	\begin{align}\label{FTP3}
		E_W(\rho_{AB})=\frac{L_\text{I}+L_\text{II}}{2G_N}\log\left[\frac{\beta}{\pi\epsilon}\cosh\left(\frac{2\pi v}{\beta}\right)\right]\,.
	\end{align}
	Note that although the expression is identical to that of phase-I, this phase gets activated only after the Page time and hence contributes non-trivially to the Page curve of the holographic reflected entropy.
	\subsection*{Phase-IV}
	In phase-IV, the EWCS terminates on the RT surfaces which cross the EOW brane. We find it easier to perform the computations in the original planar coordinates\footnote{Once again, owing to the symmetry of the setup, we are suppressing the subscripts $\text{I,II}$, which will be restored at the end of computations.} $(x_i,t_i,z_i)$ given in \cref{tilde-coordinates}, where the equations for the RT surfaces may be easily obtained as \cite{Fujita:2011fp}
	\begin{align}
		x_i^2+z_i^2=x_0^2~~,~~t=t_0\,,
	\end{align}
	where $(u_0,v)$ has been mapped to $(x_0,t_0)$ utilizing the transformations in \cref{thermal-map,tilde-coordinates}. In these coordinates, the task of finding the EWCS reduces to that for a minimal curve starting on the above RT surface and terminating on $(x_1,t_1)$. We have already performed an identical computation in the tilde coordinates for phase-II. Hence, utilizing \cref{d-FT2} we may write down the corresponding length as follows
	\begin{align}
		d_{\text{min}}&=L \log\left[\frac{\sqrt{\left(x_0^2+x_1^2-(t_0-t_1)^2\right)^2-4x_0^2x_1^2}}{\epsilon(x_1,t_1)x_0}\right]\notag\\
		&=L\log\left[\frac{2\sqrt{\left((\tilde{x}_0-\tilde{x}_1)^2-(\tilde{t}_0-\tilde{t}_1)^2\right)\left(\ell^4+2\ell^2(\tilde{t}_0\tilde{t}_1-\tilde{x}_0\tilde{x}_1)+(\tilde{t}_0^2-\tilde{x}_0^2)(\tilde{t}_1^2-\tilde{x}_1^2)\right)}}{\tilde{\epsilon}_1\left(\tilde{x}_0^2-\tilde{t}_0^2-\ell^2\right)}\right],
	\end{align}
	where in the second equality, we have used \cref{tilde-coordinates} and the fact that the cut-off in tilde and non-tilde coordinates are related by
	\begin{align}
		\epsilon(x,t)=\frac{4\ell^2\tilde{\epsilon}}{(\tilde{x}+\ell)^2-\tilde{t}^2}\,.
	\end{align}
	Finally transforming to the cylinder coordinates using \cref{thermal-map,cut-off-FT}, restoring the subscripts $\text{I,II}$ and accounting for the pair of geodesics in both copies of the geometry, we obtain the EWCS for this phase as follows
	\begin{align}\label{FTP4}
		E_W(\rho_{AB})=\frac{L_\text{I}+L_\text{II}}{2G_N}\log\left[\frac{\beta}{\pi\epsilon}\,\frac{\cosh\left(\frac{2\pi u_1}{\beta}\right)-\cosh\left(\frac{2\pi u_0}{\beta}\right)}{\sinh\left(\frac{2\pi u_0}{\beta}\right)}\right]\,.
	\end{align}
	
	\subsection*{Phase-V}
	In the final phase depicted in \cref{fig:finiteT-after-vP}$\textcolor{blue}{(c)}$, the EWCS between $A$ and $B$ consists of two geodesics connecting the shared boundary of $A$ and $B$ in CFT$^\text{I}_2$ and CFT$^\text{II}_2$, which cross the EOW brane. The length of such geodesics has already been computed in \cite{Anous:2022wqh} and hence the EWCS is given as
	\begin{align}\label{FTP5}
		E_W(\rho_{AB})=\frac{L_\text{I}+L_\text{II}}{2G_N}\log\left[\frac{\beta}{\pi\epsilon}\sinh\left(\frac{2\pi u_1}{\beta}\right)\right]+2 S_\text{int}\,,
	\end{align}
	where $S_\text{int}$ is the interface entropy defined in \cref{EW-phase1}.
	
	\subsection{Page curves}
\begin{figure}[h!]
		\centering
		\begin{subfigure}[b]{0.45\textwidth}
			\centering
			\includegraphics[width=6cm,height=3.3cm]{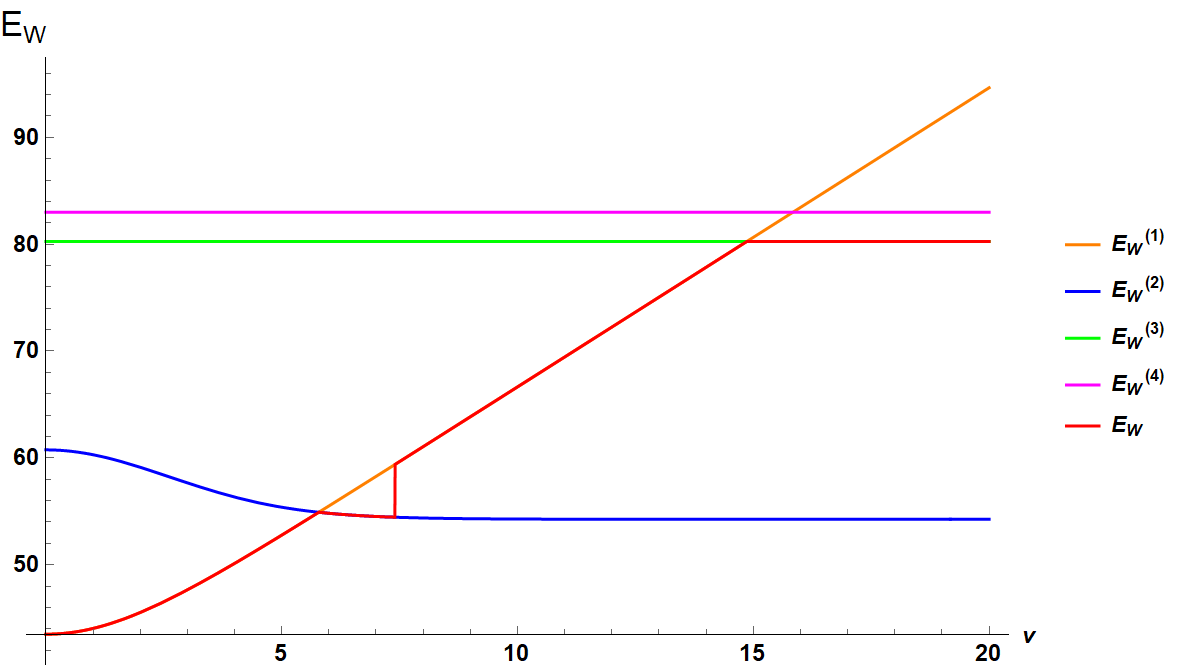}
			\caption{$\beta=5\pi,u_0=0.1,u_1=8$}
			\label{fig:PC1}
		\end{subfigure}
		\hfill
		\begin{subfigure}[b]{0.45\textwidth}
			\centering
			\includegraphics[width=6cm,height=3.3cm]{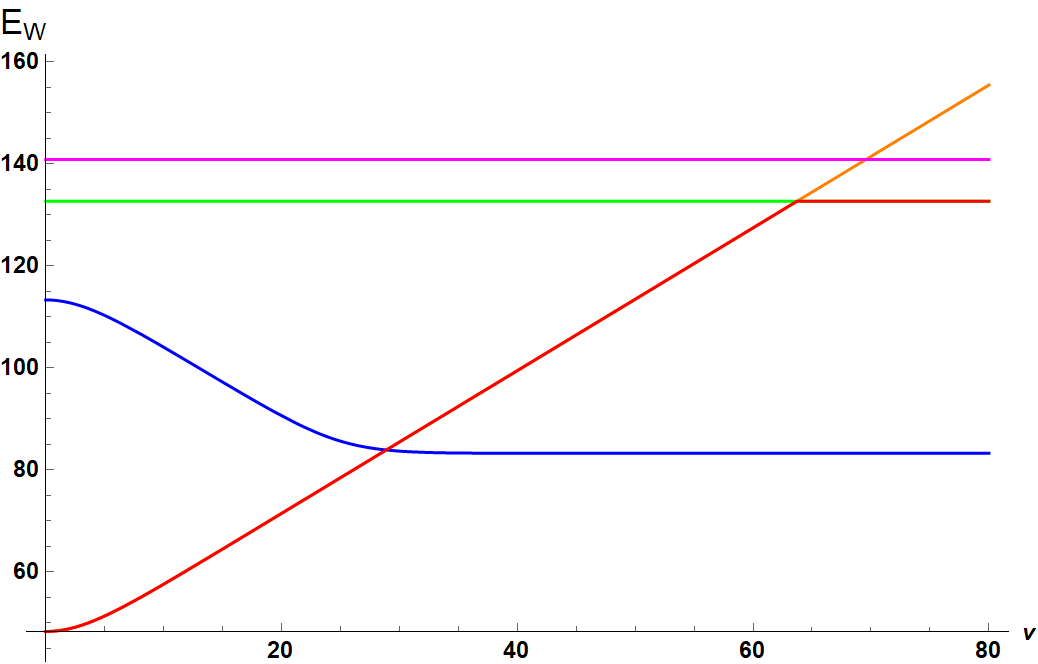}
			\caption{$\beta=10\pi,u_0=0.1,u_1=50$}
			\label{fig:PC2}
		\end{subfigure}
		\hfill
		\begin{subfigure}[b]{0.45\textwidth}
			\centering
			\includegraphics[width=6cm,height=3.3cm]{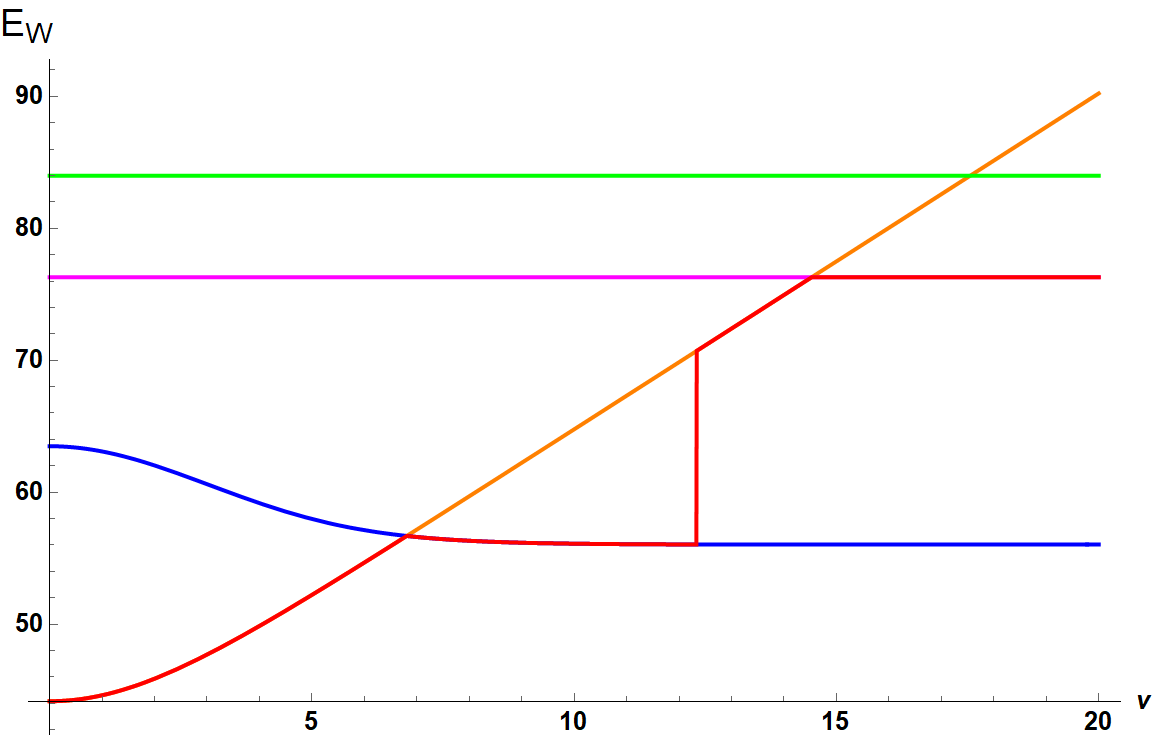}
			\caption{$\beta=5.5\pi,u_0=0.5,u_1=10$}
			\label{fig:PC3}
		\end{subfigure}
		\hfill
		\begin{subfigure}[b]{0.45\textwidth}
			\centering
			\includegraphics[width=6cm,height=3.3cm]{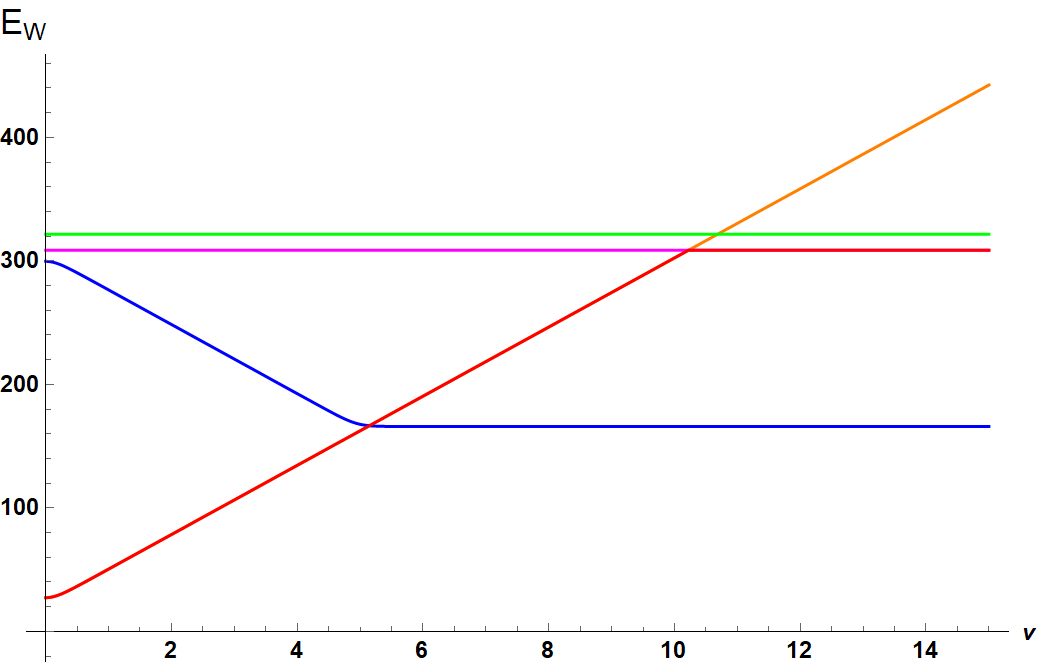}
			\caption{$\beta=0.5\pi,u_0=0.5,u_1=10$}
			\label{fig:PC4}
		\end{subfigure}
		\hfill
		\begin{subfigure}[b]{0.45\textwidth}
			\centering
			\includegraphics[width=6cm,height=3.3cm]{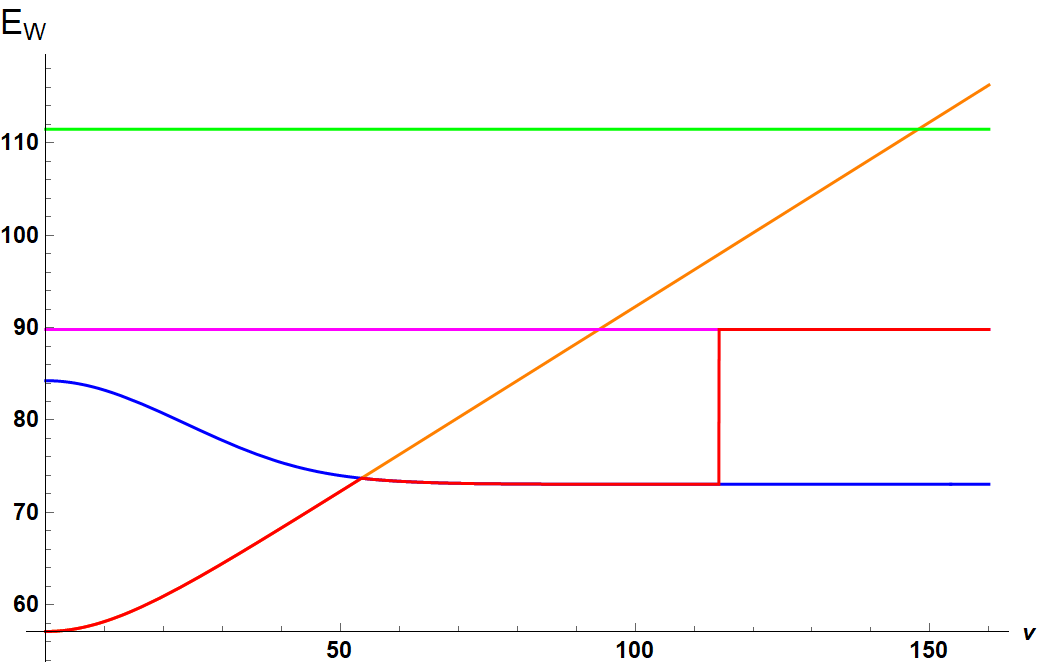}
			\caption{$\beta=35\pi,u_0=20,u_1=100$}
			\label{fig:PC5}
		\end{subfigure}
		\caption{Page curves (red) : For $c_\text{I}=1,c_\text{II}=20,L_\text{I}=0.1,L_\text{II}=0.2,\epsilon=0.01,\delta=0.01$.}
		\label{fig:Page curves}
	\end{figure}

	In this subsection, we plot the evolution of the EWCS with time in \cref{fig:Page curves} which correspond to the analogues of the Page curves for reflected entropy from the holographic duality mentioned earlier. We note that the EWCS between nearby radiation and distant radiation experiences two phases. At early times the entanglement entropy of $A\cup B$ is in the connected phase and correspondingly the EWCS is given by the minimum of phase-I and phase-II. On the other hand, after the Page time for $A\cup B$ given in \cref{Page-time}, the disconnected phase dominates and the EWCS is given by the minimum of phase-III, phase-IV and phase-V.
	
	\section{Reflected entropy from island prescription: TFD state }\label{sec:SRIsT}
	
	\begin{figure}[h!]
		\centering
		\begin{subfigure}[b]{0.45\textwidth}
			\centering
			\includegraphics[width=6cm,height=5cm]{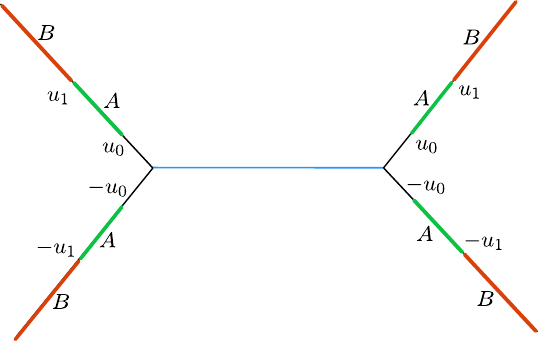}
			\caption{}
			\label{SRADJFTCFT1}
		\end{subfigure}
		\hspace{.5cm}
		\begin{subfigure}[b]{0.45\textwidth}
			\centering
			\includegraphics[width=6cm,height=5cm]{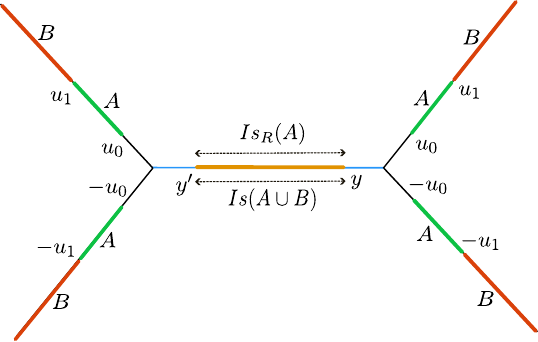}
			\caption{}
			\label{SRADJFTCFT2}
		\end{subfigure}
		\vspace{.5cm}
		\begin{subfigure}[b]{0.45\textwidth}
			\centering
			\includegraphics[width=6cm,height=5cm]{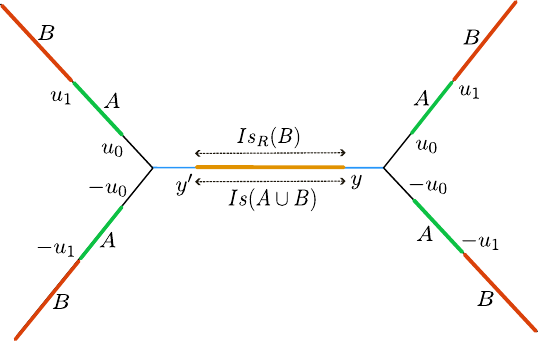}
			\caption{}
			\label{SRADJFTCFT3}
		\end{subfigure}
		\hspace{.5cm}
		\begin{subfigure}[b]{0.45\textwidth}
			\centering
			\includegraphics[width=6cm,height=5cm]{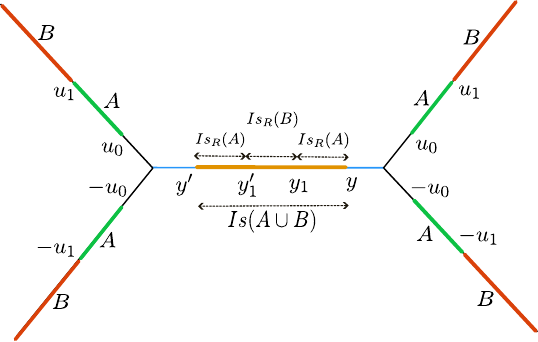}
			\caption{}
			\label{SRADJFTCFT4}
		\end{subfigure}
		\caption{Schematic for a mixed state configuration involving adjacent subsystems $A$ and $B$ in the two copies of thermal ICFT$_{\text{2}}$s in a TFD state and their corresponding reflected entropy islands on the brane in the effective two dimensional theory. Note that we have denoted  the points on brane  in the left and the right subsystem by $y'$ and $y$ respectively for clarity but their  numerical values are same.}\label{DJfinitetemp}
	\end{figure}
	
	In this section, we compute the island contributions to the reflected entropy for  the mixed state of adjacent subsystems $A$ and $B$ in  two copies of ICFT$_{\text{2}}$s in a TFD state. As depicted in \cref{DJfinitetemp} we consider one of the two intervals denoted by $B$ to be semi-infinite in order to simplify our computations.   As described in \cref{sec:EWCST}, the two copies of the ICFT$_{\text{2}}$s in a TFD state are on a cylinder which is obtained from the vacuum state on the complex plane through a  series  of conformal transformations. Note that on such a cylinder we denote the complex coordinates on the two copies ICFT$_{\text{L}}$ and ICFT$_{\text{R}}$ by  $\tilde{q}$ and  $q$ respectively. Since the intervals chosen in the CFT$_{\text{I}}$ and CFT$_{\text{II}}$ are identical, the spatial $u$ coordinates differ only by sign whereas the temporal $v$ coordinates are same i.e if 
	\begin{align}
		q_i^{\text{I}}&=u_i+i v_E, \hspace{1.3cm}\, \,	q_i^{\text{II}}=-u_i+i v_E\nonumber\\
		\tilde{q}_i^{\text{I}}&=u_i-iv_E-\frac{i \beta}{2}, \hspace{0.4cm}\, \, \tilde{q}_i^{\text{II}}=-u_i-i v_E-\frac{i \beta}{2}
	\end{align}
	where $i=0,1,2$). Note that the  points on the brane are denoted by 
	\begin{align}
q_b^{\text{I}}&=-y+i v_E,\hspace{1.3cm} \, \, q_b^{\text{II}}=y+iv_E \nonumber \\
\tilde{q}_b^{\text{I}}&=-y-iv_E-\frac{i \beta}{2}, \hspace{0.4cm} \, \, \tilde{q}_b^{\text{II}}=y-iv_E-\frac{i \beta}{2}
	\end{align}
	When we require multiple points on the brane we will denote them as $q_a,q_b,q_c..$ and $\tilde{q}_a,\tilde{q}_b,\tilde{q}_c..$. At the end of computation we will analytically continue to Lorentzian coordinates by Wick rotation $v_E=-iv$. Note that in the above definitions the Roman numerals  I,II   in the superscript indicate whether the coordinate is in CFT$_{\text{I}}$ or CFT$_{\text{II}}$ within a single copy of the ICFT$_{\text{2}}$. The two subsystems are given by $A=A_L\cup A_R$ and $B=B_L\cup B_R$ where 
	\begin{align}\label{ABTFD11}
		A_L&=[q_0^{\text{I}},q_1^{\text{I}}]\cup[q_0^{\text{II}},q_1^{\text{II}}],\nonumber\\
		 B_L&=[q_1^{\text{I}},q_2^{\text{I}}]\cup[q_1^{\text{II}},q_2^{\text{II}}],\nonumber\\
			A_R&=[\tilde{q}_0^{\text{I}},\tilde{q}_1^{\text{I}}]\cup[\tilde{q}_0^{\text{II}},\tilde{q}_1^{\text{II}}],\nonumber\\
		B_R&=[\tilde{q}_1^{\text{I}},\tilde{q}_2^{\text{I}}]\cup[\tilde{q}_1^{\text{II}},\tilde{q}_2^{\text{II}}].
	\end{align}
Note that we have chosen $B$ to be semi-infinite and therefore $q_2^{\text{I}}=[\infty,v]$, $q_2^{\text{II}}=[-\infty,v]$  and $\tilde{q}_2^{\text{I}}=[\infty,-v-\frac{i \beta}{2}]$, $\tilde{q}_2^{\text{II}}=[-\infty,-v-\frac{i \beta}{2}]$. In the rest of this article we will drop the superscripts I,II for brievity. However note that there is a sign difference between the spatial coordinate depending on whether it is occuring in the correlation function of CFT$_{\text{I}}$ or CFT$_{\text{II}}$.  As discussed in the previous section there are two phases for the bulk entanglement wedge corresponding to the  semi-infinite subsystem $AB$. These two phases correspond to the presence or absence of the entanglement entropy islands for  $AB$. For each phase of the entanglement entropy there are multiple sub-phases for the reflected entropy which we describe below.

	\subsection{Before Page Time}
	We begin by considering the case when the subsystem $AB$ does not receive any island contribution as depicted in fig. \ref{SRADJFTCFT1}. When there are no island contributions the required correlation function corresponding to the reflected entropy is given by
	\begin{align}\label{Gmn4ptFTNI}
		G^{m,n}_{\mathrm{CFT_{k}}}&=\langle \sigma_{g_A}(q_0)\sigma_{g_A^{-1}g_B}(q_1)\sigma_{g_A}(\tilde{q}_0)\sigma_{g_A^{-1}g_B}(\tilde{q}_1)\rangle 
	\end{align}
	where  $k=\text{I},\text{II}$ indicates that the correlators in CFT$_{\text{I}}$ and CFT$_{\text{II}}$ have the same form.	As earlier we have dropped the subscript $CFT_{k}^{\otimes mn}$ on the $RHS$ for brevity.
	Apart from these we would also require the following correlator which occurs in the denominator in the RHS of \cref{SRGmn0} for the effective reflected entropy
	\begin{align}\label{2ptdenf}
		G^{m}_{\mathrm{CFT_{k}}}&=\langle \sigma_{g_m}(q_0)\sigma_{g_m}(\tilde{q}_0)\rangle 
	\end{align}
	
	\subsubsection*{Phase-I}
	
	In phase-I the four point correlator above simply factorizes into product of 2 two-point correlators in the large-$c$ limit as follows
	\begin{align}
		G^{m,n}_{\mathrm{CFT_{k}}}&=\langle \sigma_{g_A}(q_0)\sigma_{g_A}(\tilde{q}_0)\rangle \langle\sigma_{g_A^{-1}g_B}(q_1)\sigma_{g_A^{-1}g_B}(\tilde{q}_1)\rangle 
	\end{align}
	From the above result and \cref{2ptdenf} we have all the correlation functions required to compute the reflected entropy in \cref{SRGmn0} in terms of the two point functions. Hence substituting them in \cref{SRGmn0}, we obtain the reflected entropy to be as follows
	\begin{align}
		S_R(A:B)=2\frac{c_\text{I}+c_\text{II}}{3}\log\bigg[\frac{\beta}{\pi \epsilon}\cosh\left(\frac{2\pi v}{\beta}\right)\bigg].
	\end{align}
Note that the above result for the reflected entropy exactly matches with twice the area of EWCS in the bulk geometry given in \cref{FTP1} upon utilizing the Brown-Henneaux relation.
	\subsubsection*{Phase-II}
	In this phase, the size of the subsystem $A$ is  considered  to be small compared to that of $B$ and in the large-c limit the four point function in \cref{Gmn4ptFTNI} receives maximum contribution from the block corresponding to the operator $\sigma_{g_B}$  ($| \sigma_{g_B}\rangle \langle\sigma_{g_B}| $ ) which leads to the following factorization\footnote{Note that this phase corresponds to the channel in which the operators $\sigma_{g_A^{-1}g_B}$ and $\sigma_{g_B}$ come close by. Since the leading operator in their OPE expansion is  $\sigma_{g_B}$, it  provides the dominant contribution in the large-c limit.} 
	\begin{align}
		G^{m,n}_{\mathrm{CFT_{k}}}&=\langle \sigma_{g_A}(q_0)\sigma_{g_A^{-1}g_B}(q_1)\sigma_{g_B}(\tilde{q}_0)\rangle \langle\sigma_{g_B}(q_0) \sigma_{g_A}(\tilde{q}_0)\sigma_{g_A^{-1}g_B}(\tilde{q}_1)\rangle \nonumber\\
		G^{m}_{\mathrm{CFT_{k}}}&=\langle \sigma_{g_m}(q_0) \sigma_{g_m}(\tilde{q}_0)\rangle 
	\end{align}
	Since the form of the three point correlators are fixed by conformal symmetry we obtain the following expression for the reflected entropy
	\begin{align}
		S_R(A:B)=2\frac{c_\text{I}+c_\text{II}}{3}\log\left[\frac{\beta\left(e^{\frac{2\pi u_1}{\beta}}-e^{\frac{2\pi u_0}{\beta}}\right)\sqrt{e^{\frac{4\pi u_0}{\beta}}+e^{\frac{4\pi u_1}{\beta}}+2e^{\frac{2\pi (u_0+u_1)}{\beta}}\cosh\left(\frac{4\pi v}{\beta}\right)}}{2\pi\epsilon\, e^{\frac{2\pi (u_0+u_1)}{\beta}}\cosh\left(\frac{2\pi v}{\beta}\right)}\right].
	\end{align}
	Once again this precisely matches with  twice the EWCS given in \cref{FTP2} verifying the holographic duality between the two.

	\subsection{After Page Time}
	
	
	We now proceed to compute the reflected entropy for phases in which the subsystem-$AB$ possesses an entanglement entropy island.
	\subsubsection*{Phase-III}
	The island phase for the entanglement entropy of the subsystem-$AB$ leads to three sub-phases for the reflected entropy (phase III, phase IV and phase V). 
	In phase III, the subsystem-$A$ is so large that the entire entanglement entropy island belongs to the reflected entropy island of $A$ as depicted in the  above figure \ref{SRADJFTCFT2}.  
	The correlator required to determine the reflected entropy  is given as
	\begin{align}\label{isSRADJFP3}
		G^{m,n}_{\mathrm{CFT_{k}}}&=\langle \sigma_{g_A}(q_0)\sigma_{g_A^{-1}g_B}(q_1)\sigma_{g_A^{-1}}(q_b)\sigma_{g_A^{-1}}(\tilde{q}_b)\sigma_{g_A}(\tilde{q}_0)\sigma_{g_A^{-1}g_B}(\tilde{q}_1)\rangle \\
		G^{m}_{\mathrm{CFT_{k}}}&=\langle \sigma_{g_m}(q_0)\sigma_{g_m^{-1}}(q_b)\sigma_{g_m^{-1}}(\tilde{q}_b)\sigma_{g_m}(\tilde{q}_0)\rangle 
	\end{align}
	In the large-c limit, each  six point  correlator in \cref{isSRADJFP3} factorizes into the product of a two point correlator of the composite operator $\sigma_{g_A^{-1}g_B}$ and a four point correlator of non-composite operators. The four point function of non-composite operators further factorizes into a product of 2 two-point correlators which leads to the following result
	\begin{align}\label{isSRADJFP352}
		G^{m,n}_{\mathrm{CFT_{k}}}&=\langle \sigma_{g_A}(q_0)\sigma_{g_A^{-1}}(q_b)\rangle \langle \sigma_{g_A^{-1}}(\tilde{q}_b)\sigma_{g_A}(\tilde{q}_0)\rangle \langle \sigma_{g_A^{-1}g_B}(q_1)\sigma_{g_A^{-1}g_B}(\tilde{q}_1)\rangle \\
		G^{m}_{\mathrm{CFT_{k}}}&=\langle \sigma_{g_m}(q_0)\sigma_{g_m^{-1}}(q_b)\rangle \langle\sigma_{g_m^{-1}}(\tilde{q}_b)\sigma_{g_m}(\tilde{q}_0)\rangle 
	\end{align}
	Since the full correlator is now expressed in terms of the two point correlators, we obtain the following expression for the reflected entropy 
	\begin{align}
		S_R(A:B)=2\frac{c_{\mathrm{I}}+c_{\mathrm{II}}}{3} \log \left[\frac{\beta}{\pi \epsilon} \cosh \left(\frac{2 \pi v}{\beta}\right)\right].
	\end{align}
	Observe that this is exactly equal to twice the EWCS in the bulk 3D geometry obtained in \cref{FTP3} which once again verifies the holographic duality between the two.
	\subsubsection*{Phase-IV}
	In phase IV, the subsystem-$A$ is so small that the entire entanglement entropy island belongs the reflected entropy island of $B$ as depicted in the above  \cref{SRADJFTCFT3}. 
	The correlation function required to determine the reflected entropy is given as
	\begin{align}\label{isSRADJFP34}
		G^{m,n}_{\mathrm{CFT_{k}}}&=\langle \sigma_{g_A}(q_0)\sigma_{g_A^{-1}g_B}(q_1)\sigma_{g_B^{-1}}(q_b)\sigma_{g_B^{-1}}(\tilde{q}_b)\sigma_{g_A}(\tilde{q}_0)\sigma_{g_A^{-1}g_B}(\tilde{q}_1)\rangle \\
		G^{m}_{\mathrm{CFT_{k}}}&=\langle \sigma_{g_m}(q_0)\sigma_{g_m^{-1}}(q_b)\sigma_{g_m^{-1}}(\tilde{q}_b)\sigma_{g_m}(\tilde{q}_0)\rangle 
	\end{align}
	In the large-c limit the above correlation functions factorizes as follows
	\begin{align}
		G^{m,n}_{\mathrm{CFT_{k}}}&=\langle \sigma_{g_A}(q_0)\sigma_{g_A^{-1}g_B}(q_1)\sigma_{g_B^{-1}}(q_b)\rangle \langle\sigma_{g_B^{-1}}(\tilde{q}_b)\sigma_{g_A}(\tilde{q}_0)\sigma_{g_A^{-1}g_B}(\tilde{q}_1)\rangle 
	\end{align}
	Utilizing the conformal transformation from the cylinder to the plane (where the form of the three point functions is known) and introducing the appropriate Weyl Factors for points on the brane, we may compute the Renyi reflected entropy by substituting the above correlators in  \cref{SRGmn0}. This leads to the following result for reflected entropy upon taking the replica limit
	\begin{align}
		S_R(A:B)=2\frac{c_{\mathrm{I}}+c_{\mathrm{II}}}{3 } \log \left[\frac{\beta}{\pi \epsilon} \frac{\cosh \left(\frac{2 \pi u_{1}}{\beta}\right)-\cosh \left(\frac{2 \pi u_{0}}{\beta}\right)}{\sinh \left(\frac{2 \pi u_{0}}{\beta}\right)}\right].
	\end{align}
	which precisely matches with twice the corresponding EWCS given in \cref{FTP4} upon using Brown Henneaux relations for the central charges of the two CFTs. 
	\subsubsection*{Phase-V}
	In this phase both the subsystems $A$ and $B$ posses their respective reflected entropy islands as depicted in the \cref{SRADJFTCFT4} above. In this case we require the following  correlators to compute the reflected entropy
	\begin{align}\label{isSRADJFbfP4}
		G^{m,n}_{\mathrm{CFT_{k}}}&=\langle \sigma_{g_A}(q_0)\sigma_{g_A^{-1}g_B}(q_1)\sigma_{g_{AB^{-1}}}(q_a)\sigma_{g_A^{-1}}(q_b)\sigma_{g_A^{-1}}(\tilde{q}_b)\sigma_{g_{AB^{-1}}}(\tilde{q}_a)\sigma_{g_A}(\tilde{q}_0)\sigma_{g_A^{-1}g_B}(\tilde{q}_1)\rangle\nonumber \\
		G^{m}_{\mathrm{CFT_{k}}}&=\langle \sigma_{g_m}(q_0)\sigma_{g_m^{-1}}(q_b)\sigma_{g_m^{-1}}(\tilde{q}_b)\sigma_{g_m}(\tilde{q}_0)\rangle 
	\end{align}
	In the large-c limit, the above correlation functions  factorize as follows
	\begin{align}\label{isSRADJFP4}
		G^{m,n}_{\mathrm{CFT_{k}}}=&\langle \sigma_{g_A}(q_0) \sigma_{g_A^{-1}}(q_b)\rangle\langle \sigma_{g_A^{-1}g_B}(q_1)\sigma_{g_{AB^{-1}}}(q_a)\rangle\nonumber\\
		&\times \langle \sigma_{g_A}(\tilde{q}_0)\sigma_{g_A^{-1}}(\tilde{q}_b)\rangle \langle \sigma_{g_{AB^{-1}}}(\tilde{q}_a)\sigma_{g_A^{-1}g_B}(\tilde{q}_1)\rangle\nonumber\\
		G^{m}_{\mathrm{CFT_{k}}}&=\langle \sigma_{g_m}(q_0)\sigma_{g_m^{-1}}(q_b)\rangle \langle\sigma_{g_m^{-1}}(\tilde{q}_b)\sigma_{g_m}(\tilde{q}_0)\rangle \nonumber\\
	\end{align}
Substituting the above correlators in \cref{SRGmn0}  and  adding the area term in \cref{area-term} to it, followed by an extremization similar to that in \cite{Anous:2022wqh}, we obtain the following expression for the reflected entropy
	\begin{align}
		S_R(A:B)=2\frac{c_{\mathrm{I}}+c_{\mathrm{II}}}{3} \log \left[\frac{2\beta}{\pi \epsilon} \sinh \left(\frac{2 \pi u_1}{\beta}\right)\right]+2\Phi_0.
	\end{align}
	Utilizing \cref{Phi0-delta} it is straightforward to verify that the above expression for the reflected entropy is exactly twice the large tension limit of the EWCS given in \cref{FTP5} which validates the holographic duality between the two.

	\section{Summary}\label{sec:summary}
	To summarize, in this article we have determined the reflected entropy of various mixed state configurations in a two dimensional holographic  ICFT$_2$  from the  dual entanglement wedge cross section (EWCS) in the three dimensional bulk which involves two AdS$_3$   geometries glued smoothly along a thin interface brane. Following this, we obtained the island contributions to the  reflected entropy for the required mixed states from  the two dimensional effective semi-classical description. We demonstrated that the results obtained from the island formula match precisely with the 3d bulk results in the large tension limit. 
	
	We started our computations by obtaining  the RT surfaces (and the entanglement wedges) corresponding to various phases of the holographic entanglement entropy for the mixed state configurations of adjacent and disjoint intervals in vacuum state of an  ICFT$_2$.  Following this, we computed the EWCS dual to the reflected entropy in  each phase of the  entanglement wedge.  We  showed that for each phase of the entanglement wedge there are multiple phases of the EWCS which we evaluated from the bulk AdS$_3$   geometry. Subsequently, we evaluated the island contribution to the reflected entropy in the effective two dimensional semi-classical theory utilizing the replica technique and established that  the results obtained match exactly with  twice the EWCS in the bulk $AdS_3$ geometry in the large tension limit of the interface brane.
	Quite interestingly, we showed that there are numerous phases of EWCS where the corresponding RT surface  crosses the interface brane and returns back to the original geometry. Such a phase for the entanglement entropy was shown to correspond to certain induced islands in the two dimensional effective field theory and are described by novel replica wormhole saddles \cite{Afrasiar:2023nir}. We established that there are several phases of  the EWCS for each  double crossing RT surface and obtained the corresponding expressions. We also showed that in semi-classical picture these novel phases result in induced islands for the reflected entropy. 
	
	Subsequently, we considered a holographic  ICFT$_2$  in a Thermofield double state which is dual to an eternal black hole geometry  with a thin interface brane. We determined the reflected entropy for the mixed state configuration involving two adjacent intervals (with one of the intervals being semi-infinite) from the corresponding bulk dual involving the EWCS in the 3d geometry. Following this, we obtained the island contributions to the reflected entropy from the effective two dimensional theory and demonstrated that the results determined precisely match with the corresponding bulk results in the large tension limit. Quite interestingly, the reflected entropy exhibits a much richer phase structure than the entanglement entropy. Finally we obtained the analogues of the Page curves for the reflected entropy of the mixed state in question. 
	
	Several possible interesting directions may be explored in the near future. It would be exciting to explore the rich   phase structure of entanglement negativity in the AdS/ICFT framework utilizing the holographic proposals described in \cite{Chaturvedi:2016rcn,Malvimat:2018txq,Kudler-Flam:2018qjo} and the corresponding island formulation \cite{KumarBasak:2020ams,KumarBasak:2021rrx,Basu:2022reu,Shao:2022gpg}. It would also be interesting to compare the behaviour of the reflected entropy to other quantum information theoretic measures with holographic description such as the entanglement of purification \cite{Takayanagi:2017knl}, the odd entanglement entropy \cite{Tamaoka:2018ned}, and the balanced partial entanglement \cite{Wen:2021qgx,Basu:2023wmv}. Furthermore it would be fascinating to explore how the rich phase structure of the reflected entropy and other such measures are useful in the information recovery from the Hawking radiation emitted during the black hole evaporation process. We hope to come back to these exciting issues soon.

	\section*{Acknowledgements}
	
	The research work of JKB supported by the National Science and Technology Council of Taiwan with the grant 112-2636-M-110-006. The work of VM was supported by the NRF grant funded by the Korea government (MSIT) (No. 2022R1A2C1003182) and by the Brain Pool program funded by the Ministry of Science and ICT through the National Research Foundation of Korea (RS-2023-00261799). HP acknowledges the support of this work by the NCTS, Taiwan. The work of GS is partially supported by the Dr Jagmohan Garg Chair Professor position at the Indian Institute of Technology, Kanpur.

	\bibliographystyle{JHEP}
	
	\bibliography{SRICFT}
	
\end{document}